\documentclass[3p]{elsarticle}
\usepackage{array}
\usepackage{graphicx}
\usepackage[table]{xcolor}
\usepackage{subcaption}
\usepackage{wrapfig}
\usepackage{arydshln}
\usepackage{ulem}
\usepackage{textgreek}
\usepackage{pgfplots}
\usepackage{commath}
\usepackage{multirow}
\usepackage{makecell}
\usepackage[most, breakable]{tcolorbox}
\usepackage{times}
\usepackage{amsfonts}
\usepackage{amssymb}
\usepackage[pdfauthor={Federico Quin, Danny Weyns, Omid Gheibi}]{hyperref}
\usepackage{todonotes}
\usepackage{xfrac}
\usepackage{hhline}

\def\cleanversion{1}

\usetikzlibrary{calc}
\pgfplotsset{compat=1.13}

\definecolor{darkgreen}{RGB}{21,176,26}
\definecolor{cRed}{rgb}{0.648, 0.109, 0}
\definecolor{cGreen}{rgb}{0.0, 0.496, 0.05}
\definecolor{hypothesis-color}{RGB}{136, 182, 209}

\newcommand{\ex}[1]{\noindent \emph{Example:} #1}

\newcommand{\exampleHighlight}[1]{\textcolor{blue}{#1}}
\newcommand{\techniqueName}{Machine Learning To Adaptation Space Reduction}
\newcommand{\techniqueNameShort}{ML2ASR+}
\newcommand{\sbs}{Service-Based System} 

\newcommand{\MLM}{Machine Learning Module}
\newcommand{\MLMi}{\textit{\MLM{}}}

\newcolumntype{P}[1]{>{\centering\arraybackslash}p{#1}}
\newcolumntype{M}[1]{>{\centering\arraybackslash}m{#1}}
\newcolumntype{N}{@{}m{0pt}@{}}

\newcommand{\specialcell}[2][c]{%
    \begin{tabular}[#1]{@{}c@{}}#2\end{tabular}}

\newcommand{\thresholdtext}[2]{$\threshold[ #2]{#1}$}
\newcommand{\setpointtext}[2]{$\setpoint[ #2]{#1}$}
\newcommand{\optimizationtext}[2]{$\optimization[ #2]{#1}$}

\newcommand{\threshold}[2][]{\mathcal{T}_{#2}^{\text{#1}}}
\newcommand{\setpoint}[2][]{\mathcal{S}_{#2}^{\text{#1}}}
\newcommand{\optimization}[2][]{\mathcal{O}_{#2}^{\text{#1}}}

\newcommand{\interval}[2]{[{\footnotesize #1, #2}]}

\newcommand{\dlaser}[0]{DLASeR}

\def\arrow{$\,\to\,$}
\newcommand{\abbrev}[2]{#1\arrow #2}

\newcommand{\abbrevScenarios}{\abbrev{S1}{Scenario 1}, \abbrev{S2}{Scenario 2}}

\newcommand{\abbrevML}{\abbrev{F1}{F1-score}, \abbrev{MCC}{Matthews Correlation Coefficient}, \abbrev{R2}{R2-score}, \abbrev{MSE}{Mean Squared Error}, \abbrev{MAE}{Median Average Error}, \abbrev{ME}{Maximum Error}}

\newcommand{\abbrevDeltaIoT}{\abbrev{Pl}{Packet loss}, \abbrev{La}{Latency}, \abbrev{Ec}{Energy Consumption}, \abbrevScenarios{}}

\newcommand{\abbrevSbs}{\abbrev{Fr}{Failure rate}, \abbrev{Rt}{Response time}, \abbrev{C}{Cost}, \abbrevScenarios{}}

\newcommand{\scenarioTable}[3][0pt]{\multirow{#2}{*}[#1]{\rotatebox[origin=c]{45}{\textbf{#3}}}}

\definecolor{scrap}{RGB}{255,0,0}

\ifx\cleanversion\undefined

    \definecolor{review}{RGB}{200, 22, 21}
    \newcommand{\review}[1]{{\color{review}#1}}
    \newcommand{\ctodo}[1]{\todo[inline, color=darkgreen]{#1}}
    \newcommand{\remove}[1]{{\textcolor{scrap}{\sout{#1}}}}

    \newenvironment{removeEnv}
    {\color{scrap}}
    {\color{black}}
\else
    \definecolor{review}{RGB}{0,0,0}
    \newcommand{\review}[1]{#1}
    \newcommand{\ctodo}[1]{}
    \newcommand{\remove}[1]{}
    
    \usepackage{verbatim}

\fi

\graphicspath{{figures/}}

\begin{document}

\begin{frontmatter}
\title{Reducing Large Adaptation Spaces in Self-Adaptive Systems \\Using Machine Learning}

\author[1]{Federico Quin\corref{cor1}}%
\ead{federico.quin@kuleuven.be}

\author[1,2]{Danny Weyns}
\ead{danny.weyns@kuleuven.be}

\author[1]{Omid Gheibi}
\ead{omid.gheibi@kuleuven.be}

\cortext[cor1]{Corresponding author}
\address[1]{Katholieke Universiteit Leuven, Leuven, 3000 Belgium}
\address[2]{Linnaeus University, V\"axj\"o, 351 06 Sweden}




\begin{abstract}
	Modern software systems often have to cope with uncertain operation conditions, such as changing workloads or fluctuating interference in a wireless network.
	To ensure that these systems meet their goals these uncertainties have to be mitigated. One approach to realize this is self-adaptation that equips a system with a feedback loop. The feedback loop implements four core functions -- monitor, analyze, plan, and execute -- that share knowledge in the form of runtime models.
	For systems with a large number of adaptation options, i.e., large adaptation spaces, deciding which option to select for adaptation may be time consuming or even infeasible within the available time window to make an adaptation decision. This is particularly the case when rigorous analysis techniques are used to select adaptation options, such as formal verification at runtime, which is widely adopted.
	One technique to deal with the analysis of a large number of adaptation options is reducing the adaptation space using machine learning. State of the art has showed the effectiveness of this technique, yet, a systematic solution that is able to handle different types of goals is lacking. 
	In this paper, we present \techniqueNameShort{}, short for Machine Learning to Adaptation Space Reduction Plus. Central to \techniqueNameShort{} is a configurable machine learning pipeline that supports effective analysis of large adaptation spaces for threshold, optimization, and setpoint goals. 
	We evaluate \techniqueNameShort{} for two applications with different sizes of adaptation spaces: an Internet-of-Things application and a service-based system. The results demonstrate that \techniqueNameShort{} can be applied to deal with different types of goals and is able to reduce the adaptation space and hence the time to make adaptation decisions with over 90\,\%, with negligible effect on the realization of the adaptation goals.
\end{abstract}




\end{frontmatter}

\section{Introduction}\label{sec:introduction}

Engineering modern software systems is complex. 
One of the important factors that underlies this complexity is the dynamic and complex environment in which systems need to operate, requiring the systems to deal with uncertain conditions that are often difficult to predict before they are in operation~\cite{Ghezzi2011}. These uncertainties may jeopardize the system goals. Network interference can for example affect the availability of the system if not properly dealt with. 

To mitigate such uncertainties, self-adaptation has become prevalent in modern software systems~\cite{Cheng2009,Salehie2009,weyns2020book}. Self-adaptation enhances a software system with a feedback loop mechanism that monitors the system and its environment, resolves the uncertainties, and adapts the system to maintain its goals, or degrades gracefully if necessary. Hence, self-adaptive systems consider system goals as first-class runtime entities; we refer to these goals as \textit{adaptation goals}. Adaptation goals commonly refer to quality properties of the system~\cite{Weyns2012}. 

In this paper, we apply architecture-based adaptation~\cite{Oriezy1999,Rainbow,KramerMagee,FORMS,MAHDAVIHEZAVEHI20171}, where the feedback loop implements four functions: Monitor-Analyze-Plan-Execute (MAPE in short)~\cite{Kephart}. The MAPE functions are centered around Knowledge that typically include various forms of runtime models~\cite{Blair2009}, such as architectural models of the managed system and environment, goal models, and parameterized quality models that allow predicting qualities of different system configurations. We focus on uncertainties that can be represented as parameters of runtime models, e.g., stochastic automata or Markov models. The values of the uncertainty parameters are updated by the monitor function that monitors the system and its environment. We consider three types of adaptation goals: \textit{threshold goals} that require a system to keep a system property above/below a given threshold, \textit{optimization goals} that require a system to minimize or maximize a system property, and \textit{setpoint goals} that require a system to keep a system property at a given value or as close as possible to it. An example of a threshold goal for a client-server system is to keep the failure rate of service invocations below a given threshold, an example of an optimization goal is to minimize the cost of operation, and an example of a setpoint goal is to keep the response time of service invocations at a required level. 

Our particular focus is on the analysis function of the MAPE loop that (1) determines whether the current configuration complies with the adaptation goals, and if this is not the case, (2) predicts the qualities of alternative configurations. An alternative configuration is a configuration that can be reached from the current configuration by applying one or more adaptation actions. We refer to the alternative configurations as \textit{adaptation options}, and the set of all  adaptation options as the \textit{adaptation space}. 
A common technique used to analyze the adaptation space is formal modeling and verification. Formal models represent the system and its environment from the angle of one or more quality properties. These quality models are parameterized. One set of parameters allow the instantiation of the models for a particular configuration of the system. Another set of parameters represent uncertainties that are instantiated based on the actual conditions of the system. During analysis the parameters of the quality models are instantiated. Commonly used analysis techniques are model checking~\cite{Calinescu2010,Camara2016SMGs} and runtime simulation~\cite{ActivFORMS,Weyns2016,Weyns2022}. Based on the analysis results, a decision can then be made to adapt the system compliant with the quality goals. Recently, there is an increasing use of machine learning techniques to support the adaptation functions~\cite{Gheibi2021}.  

For systems with a limited number of adaptation options, i.e., small adaptation spaces, the analysis can be done fairly quickly ensuring that adaptation decisions are made within the available time frame to handle the dynamics of the system properly. However, for larger and more complex self-adaptive systems, the time required for analysis may dramatically increase and formal assessment of the whole adaptation space may not be feasible in such situations.

Different techniques have been proposed to deal with the problem of analyzing large adaptation spaces. E.g., Cheng et al.~\cite{Cheng2013} applied search-based software engineering techniques to generate and analyze models of dynamically adaptive systems in order to deal with uncertainties both at development time and runtime. Our particular focus in this paper is on a conceptually different technique that relies on machine learning to support the reduction of  adaptation spaces, see e.g.,~\cite{Quin2019,VanDerDonckt2020,Diallo2021}. While promising, current approaches do not provide a systematic solution with first-class support for reducing large adaptation spaces during operation that is able to handle different types of goals. 

This paper contributes \techniqueNameShort{}, short for ``Machine Learning to Adaptation Space Reduction Plus'', a novel approach for reducing large adaptation spaces.\footnote{An initial version of the approach that was limited to only one type of adaptation goal was denoted ML2ASR~\cite{Quin2019}. The ``+'' emphasizes that \techniqueNameShort{} significantly extends ML2ASR. We elaborate on this in the state of the art overview in Section~\ref{sec:related-work}.} \techniqueNameShort{} relies on classic supervised machine learning techniques, particularly classification and regression. We evaluate \techniqueNameShort{} on two self-adaptive systems in distinct domains with varying sizes of adaptation spaces. We compare the approach with: a reference approach that exhaustively analyzes the whole adaptation space, 
\review{and a state of the art learning-based approach that we developed in previous work~\cite{VanDerDonckt2020}, called \dlaser{}, which exploits deep neural networks to achieve adaptation space reduction. In addition, we perform a sanity check where we compare \techniqueNameShort{} with an approach that randomly selects a subset of adaptation options in each adaptation cycle.}


The remainder of this paper is structured as follows. Section~\ref{sec:related-work} presents the state of the art and pinpoints the problem we tackle in this paper. In  Section~\ref{sec:background}, we explain the model we use for self-adaptation in this paper, we elaborate on the different types of adaptation goals, and we introduce a running example. Section~\ref{sec:approach} then describes the core contribution:  \techniqueNameShort{}, with its runtime architecture and workflow. 
Section~\ref{sec:metrics} explains the metrics that we use for evaluating \techniqueNameShort{}. 
In Section~\ref{sec:evaluation}, we evaluate \techniqueNameShort{} for two application domains. Section~\ref{sec:discussion} elaborates on the results, presents insights, and discusses threats to validity. Finally, we wrap up and conclude in Section~\ref{sec:conclusions}.

\section{State of the Art and Problem Description}\label{sec:related-work}

We have divided the state of the art into three main areas of research. For each area, we summarize a number of representative efforts and we conclude with the open problems in the area. From this analysis, we pinpoint the research problem we tackle in this paper. 

\subsection{Machine Learning to Support the Analysis of Large Adaptation Spaces}

We start with approaches that apply machine learning to deal with the analysis of large adaptation spaces. 

The FUSION framework learns the impact of adaptation decisions on the system's goals~\cite{elkhodary2010fusion}. The approach utilizes M5 decision trees to learn the utility functions that are associated with the qualities of the system. The results show a significant improvement in analysis. FUSION targets the feature selection space, focusing on proactive latency-aware adaptations relying on a separate model for each utility. 
%
Chen et al.~\cite{chen2016self} study feature selection and show that different learning algorithms perform significantly different depending on the types of quality of service attributes considered and the way they fluctuate. The work is centered on an adaptive multi-learners technique that  dynamically selects the best learning algorithms at runtime. The focus of this work is also on features instead of adaptation options.  
Metzger et al.~\cite{Metzger2019} apply 
online learning to explore the adaptation space of self-adaptive systems using feature models with an emphasis on the adaptation and evolution of adaptation rules.

Jamshidi et al.~\cite{jamshidi2019machine} present an approach that learns a set of Pareto optimal configurations offline that are then used at runtime to generate adaptation plans. The approach reduces adaptation spaces, while the system can still apply model checking to quantitatively reason about adaptation decisions. 
%
Camara et al.~\cite{camara2020quantitative}
use reinforcement learning to select an adaptation pattern relying on two long short-term memory (LSTM) deep learning models. The focus is on the use of runtime quantitative verification, with support for threshold goals. %
Thallium exploits a combination of automated formal modeling techniques to significantly reduce the number of states that need to be considered with each adaptation decision~\cite{stevens2020reducing}. Thallium addresses the adaptation state explosion by applying utility bounds analysis.
Diallo et al.~\cite{Diallo2021} present a framework consisting of a MAPE-K feedback loop with an explainable AI module to tackle the issue reducing adaptation spaces. Their framework leverages convolutional neural networks to efficiently reduce adaptation spaces, alongside using explainable AI to build trust in the system.   

In our initial work~\cite{Quin2019} we applied classification and regression to reduce large adaptation spaces. The work also only considered threshold goals. In~\cite{VanDerDonckt2020}, we investigated the use of deep learning to reduce the adaptation space of self-adaptive systems. That work focused on handling threshold and optimization goals only. 
\vspace{5pt}\\
\textit{Open problems.} Several approaches that apply machine learning to enhance the runtime analysis of self-adaptive systems look at a coarse-grained level of system features rather than a fine-grained level of adaptation options. The approaches that look at the reduction of large adaptation spaces propose solutions that inherently mix the reduction of the adaptation space with the way analysis is performed, while other approaches (including our own earlier work) only consider specific types of adaptation goals. In conclusion: existing approaches in this area do not provide explicit support for adaptation space reduction, or they cover only specific types of adaptation goals.  

\subsection{Reinforcement Learning to Support Decision-making in Self-Adaptation}

We look now at reinforcement learning techniques used to support decision-making in self-adaptation.

Porter et al.~\cite{Porter2016} study the dynamic composition of software elements using a reinforcement learning algorithm, covering the analysis and planning stages in the self-adaptation process. The approach reduces the adaptation space to a single option, hence integrating adaptation space reduction and decision-making. 
%
Idziak et al.~\cite{Idziak2014} study different machine learning algorithms to deal with the so called virtual machine placement problem. These algorithms similarly take over the analysis and planning stages of the self-adaptation process. 
Lui et al.~\cite{Liu2018} use a reinforcement learning algorithm to improve resource efficiency in autonomous electrified vehicles. Similarly to the previous two works, the approach reduces the adaptation space to a single option that is used for decision-making.
Bu et al.~\cite{Bu2013} and Metzger et al.~\cite{Metzger2020} propose strategies to explore the adaptation options in reinforcement learning algorithms. 
\vspace{5pt}\\
\textit{Open problems.} While relying on different learning techniques compared to the approaches discussed above, the approaches proposed in this area also inherently integrate the reduction of adaptation spaces with the decision-making to select the best adaptation options for the goals at hand. An advantage of relying on reinforcement learning to realize this integration is that it does not require a (formal) model of the system, which may be a benefit if creating such a model is problematic. In conclusion: the proposed approaches do not support a separation of concerns between an explicit and tune-able reduction of adaptation spaces and the decision-making of selecting the best option.

\subsection{Efficient Analysis in Self-Adaptive Systems}

A number of approaches have been proposed to enhance the efficiency of analysis in self-adaptive systems. 

%



Filieri et al.~\cite{Filieri2011} propose an approach to generate a static set of expressions from a reliability model with a set of requirements. By using these expressions more efficient analysis is possible at runtime. That approach targets formal models based on PCTL (Probabilistic Computation Tree Logic).  
%
Calinescu et al.~\cite{Calinescu2012} combine compositional verification with model checking to effectively adapt large-scale systems. The authors employ assume-guarantee reasoning to reduce the cost of analyzing system properties, compared to infeasible exhaustive model checking approaches. 
%
Gerasimou et al.~\cite{Gerasimou2014} explore caching, lookahead, and nearly-optimal reconfiguration techniques to optimize the response time and overhead of Runtime Quantitative Verification to enhance scalability. 

Ghahremani et al.~\cite{Ghahremani2017} look at ways of reducing the cost of realizing self-adaptation in self-healing systems by combining utility-driven approaches with rule-based adaptation. 
Moreno et al.~\cite{moreno2018flexible}  present an approach for proactive latency-aware adaptation that relies on stochastic dynamic programming to enable more efficient decision-making. Experimental results show that this approach is close to an order of magnitude faster than runtime probabilistic model checking to make adaptation decisions, while preserving the same effectiveness. 

El-Kassabi et al.~\cite{ElKassabi2019} use a deep neural network to support proactive system adaptation by providing predictions of cloud resource usage. The predictions enable the suggestion of adaptation decisions to anticipate future quality of service violations. 
Di Sanzo et al.~\cite{DiSanzo2015} equip a client-server application with a framework that provides proactive management of the application. The framework exploits a multitude of machine learning methods such as linear regression and support vector machines to build and use failure prediction models at runtime. The predictions are then used to proactively adapt the system before failures take place. 
Ghahremani et al.~\cite{Ghahremani2018} evaluate machine learning algorithms for the prediction of system utility in adaptive systems, without relying on detailed system information. 
\vspace{5pt}\\
\textit{Open problems.} The approaches proposed into this area can be structured in three groups. A first group focuses on improving the verification process. These approaches do not deal with the problem of adaptation space reduction but can be combined with an approach for adaptation space reduction. A second group focuses on alternative solutions to enhance the efficiency of decision-making in self-adaptive systems. Yet, as with other related approaches discussed above, these approaches inherently integrate an implicit reduction of adaptation spaces with the decision-making to select an adaptation option. A third group applies machine learning techniques to make predictions of qualities and other properties to support the decision-making process. These solutions are complementary to the problem of adaptation space reduction. In conclusion: two groups of related efforts do not solve the problem of adaptation space reduction, but can be combined with approaches to reduce the adaptation space in order to enhance the efficiency of analysis; another group of related efforts do not separate the reduction of adaptation spaces with decision-making.   

\subsection{Research Problem}\label{sec:problem-description}

The analysis of the related work highlights the
need for systematic approaches that provide explicit first-class support for adaptation space reduction while covering different types of goals. To that end, we formulate the following research question that we tackle in this work: 

\noindent 
\begin{quote}
   \textit{How can machine learning be used to reduce large adaptation spaces of self-adaptive systems with different types of adaptation goals to perform more efficient analysis without compromising the goals?} 
\end{quote}

To answer the research question, we propose \techniqueNameShort{}, a novel approach for adaptation space reduction. 
Leveraging on classification and regression, \techniqueNameShort{} offers a modular approach for efficient reduction of adaptation spaces for self-adaptive systems with threshold, optimization, and setpoint goals. 
We translate the research question to six requirements for \techniqueNameShort{} that serve as drivers for devising the solution and evaluating it.

The first four requirements -- reusability, automatic operation at runtime, modularity adaptation goals, and granularity of adaptation space reduction -- are of a qualitative nature. 
The last two requirements -- negligible utility penalty and efficiency -- are of a quantitative nature.  
\vspace{5pt}\\
\textit{Reusability.} As a first requirement, the solution should be reusable, i.e., the solution should offer distinct functionalities and modules that can be instantiated and applied across application domains. 
We evaluate this requirement by demonstrating that the proposed solution can be applied to applications in two different domains.
\vspace{5pt}\\
\textit{Automatic Operation at Runtime.}
As a second desirable requirement, we want the solution to operate at runtime without human involvement. 
We evaluate this requirement by demonstrating that the proposed solution fully automatically reduces adaptation spaces at runtime for different application domains. 
\vspace{5pt}\\
\textit{Modularity Adaptation Goals.}
As a third requirement, we want the solution to be able to handle different types of adaptation goals. The approach should be able to handle independent types of adaptation goals, as well as a combination of different types of goals in one system. 
We evaluate this requirement by demonstrating that the proposed solution \mbox{can be applied to instances of the same applications with different types and combinations of adaptation goals.}
\vspace{5pt}\\
\textit{Granularity of Adaptation Space Reduction.}
As a fourth requirement, we want our solution to have the option to specify the granularity of adaptation space reduction, i.e., the degree to which the solution reduces the adaptation space. Granularity applies to optimization and/or setpoint goals, enabling to determine which adaptation options to include based on well-defined criteria. 
E.g., for a setting with a setpoint goal, we may require the solution to find all the adaptation options within a given window around the setpoint value. 
Differentiating the granularity offers flexibility when the available adaptation time may be different under different conditions. 
We evaluate this requirement by demonstrating \mbox{that the proposed solution can be applied for different levels of granularity of adaptation space reduction.}
\vspace{5pt}\\
\textit{Negligible Utility Penalty.}
As a fifth requirement, we desire that the solution reduces the adaptation space with little or no penalty on the quality properties that are the subject of adaptation compared to an ideal solution where no adaptation space reduction is applied. Utility denotes here the effect on the quality properties due to the adaptation decisions made by using learning. 
We evaluate this requirement by comparing the differences in mean values of the relevant quality properties over time with and without learning. 
%
Depending on the type of goal (elaborated in Section~\ref{sec:background}) we either compare the satisfaction of the goal or compare the difference of the quality tied to that specific goal.
We provide a concrete metric for the evaluation of utility penalty in Section~\ref{sec:metrics}.
\vspace{5pt}\\
\textit{Efficiency.}
As a sixth and final requirement, the solution should be efficient, i.e., the adaptation space should be reduced such that the analysis can be performed within the time window available to make adaptation decisions. We evaluate this requirement by demonstrating that the proposed solution effectively reduces the adaptation space in two different domains. We use three metrics to judge the efficiency of the adaptation space reduction: (1) the Average Adaptation Space Reduction (AASR in short) that compares the average number of adaptation options selected by learning over multiple adaptation cycles with the average of the total number of adaptation options over these adaptation cycles; (2) the total percentage of time saved as a result of the space reduction; and (3) the percentage of overhead in time of \techniqueNameShort{} due to learning compared to the verification time required to verify the reduced adaptation space. 
We provide concrete metrics for the evaluation of efficiency in Section~\ref{sec:metrics}.

\section{Model of Self-Adaptive System with Adaptation Goals and Running Example}\label{sec:background}

We briefly outline the model for self-adaptation that we use in this research. Then, we give a simple example of a self-adaptive system that we use as a running case in the paper. Finally, we explain different types of adaptation goals.

\subsection{Model of Self-Adaptive System}

\begin{figure}[!thb]
    \centering
    \includegraphics[width=0.47\linewidth]{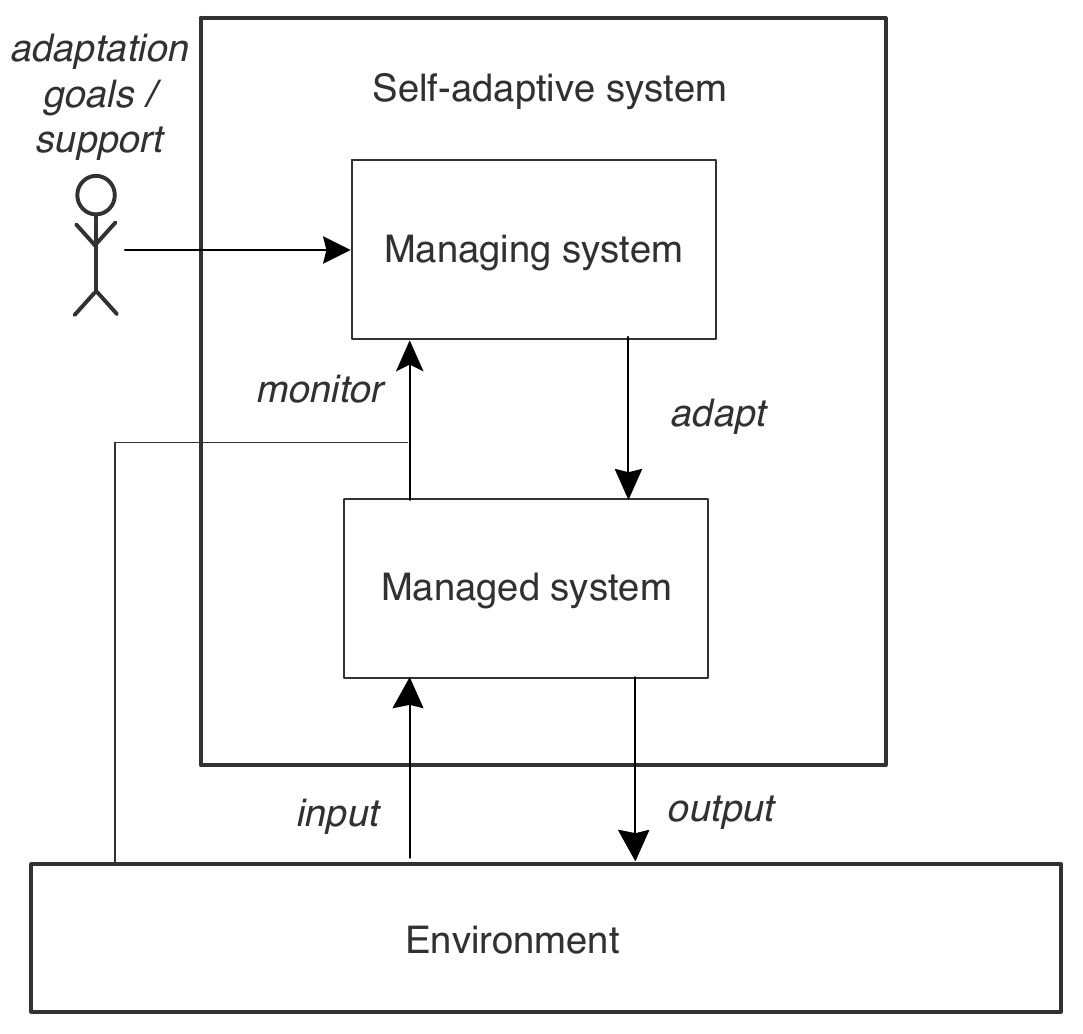}
    \caption{Model of a self-adaptive system}
    \label{fig:overview_systems}
\end{figure}

Figure~\ref{fig:overview_systems} shows a high-level model of a self-adaptive system as we follow in this paper, leveraging on~\cite{weyns2020book}.  

A self-adaptive system consists of two parts: a \textit{managed system} and a \textit{managing system}. The managed system can be any regular software-intensive system or a part of it. Hence, the managed system may refer to an entire system, a subsystem, one or more components, just a particular feature of a larger system, infrastructure or resources used by a system, etc. Other terminology used to refer to self-adaptive system are auto-tuned system, elastic system, controlled system, context controlled system, autonomic system, among others. 

The managed system takes \textit{input} from an \textit{environment} and produces \textit{output} to the environment. While the managed system can be controlled, the elements in the environment cannot. The environment may include other software systems, hardware, communication networks, users, the operating context, and so forth.  The managing system acts upon the managed system with a particular purpose, for instance to improve its performance when operating conditions change or to deal with errors that may suddenly appear. The purpose is provided by stakeholders in the form of \textit{adaptation goals}. The managing system \textit{monitors} the managed system and/or its environment during operation, resolves uncertainties, and based on the adaptation goals \textit{adapts} the managed system or parts of it when needed. 
A common approach to realize the managing system is by means of combining four basic functions: Monitor-Analyze-Plan-Execute that share a common Knowledge, which is often referred to as MAPE-K or MAPE in short~\cite{Kephart}. The types of adaptations of the managed system may range from adjusting parameter settings, up to architectural re-configurations. Hence, the managed system needs to provide the necessary support to be monitor-able and adapt-able. Operators or other stakeholders may \textit{support} the managing system in its tasks, but this is optional. 

\subsection{Running Example}

We introduce a small example of a self-adaptive system that we use as a running case to illustrate \techniqueNameShort{}. The managed system in this example is a simple service-based system that handles service requests of clients through the invocation of a series of services. These services are deployed on two machines named \textbf{M1} and \textbf{M2}. The system has to deal with two uncertainties: fluctuations in network bandwidth and the workload of both machines respectively. These uncertainties affect three qualities that form the adaptation goals: the \textit{failure rate}, \textit{response time}, and the \textit{cost} of service requests. To make sure that the qualities comply with the service level agreements of users, the system is equipped with a managing system. This managing system realizes a feedback loop that monitors the service system and has the ability to adapt the distribution of service requests between \textbf{M1} and \textbf{M2}.

\subsection{Adaptation Goals}

One of the requirements and a distinct feature of \techniqueNameShort{} is support for different types of adaptation goals. We start with describing the types of adaptation goals \techniqueNameShort{} supports one by one, and then we explain how multiple types of goals can be combined. We illustrate the goals with the running example.

\subsubsection{Threshold Goals} 
The first type of adaptation goal that we cover in this work is a threshold goal. A threshold goal imposes a restriction on one of the system's quality properties in the form of a threshold value that should not be exceeded. Exceeded in this context can refer to either an upper bound value that the quality property should not cross, or a lower bound value that acts as a minimum requirement for the quality property. We define the satisfaction of a threshold goal $\threshold{} \in \mathbb{T}$ with a threshold value $\bar{x}$ for any value of the quality property $q$ (or quality value in short) as follows:

\begin{equation}
    \threshold{< \bar{x}}(q) = \left\{
    \begin{array}{@{}l@{\thinspace}l}
         True &: q < \bar{x}  \\
         False &: q \geqslant \bar{x} \\ 
    \end{array}
    \right.
\end{equation}

\begin{equation}
    \threshold{> \bar{x}}(q) = \left\{
    \begin{array}{@{}l@{\thinspace}l}
         True &: q > \bar{x}  \\
         False &: q \leqslant \bar{x} \\ 
    \end{array}
    \right.
\end{equation}

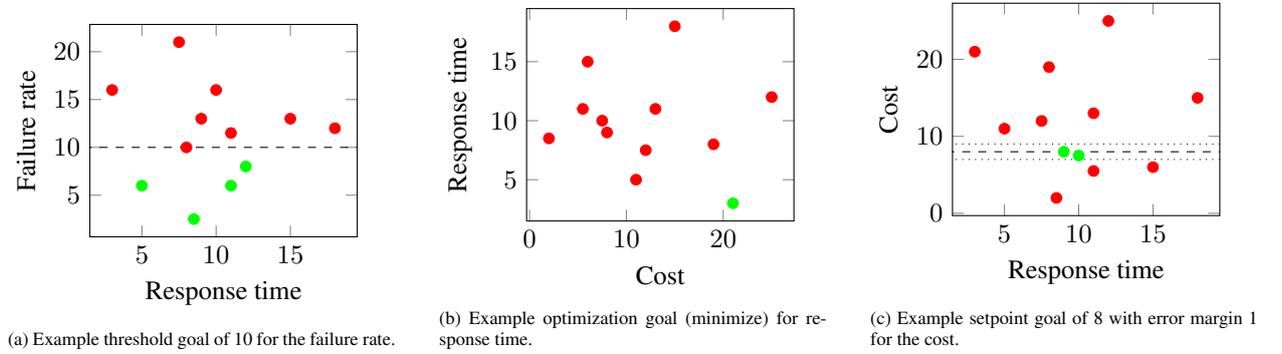
\begin{figure}[!bht]
\begin{subfigure}{.31\linewidth}
	\begin{tikzpicture}
		\begin{axis}[%
		scatter/classes={%
			a={green},%
			b={red}},
		xlabel={Response time},
		ylabel={Failure rate},
		width=1\linewidth]
		\addplot[scatter,only marks,%
			scatter src=explicit symbolic]%
		table[meta=label] {
			x     y      label
			5     6      a
			3     16     b
			9     13     b
			8     10     b
			7.5   21     b
			8.5   2.5    a
			11    6      a
			15    13     b
			18    12     b
			12    8      a
			11    11.5   b
			10    16     b
		};
		\draw[black, dashed] (0, 10) -- (30, 10);
		\end{axis}
	\end{tikzpicture}
	\caption{Example threshold goal of 10 for the failure rate.}
	\label{fig:threshold_goal}
\end{subfigure}\hfill%
\begin{subfigure}{.31\linewidth}
	\begin{tikzpicture}
		\begin{axis}[
		scatter/classes={
			a={green},
			b={red}},
		xlabel={Cost},
		ylabel={Response time},
		width=1\linewidth]
		\addplot[scatter,only marks,
			scatter src=explicit symbolic]
		table[meta=label] {
			x     y      label
			11 	  5      b
			21 	  3      a
			8 	  9      b
			19 	  8      b
			12 	  7.5    b
			2 	  8.5    b
			13 	  11     b
			6 	  15     b
			15 	  18     b
			25 	  12     b
			5.5   11     b
			7.5   10     b
		};
		\end{axis}
	\end{tikzpicture}
	\caption{Example optimization goal (minimize) for response time.}
	\label{fig:optimization_goal}
\end{subfigure}\hfill%
\begin{subfigure}{.31\linewidth}
	\begin{tikzpicture}
		\begin{axis}[%
		scatter/classes={%
			a={green},%
			b={red}},
		xlabel={Response time},
		ylabel={Cost},
		width=1\linewidth]
		\addplot[scatter,only marks,%
			scatter src=explicit symbolic]%
		table[meta=label] {
			x     y      label
			5     11     b
			3     21     b
			9     8      a
			8     19     b
			7.5   12     b
			8.5   2      b
			11    13     b
			15    6      b
			18    15     b
			12    25     b
			11    5.5    b
			10    7.5    a
		};
		\draw[black, dashed] (0, 8) -- (20, 8);
		\draw[black, dotted] (0, 7) -- (20, 7);
		\draw[black, dotted] (0, 9) -- (20, 9);
		\end{axis}
	\end{tikzpicture}
	\caption{Example setpoint goal of 8 with error margin 1 for the cost.}
	\label{fig:setpoint_goal}
\end{subfigure}
\caption{Example scenarios for each type of adaptation goal.}
\label{fig:adaptation_goals}
\end{figure}

A threshold goal allows a self-adaptive system to categorize adaptation options in two distinct classes: compliant with the threshold goal or in violation of the threshold goal. Hence, threshold goals form a perfect candidate for classification of adaptation options, a classic supervised machine learning technique.
\vspace{5pt}\\
\ex{Applied to the running example, we can define a threshold goal for the system to keep the failure rate below a given time percentage, say 10\%, as shown in Figure~\ref{fig:threshold_goal}. In this case, the set of quality values $q$ that satisfy the threshold goal, i.e.,  $\threshold{<10\%}(q) = True$, correspond to classification class 1, while the set of quality values $q$ that do not satisfy the threshold goal, i.e., $\threshold{<10\%}(q) = False$, correspond to classification class 0.}

\subsubsection{Optimization Goals} 
The second type of adaptation goal that we cover is an optimization goal. As the name suggests, an optimization goal aims to optimize a quality property of the system, which can be either maximize or minimize the value of the quality property. We define the satisfaction of an optimization goal $\optimization{} \in \mathbb{O}$ for any quality value $q$ as follows:

\begin{equation}
    \optimization{min}(q) = \left\{
    \begin{array}{@{}l@{\thinspace}l}
         True &: q = min(\{q_1, q_2, ..., q_n\})  \\
         False &: \textrm{otherwise} \\ 
    \end{array}
    \right.
\end{equation}
\begin{equation}
    \optimization{max}(q) = \left\{
    \begin{array}{@{}l@{\thinspace}l}
         True &: q = max(\{q_1, q_2, ..., q_n\})  \\
         False &: \textrm{otherwise} \\ 
    \end{array}
    \right.
\end{equation}

with $\{q_1, q_2, ..., q_n\}$ the set of quality values of all the adaptation options in the adaptation space.

The natural approach to predict the values of the quality property and judge the adaptation options accordingly is regression. After the prediction, different strategies can be applied to perform the analysis. One strategy is selecting and analyzing a subset of adaptation options that were predicted to have quality values close to optimal. This way a small margin of error for the applied regression technique is taken into account. Another strategy is to restrict the analysis to only the adaptation option with the optimally predicted value of the quality property. This strategy can be applied if the time for computing the adaptation option is critical; yet, it may miss the best adaptation option since the predictions with regression are subject to errors. The strategy chosen represents the requirement of granularity of adaptation space reduction, see  Section~\ref{sec:problem-description}.
\vspace{5pt}\\
\ex{For the running example, we can define an optimization goal that minimizes the response time of service requests to the system, i.e., $\optimization{min}(q)$. Here we reduce the adaptation space by looking at the top 10 adaptation options in terms of predicted response time.\footnote{In this paper we define granularity in terms of an absolute number. An alternative approach could be to define granularity as a percentage of the total number of adaptation options. Using a number has the advantage that the time and resources required for analysis can be estimated; the advantage of using percentages is that the relative number of options considered is fixed. So there is a tradeoff between the two options. The proposed solution can be easily adjusted for both options.} Alternatively, we could opt to reduce the adaptation space to just a single option, when choosing a more strict granularity. Figure~\ref{fig:optimization_goal} shows the optimization goal when we choose to reduce the adaptation space to just one option.}

\subsubsection{Setpoint Goals}
The third and final type of adaptation goal covered in this paper is a setpoint goal. The aim of a setpoint goal is to keep the quality property of interest at (or close to) a given target value (i.e., the setpoint value or just the setpoint). We define the satisfaction of a setpoint goal $\setpoint{} \in \mathbb{S}$ with target $\mu$ and error margin $\epsilon$ for any quality value $q$ as follows: 

\begin{equation}
    \setpoint{\mu, \epsilon}(q) = \left\{
    \begin{array}{@{}l@{\thinspace}l}
         True &:  \abs{q - \mu} < \epsilon \\
         False &: \textrm{otherwise} \\ 
    \end{array}
    \right.
\end{equation}

For this type of goal, both classification and regression are candidates to predict quality values. Regression allows the identification of adaptation options with predicted quality values close to the setpoint value. Classification on the other hand enables the classification of adaptation options as either (1) being inside the specified epsilon window around the setpoint value or (2) outside the window. 
\vspace{5pt}\\
\ex{For the running example, we can specify a setpoint goal to keep the average cost of service invocations in the system at 8 cents with an error margin of 1 cent, i.e., $\setpoint{8c, 1c}(q)$, as shown in Figure~\ref{fig:setpoint_goal}. Depending on the granularity set for adaptation space reduction, the adaptation space is reduced to adaptation options within a limited window around the setpoint value.} 

\subsubsection{Combination of Multiple Goals}
In practice, self-adaptive systems usually have to deal with multiple adaptation goals. \techniqueNameShort{} supports adaptation space reduction for an arbitrary set of adaptation goals. However, in this paper we restrict ourselves to combinations of multiple threshold goals $\mathbb{T}$, multiple setpoint goals $\mathbb{S}$, and a single optimization goal $\optimization{}$, representing a large class of practical systems, as illustrated with the running example and the cases used for the evaluation of \techniqueNameShort{} in Section~\ref{sec:evaluation}. The combined set of goals, denoted as $\mathbb{G}$, is defined as:   

\begin{equation}
    \mathbb{G} =\,\,\,<\mathbb{T}, \mathbb{S}, \{\optimization{}\}>
\end{equation}

Hence, self-adaptive systems that rely on multi-objective optimization of adaptation goals to make adaptation decisions are not in scope of the work presented in this paper. 
The following sections explain in detail how \techniqueNameShort{} reduces adaptation spaces when a combination of goals $\mathbb{G}$ needs to be satisfied.
\vspace{5pt}\\
\ex{For the running example, we can combine different types of goals as specified above, for instance keeping the failure rate below a given threshold while minimizing the response time of service requests to the system.}

\section{\techniqueName{}}\label{sec:approach}

We now present \techniqueNameShort{}, addressing the research question we presented in Section~\ref{sec:problem-description}. \techniqueNameShort{} is a modular approach for adaptation space reduction in self-adaptive systems, meaning it can be instantiated in multiple ways, depending on the needs of the domain at hand. We focus specifically on the use of two classic supervised machine learning methods: classification and regression, applied to systems with different types of adaptation goals. %

We start with presenting the runtime architecture of \techniqueNameShort{} that integrates a machine learning module in the architecture of a self-adaptive system. Then, we give a high-level overview of the workflow of \techniqueNameShort{}. Finally, we zoom in on the design time and runtime stages of the workflow.

\subsection{Runtime Architecture of \techniqueNameShort{}}\label{subsec:runtime-architecture}

\begin{figure}
    \centering
    \includegraphics[width=.7\linewidth]{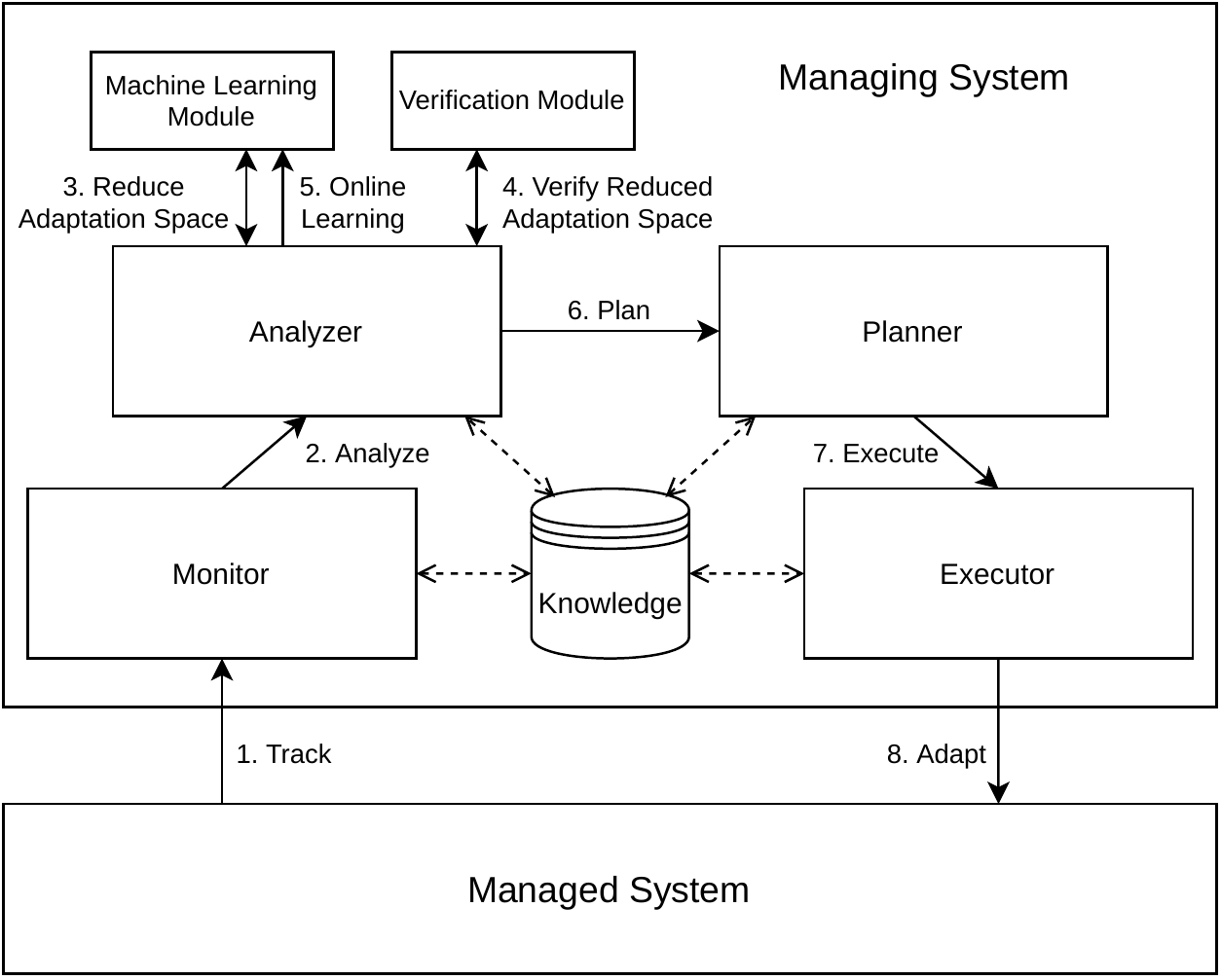}
    \caption{General MAPE-K architecture extended with a \MLM{} that reduces adaptation spaces.}
    \label{fig:architecture-MAPE-MLM}
\end{figure}

Figure~\ref{fig:architecture-MAPE-MLM} shows the high-level runtime architecture of a MAPE-based self-adaptive system extended with a \MLMi{} that realizes adaptation space reduction. The \textit{Monitor} tracks the uncertainties and properties of the underlying managed system (1) and updates the information in the \textit{Knowledge} repository. The \textit{Analyzer} then evaluates the need for adaptation, based on the current conditions (2). When this is the case, the analyzer composes a set of possible adaptation options, i.e., the configurations that can be reached from the current configuration by applying adaptation. This set is then passed to the \MLMi{} (3) that makes predictions of the adaptation options using the machine learning models. Based on these predictions and the adaptation goals, the \MLMi{} filters the options, reducing the set of adaptation options. These adaptation options are verified by the \textit{Verifier Module} using a set of runtime models of the quality properties that correspond with the adaptation goals (4). The resulting estimates of the quality properties per adaptation option are then used by the \MLMi{} to further train its internal learning models (5), resembling the online learning part of \techniqueNameShort{}. The \textit{Planner} then evaluates the verified adaptation options, determines the best adaptation option available based on the adaptation goals, and composes a plan to adapt the managed system (6). Finally, the \textit{Planner} triggers the \textit{Executor} (7) that executes the steps of the plan adapting the managed system (8).

In the remainder of this section, we explain how the \MLM{} is designed for a problem at hand (design stage of the \techniqueNameShort{} workflow) and how the module reduces adaptation spaces at runtime (runtime stage). 

\subsection{High-level Overview of the \techniqueNameShort{} Workflow}

\begin{figure}
    \centering
    \includegraphics[width=.7\linewidth]{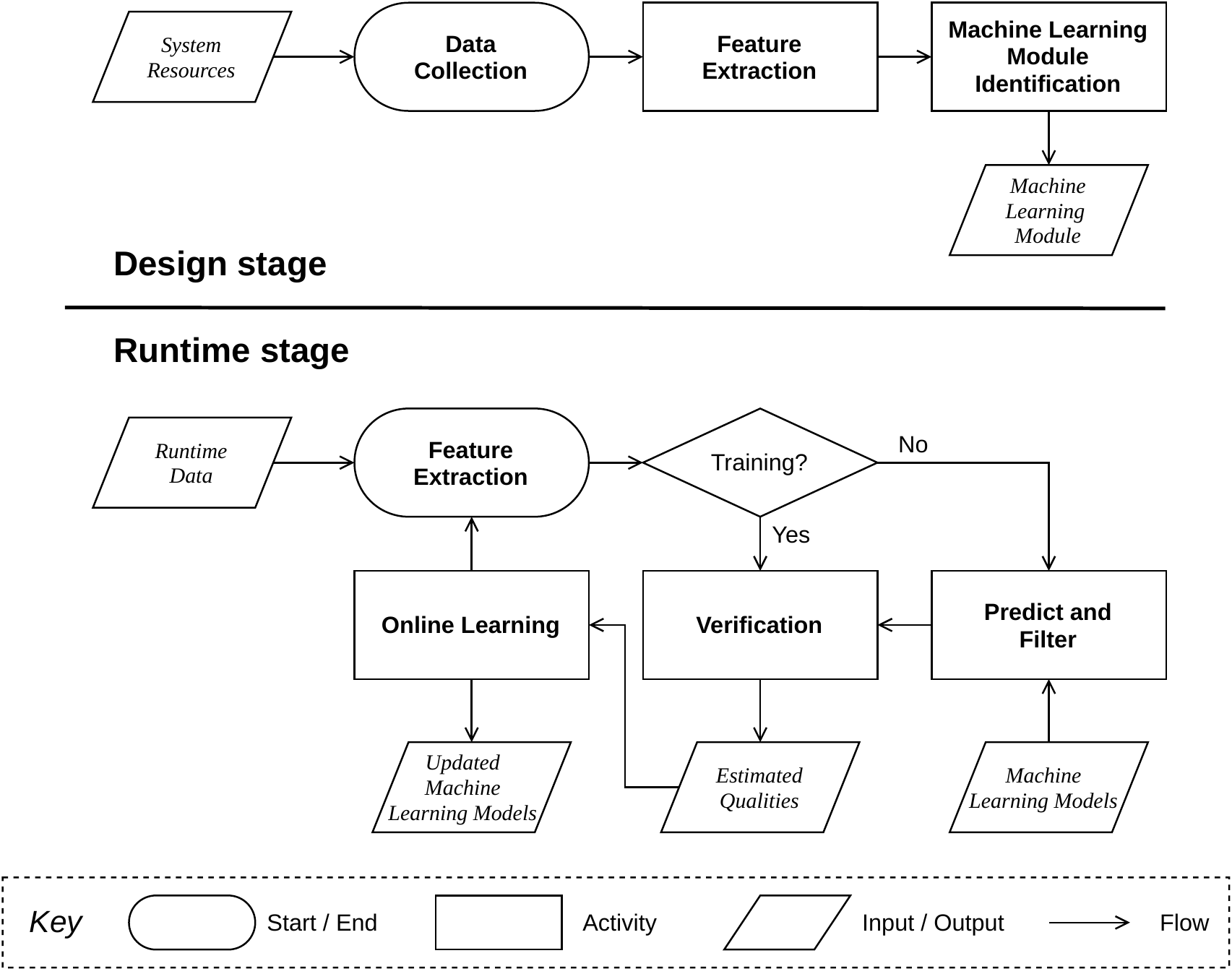}
    \caption{High level overview of the workflow of \techniqueNameShort{}.}
    \label{fig:high-level-overview}
\end{figure}

We start with a high-level overview of the workflow of \techniqueNameShort{}, shown in Figure~\ref{fig:high-level-overview}. We explain the two stages of the workflow in general here and discuss them in detail in the next sections. 

The design stage starts with the collection of data from the managed system and its environment. This data captures information relevant to the adaptation of the system over a period of time. This includes properties in the environment that affect the behavior of the system (e.g., actual workloads of the machines in the running example), system configurations (e.g., the distribution of service requests between the machines), and quality properties (e.g., the response time of service requests). 
Besides the system in operation, other suitable resources can be used to collect the data, such as a simulator or files with historical data. 
Next, features are extracted from the data. Features are measurable properties of the system and its environment that are relevant for self-adaptation. Uncertainties in the running example are the fluctuating workload of the machines and the bandwidth of the network. 
The extracted features are then used for the identification of the \MLM{}. To that end, different configurations of the \MLM{} (based on different types of learning models and other attributes such as scalers that are used to normalize the collected data) are compared and the best configuration is selected. The output of the design stage is a \MLM{} that comprises machine learning models with a set of attributes (for instance scalers), and a predictor with a filter that allows predicting the qualities of adaptation options that can then be filtered to reduce the adaptation space. The \MLM{} is then ready for deployment and use at runtime.

The runtime stage works in cycles, each  representing an opportunity for the system to perform adaptation. The workflow starts with gathering runtime data from the managed system and its environment that is relevant for adaptation. 
An example for the running example is the actual value of the workload of the two machines used in the service-based system. 
From this data, features are extracted, similarly to the design stage, yet now based on the data collected at runtime. Then two sub-stages are distinguished: training and testing. Immediately after deployment of the \MLM{}, the machine learning models need to be trained to make accurate predictions about quality properties in the system, filter the adaptation options, and reduce the adaptation space. The adaptation options in the running example are determined by the different settings that are available for distributing service requests between the two machines. In the training sub-stage, the system does not make any predictions yet. Consequently, as many adaptation options as possible are analyzed (i.e., the qualities are estimated using a verifier). Different heuristics can be used to select adaptation options from the total set, for instance options may be selected randomly, or the options may be divided into batches that are analyzed in subsequent slots. The number of cycles that are used for the training sub-stage is a parameter that is determined during the design stage. 

The second sub-stage of the runtime workflow is called testing. During testing, the trained machine learning models are effectively used to reduce the adaptation space. In addition, the new verification results for adaptation options of the reduced adaptation space are used to continue the learning of the machine learning models.  
In the testing sub-stage, the \MLM{} predicts the quality properties of the adaptation options, and based on these results and the adaptation goals set for the system, a subset of adaptation options is selected for verification. The verification results, i.e., estimates of the quality properties of the adaptation options of the reduced adaptation space, are used for online learning. 
The updated machine learning models are then ready to perform adaptation space reducing for the next cycle, and the verification results can be used by the planner to make an adaptation decision.  

In the following sections, we elaborate on the different steps of the two stages of the workflow. To precisely describe the different activities, we use a lightweight formalization. 

\subsection{Design Stage of the \techniqueNameShort{} Workflow in Detail}\label{subsec:design-stage}

\begin{figure}
    \centering
    \includegraphics[width=0.8\textwidth]{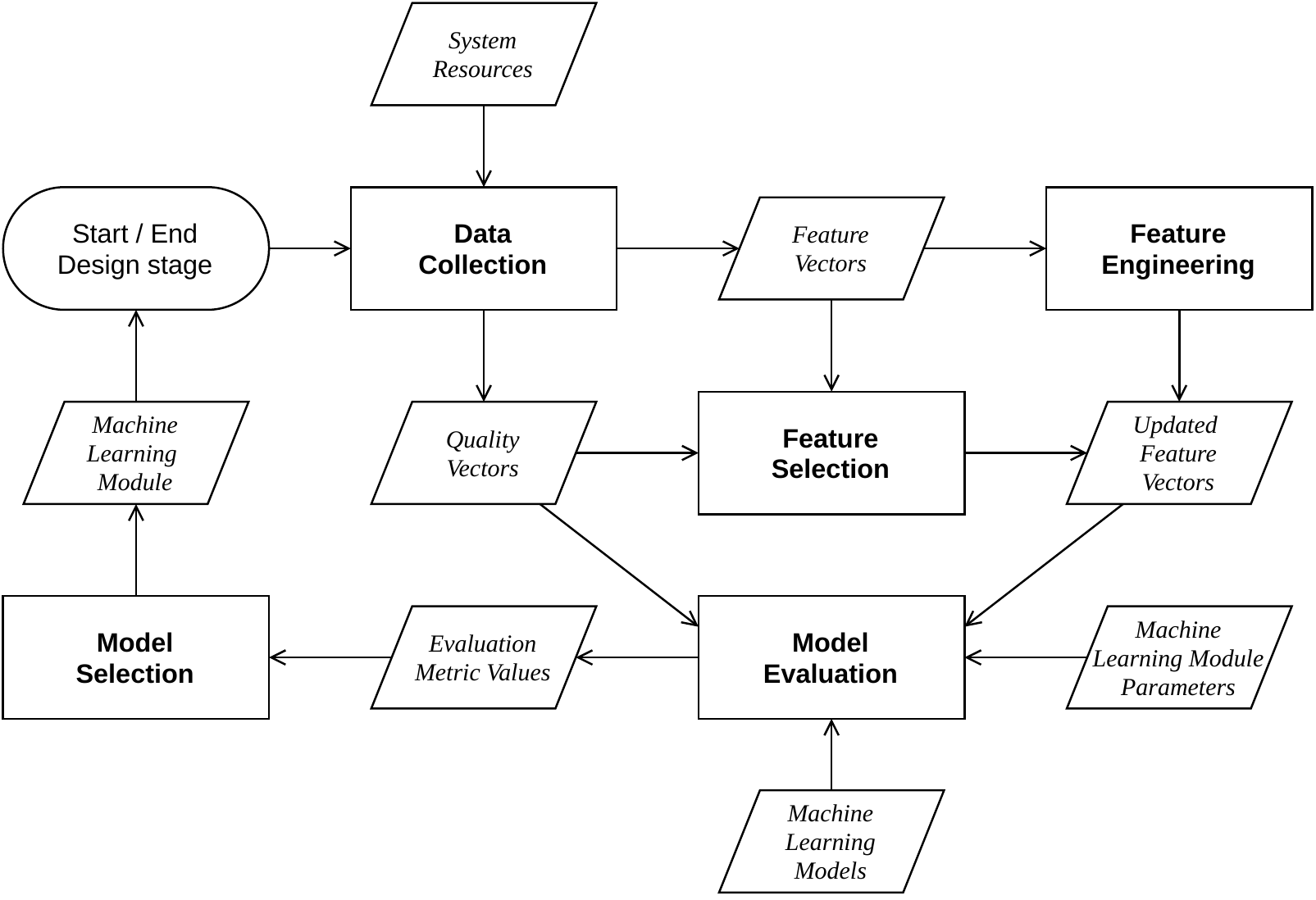}
    \caption{Workflow of the design stage activities for \techniqueNameShort{}.}
    \label{fig:workflow-design-MLM}
\end{figure}

Figure~\ref{fig:workflow-design-MLM} describes the workflow of the design stage activities in detail. 
The design stage comprises five distinct activities: \textit{Data Collection}, \textit{Feature Selection}, \textit{Feature Engineering}, \textit{Model Evaluation}, and \textit{Model Selection}. The output of the design stage is a configuration for the \MLM{} that can then be deployed and used to support a self-adaptive system with reducing large adaptation spaces at runtime. 

Before explaining the activities in detail, we highlight the software artifacts used for each activity and the responsibilities of the engineer; the various software artifacts are at the disposal of the engineer to perform the different activities. Table~\ref{tab:human-input-design} gives an overview of the artifacts used for the activities with the responsibilities of the engineer. 

Data collection is initiated by an engineer who selects a system resource and configures an artifact that is then used to collect data.  
Feature selection uses the collected data as input in a feature importance function that is used to filter out unimportant features.
Feature engineering is automatically initiated after feature selection which adjusts individual feature values according to the feature scaling algorithm.
Model evaluation automatically initiates after feature engineering by taking the updated features and collected system qualities to run and evaluate different machine learning algorithms.
Lastly, model selection is performed by an engineer who inspects the evaluation metrics from model evaluation to make a final decision about the configuration of the \MLM{}.

\begin{table}[ht!]
    \centering
    \caption{Responsibilities engineer with supporting software artifacts for the design stage activities.}
    \begin{tabular}{|l|p{4.2cm}|p{5.5cm}|}
        \hline
        \textbf{Activity} 
        & \textbf{Artifacts} & \textbf{Responsibilities engineer} \\\hline
        Data collection & System resource with data\newline collection artifact & Determine system resource and\newline configure data collection artifact \\\hline
        Feature selection & Feature importance function & Select feature importance function  \\\hline
        Feature engineering & Feature scaling algorithm\newline implementations & Choose feature scaling algorithm to use \\\hline
        Model evaluation & Machine learning algorithm implementations & Determine machine learning algorithms to evaluate and evaluation metrics to measure \\\hline
        Model selection & 
        & Make final decision about configuration by inspecting evaluation metrics  \\\hline
    \end{tabular}
    \label{tab:human-input-design}
\end{table}

\subsubsection{Data Collection}

During \textit{Data collection}, data concerning adaptation is gathered from the managed system and the environment in which the system operates. We categorize the data into two categories: potential features and system qualities. A potential feature $f$ is any type of property of the system or the environment that could have an influence on at least one quality property of the system. A system quality $q$ represents a non-functional property of the system. Data is collected for a period of time. At each time instance, the potential features and the associated qualities are collected.

We introduce the following definitions\footnote{We use the variable $n$ in multiple definitions to denote the number of elements in a set. However, each of the scope of these numbers is a single definition and values of $n$ can be different in different definitions.}:


\begin{itemize}
    \item[] $\Pi_i = \{\pi_1, ..., \pi_n\}$: A set of adaptation options in the system.
    \item[] $\Pi = \{\Pi_1, ..., \Pi_n\}$: The set of all sets of adaptation options.
    \item[] $U_i = \{u_1, ..., u_n\}$: A set of uncertainties that can be monitored. 
    \item[] $U = \{U_1, ..., U_n\}$: The set of all possible sets of uncertainties that can be monitored.
    \color{black}

	\item[] $\lambda_i = \{f_1, ..., f_{n+m}\}$: A set of features comprising $n$ features that represents a system configuration and $m$ features that represent uncertainties.\footnote{We use the term \textit{potential feature} to refer to a feature that may have an effect on any quality property of the system. A potential feature becomes a \textit{feature} if it has an actual effect on any quality property of the system, which is determined during feature selection. We do not distinguish potential features and features in the formal definition of feature sets.} We call $\lambda_i$ a feature vector. 
	\item[] $\Lambda_i = \{\lambda_1, ..., \lambda_n\}$: The set of possible feature vectors. 
	\item[] $\Lambda = \{\Lambda_1, ..., \Lambda_n\}$: The set of possible sets of feature vectors.
	
	\item[] $\phi_i = \{q_1, ..., q_n\}$: A set of qualities of a system. We call $\phi_i$ a quality vector. 
	\item[] $\Phi_i = \{\phi_1, ..., \phi_n\}$: The set of possible quality vectors.
	\item[] $\Phi = \{\Phi_1, ..., \Phi_n\}$: The set of possible sets of quality vectors.
	
	\item[] $s \in S$: A system resource, with $S$ the set of all resources of managed systems. 
\end{itemize}

The standard resource used for data collection is the system deployed in its real world setting. This resource ensures that the most accurate data is collected to design and configure the \MLM{}. However, collecting real-world data may be hard, for instance for large-scale distributed systems, or it may be an expensive and time-consuming process. Alternative approaches can then be applied, such as simulating the system or using historical data collected from the system. Such techniques may be more convenient to generate large amounts of data covering a wide range of different system states. 
We formally define the $CollectData$ function as follows:

\begin{itemize}
	\item[] $CollectData: S \rightarrow \Lambda \times \Phi$
	\item[] $CollectData(s) = <\{\lambda_1, ..., \lambda_n\}, \{\phi_1, ..., \phi_n\}>$
\end{itemize}

\textit{Data collection} from resource $s$ results in a list of feature vectors $\{\lambda_1, ..., \lambda_n\}$ and quality vectors $\{\phi_1, ..., \phi_n\}$. The potential features of $\lambda_i$ correspond with quality values $\phi_i$. The feature vectors and quality vectors provide the input to the next design stage activities.
\vspace{5pt}\\
\ex{Table~\ref{tab:approach-example-data-collection-sbs} shows an excerpt of data collected for the running example application. Each row defines a feature vector with values for  $\{$\textit{Distribution, Workload M1, ..., ABW M2}$\}$ and a quality vector with values for $\{$\textit{Response time}$\}$. Note that only a subset of potential features and qualities are listed in the table for the sake of clarity. We also consider a small set of feature vectors to keep the example simple. The full data set includes all the features that may have an impact on the qualities of the system, as well as all the associated quality values of the system.}

\begin{table}
	\centering
	\caption{Excerpt of data collected for the service-based application (abbreviations: \abbrev{M}{Machine}, \abbrev{ABW}{Available Bandwidth}).}
	\begin{tabular}{|c|c|c|c|c:c|} \hline
		\textbf{Distribution} & \textbf{Workload M1} & \textbf{Workload M2} & \textbf{ABW M1} & \textbf{ABW M2} & \textbf{Response time} \\ \hline
		40 & 75 & 20 & 10 & 10 & 16ms \\ \hline
		50 & 40 & 15 & 50 & 50 & 8ms \\ \hline
		20 & 60 & 80 & 25 & 30 & 14ms \\ \hline
		80 & 5 & 25 & 50 & 75 & 3ms \\ \hline
		50 & 25 & 20 & 70 & 60 & 5ms \\ \hline
		60 & 70 & 40 & 40 & 60 & 11ms \\ \hline
	\end{tabular}
	\vspace{5pt}
	\label{tab:approach-example-data-collection-sbs}
	\vspace{-15pt}
\end{table}

\subsubsection{Feature Selection}

During the next two activities relevant features are extracted from the collected data, defined as follows: 
\begin{itemize}
    \item[] $ExtractFeatures = EngineerFeatures \circ SelectFeatures$
\end{itemize}

During \textit{Feature selection}, the potential features and their respective quality values are evaluated using a feature selection algorithm. This algorithm analyzes the impact of individual features on the quality values associated with them. Irrelevant features, i.e., features that do not (or only marginally) influence the qualities, can be filtered out. This will simplify the machine learning model and enhance the performance of the \MLM{}.

It is important to note that \textit{Feature selection} carries an inherent risk. In fact, the algorithm determines the relevance of each feature based on its influence on the quality values, yet this evaluation is based on the data collected during \textit{Data collection}. If this data does not cover the scenarios where the feature has an influence on the qualities of the system, this feature will not be selected. For this reason, we leave \textit{Feature selection} as an optional activity in the design stage. It is the task of the engineer to carefully evaluate the data collected from the system to determine whether or not feature selection should be included. 
Feature selection is formally defined as follows:

\begin{itemize}
    \item[]
    $IND = 2^{\{1, ..., n\}}$: The set of all possible subsets of indices in the range $[1\ldots n]$.
    \item[] $ind \in IND$: The set of indices of features that are deemed relevant. \color{black}
    \item[] $SelectFeatures: \Lambda_i \times IND \rightarrow \Lambda_j$
    \item[] $SelectFeatures(\{\lambda_1, ..., \lambda_n\}, ind)$ 
    $= \{\,\lambda_j^{sel} \subseteq \lambda_i \mid \forall f_n \in \lambda_j^{sel}: Relevant(f_n, ind) = True\}$
\end{itemize}

\textit{Feature selection} uses the $Relevant$ function which uses the set of indices (denoted as \texttt{ind}) to decide whether individual features of a feature vector should be included or filtered out. Hence, the features in the resulting feature vector are the subset of the features in the original feature vector that are relevant.
\vspace{5pt}\\
\ex{The results of applying feature selection on the excerpt of the data collected from our example system (shown in Table~\ref{tab:approach-example-data-collection-sbs}) are shown in  Table~\ref{tab:approach-example-feature-selection-sbs}. In this case, feature selection determined that feature \textit{ABW 2} has no influence on the response time and consequently, this feature is excluded.}

\begin{table}[!htb]
	\centering
		\caption{Example of performing \textit{Feature selection} on the data from Table \ref{tab:approach-example-data-collection-sbs}.}
	\begin{tabular}{|c|c|c|c|c:c|} \hline
		\textbf{Distribution} & \textbf{Workload M1} & \textbf{Workload M2} & \textbf{ABW M1} & {\color{cRed} \sout{\textbf{ABW M2}}} & \textbf{Response time} \\ \hline
		40 & 75 & 20 & 10 & {\color{cRed} \sout{10}} & 16ms \\ \hline
		50 & 40 & 15 & 50 & {\color{cRed} \sout{50}} & 8ms \\ \hline
		20 & 60 & 80 & 25 & {\color{cRed} \sout{30}} & 14ms \\ \hline
		80 & 5 & 25 & 50 & {\color{cRed} \sout{75}} & 3ms \\ \hline
		50 & 25 & 20 & 70 & {\color{cRed} \sout{60}} & 5ms \\ \hline
		60 & 70 & 40 & 40 & {\color{cRed} \sout{60}} & 11ms \\ \hline
	\end{tabular}
	\label{tab:approach-example-feature-selection-sbs}
	\vspace{-8pt}
\end{table}

\subsubsection{Feature Engineering} \label{subsubsec:approach-featureengineering}

During \textit{Feature engineering},
the concrete values of the features are inspected and adjusted if this benefits the quality of predictions. As such feature engineering ties in closely with the next activity: \textit{Model selection}. A well known example of feature engineering is scaling that is used for features with values of varying magnitude, range, and units. One scaling technique is normalization where values of features are shifted and re-scaled to fit in a range between 0 and 1 (known as Min-Max scaling). Another scaling technique is standardization where values of features are centered around the mean with a unit standard deviation. Formally,  feature engineering is defined as follows:

\begin{itemize}
    \item[] $Transform: \lambda_i \rightarrow \lambda_j$
    \item[] $EngineerFeatures: \Lambda_i \rightarrow \Lambda_j$
    \item[] $EngineerFeatures(\{f_1, ..., f_n\}) = \{\,f_j \mid f_j = Transform(f_i),\: f_i \in \{f_1, ..., f_n\}\,\}$
\end{itemize}

\textit{Feature engineering} is centered around $Transform$ that transforms the values of the features according to a  concrete engineering method that is used (e.g. scaling with normalization or with standardization). The result of feature engineering is a set of normalized features. 
\vspace{5pt}\\
\ex{Table~\ref{tab:approach-example-feature-engineering-sbs} shows an example of feature engineering applied to the selected features of our running example shown in Table~\ref{tab:approach-example-feature-selection-sbs}. In this particular case, the values of the distribution, workload and available bandwidth are normalized, i.e.,  the values are rescaled to a range between 0 and 1 instead of original values between 0 and 100.} 

\begin{table}[!htb]
	\centering
		\caption{Example of performing \textit{Feature engineering} on the data from Table \ref{tab:approach-example-feature-selection-sbs}. The engineered feature values are marked in blue.}
	\begin{tabular}{|c|c|c|c:c|} \hline
		\textbf{Distribution} & \textbf{Workload M1} & \textbf{Workload M2} & \textbf{ABW M1} & \textbf{Response time} \\ \hline
		\exampleHighlight{0.4} & \exampleHighlight{0.75} & \exampleHighlight{0.2} & \exampleHighlight{0.1} & 16ms \\ \hline
		\exampleHighlight{0.5} & \exampleHighlight{0.4} & \exampleHighlight{0.15} & \exampleHighlight{0.5} & 8ms \\ \hline
		\exampleHighlight{0.2} & \exampleHighlight{0.6} & \exampleHighlight{0.8} & \exampleHighlight{0.25} & 14ms \\ \hline
		\exampleHighlight{0.8} & \exampleHighlight{0.05} & \exampleHighlight{0.25} & \exampleHighlight{0.5} & 3ms \\ \hline
		\exampleHighlight{0.5} & \exampleHighlight{0.25} & \exampleHighlight{0.2} & \exampleHighlight{0.7} & 5ms \\ \hline
		\exampleHighlight{0.6} & \exampleHighlight{0.7} & \exampleHighlight{0.4} & \exampleHighlight{0.4} & 11ms \\ \hline
	\end{tabular}
	\label{tab:approach-example-feature-engineering-sbs}
	\vspace{-4pt}
\end{table}

\subsubsection{Evaluation of Models} \label{subsubsec:approach-modelevaluation}

During the last two activities of the design stage, we identify the machine learning models of the \MLM{}, which is defined as follows: 
\begin{itemize}
    \item[] $IdentifyModels$ $= SelectModels \circ EvaluateModels$
\end{itemize}

We start with \textit{Evaluation of models} that uses the features extracted from the data of the system to determine a set of metrics that can be used to evaluate the performance of different machine learning models (in the next activity). Such metrics are determined to evaluate learning models per adaptation goal. To that end, a list of potential machine learning models are composed that combine different learning algorithms with variations on their internal loss and penalty functions\footnote{Whereas loss corresponds to the inaccuracy of predictions as explained above; a penalty expresses the degree of impact that the loss will have on the model of the learner.}. It is important to note that the selected algorithms need to support online learning, i.e., have the ability to continue training and thus updating machine learning models after deployment. 

Besides the list of machine learning models, two internal parameters of the \MLM{} are evaluated during model evaluation. The first parameter, \textit{exploration rate}, represents the percentage of extra adaptation options that are selected for analysis (by the self-adaptive system) on top of the adaptation options that are predicted by the \MLM{} as being compliant with the adaptation goals. Exploring an additional percentage of adaptation options ensures that the \MLM{} also relearns a sample of options that may otherwise be ignored. The second parameter that we evaluate is called \textit{warm-up count}. This parameter gives an indication for the number of training cycles that the \MLM{} should consider before it can be used to make meaningful predictions during operation (i.e., switch from training to testing).

We introduce the following definitions:

\begin{itemize}
    \item[] $M = \{m_1, ..., m_n\}$: The set of machine learning models to be evaluated.
    \item[] $\mathbb{M}$: The set of all sets of machine learning models.
    \item[] $E = \{e\}$: The \textit{exploration rate} internal to the \MLM{}.
    \item[] $W = \{w\}$: The \textit{warm-up count} internal to the \MLM{}.
    \item[] $\theta$: A set of metrics for the evaluation of machine learning models. 
    \item[] $\Theta$: The complete set of possible evaluation metrics.
    \item[] $\mathbb{E}$: The set of all sets of evaluation metrics for machine learning models. 
    
\end{itemize}

For the evaluation of the machine learning model we use train-test split. Train-test split is an efficient procedure to estimate the performance of classification or regression models. The method can be used if a sufficient large labeled dataset is available~\cite{Ayoub2018,Maimo2018}, which applies to our case where such dataset can be obtained from the system or a simulation as explained above.\footnote{The dataset of DeltaIoT, the first application used in the evaluation, comprises in total 64800 data points; the dataset of SBS, the second application, comprises 1350000 data points.} 
The evaluation of a machine learning model involves two steps: (1) training the model with a set of feature- and quality vectors and (2) testing the efficacy of the model by making predictions over a different set of feature vectors, examining the predictions through analyzing the according machine learning metrics. This process can range from splitting up the complete data set into two partitions (a train- and test dataset) to dividing the complete data set into multiple pairs of train- and test datasets (cross validation). For the interested reader we refer to~\ref{app:split} where we present a formal foundation for the former.
\textit{Model evaluation} is then defined as follows:

\begin{itemize}
    \item[] $EvaluateModels: \mathbb{M} \times \Lambda \times \Phi \times E \times W \rightarrow  \mathbb{M} \times \mathbb{E}$
     \item[] $EvaluateModels(M_i, ExtractFeatures(\Lambda_i, \{ind\}), \Phi_i, e, w) = 
     \ < \{m_1, ..., m_n\}, \{\theta_1, ..., \theta_n\} >$
\end{itemize}


Model evaluation results in a set of metrics sets, 
one set per machine learning model.
Recall that metrics are determined per adaptation goal. Hence, we repeat model evaluation per goal, resulting in a set of evaluation metrics sets for each adaptation goal. After model evaluation, the metrics are used in the final design stage activity to select the learning models of the \MLM{} that will be used for adaptation space reduction at runtime. 

For model evaluation of threshold goals, we apply classification using two evaluation metrics: F1-score, and Matthews correlation coefficient. 
For model evaluation of setpoint and optimization goals, we apply regression using four evaluation metrics: the R2-score, mean squared error, median absolute error, and maximum error. 
We elaborate on all aforementioned metrics further in Section~\ref{sec:metrics}.

\subsubsection{Selection of Models}\label{subsubsec:approach-modelselection}
In the last activity of the design stage, we select a learning model from the evaluated learning models relying on the metrics derived from the evaluation of these models.  
\textit{Selection of models} is formally defined as follows:

\begin{itemize}
    \item[] $SelectModels: \mathbb{M} \times \mathbb{E} \rightarrow M$
    \item[] $SelectModels(\{m_1, ..., m_n\}, \{\theta_1, ..., \theta_n\}) = m^{selected}$
\end{itemize}

During model selection the designer evaluates the metrics for the different machine learning models to make an informed decision about which model to use at runtime. This is repeated for each adaptation goal. Once the learning models are selected the \MLM{} can be configured and deployed to be used at runtime (we explain the elements of a \MLM{} configuration below). 
\vspace{5pt}\\
\ex{Table~\ref{tab:approach-model-selection-sbs} illustrates model evaluation and model selection for our running case. The data in the table builds on the previous examples and considers three machine learning models for classification, denoted with Model 1, 2 and 3. To keep it simple, we restrict the evaluation to a single threshold goal: the response time of service requests should not exceed 10ms. We also consider the accuracy of the model as a single evaluation metric\footnote{Evidently, in practice, multiple metrics will be used, but we simplify this here for illustration purposes.}. The table at the top shows the predictions of the different learning models. E.g., the first line for the features with response time 16ms exceeds the threshold goal (of 10ms) and should be classified as 0. This is correctly done by Model 1 and Model 2, but not by Model 3. The table at the bottom shows the accuracy of each Model. Based on these results, selecting a model is straightforward: the engineer selects model 3 in this example which has the highest accuracy.\footnote{ Accuracy is computed as the fraction of correct predictions expressed as  a percentage.} In case, multiple metric are used, the engineer needs to make an informed decision taking into account the different results.}


\newcommand{\correct}[1]{{\color{cGreen} #1}}
\newcommand{\wrong}[1]{{\color{cRed} \underline{#1}}}
\begin{table}[!htb]
	\centering
		\caption{Excerpt of \textit{Model selection} and \textit{Model evaluation} in the service-based application. Values in columns Model 1, 2, and 3 represent the classes of predictions of the models. Class 0 means that the response time goal is violated, class 1 means that the goal is achieved. The values of the classes that are predicted correctly are marked in green; the values of the classes that are predicted incorrectly are marked in red and are underlined. }
	\begin{tabular}{|c;{0.2pt/2pt}M{2.5cm}:c|c|c|} \hline
		\textbf{Response time} & \textbf{Response time \newline \textless\,10ms?} & \textbf{Model 1} & \textbf{Model 2} & \textbf{Model 3} \\ \hline
		16ms  & 0 & \correct{0} & \correct{0} & \wrong{1} \\ \hline
		8ms   & 1 & \correct{1} & \correct{1} & \correct{1} \\ \hline
		14ms  & 0 & \correct{0} & \wrong{1} & \correct{0} \\ \hline
		3ms   & 1 & \correct{1} & \correct{1} & \correct{1} \\ \hline
		5ms   & 1 & \wrong{0} & \wrong{0} & \correct{1} \\ \hline
		11ms  & 0 & \wrong{1} & \correct{0} & \correct{0} \\ \hline
	\end{tabular}
	\vspace*{.4cm}
	\newline
	\begin{tabular}{|l|c|c|c|}
		\cline{2-4}
		\multicolumn{1}{l|}{} & \textbf{Model 1} & \textbf{Model 2} & \textbf{\underline{Model 3}} \\ \hline 
		Accuracy & 50\% & 66.6\% & 83.3\% \\ \hline
	\end{tabular} \vspace{5pt}
	\label{tab:approach-model-selection-sbs}
	\vspace{-5pt}
\end{table}

\subsubsection{Setup and Configuration of the \MLM{}}

\begin{figure}
    \centering
    \includegraphics[width=1\linewidth]{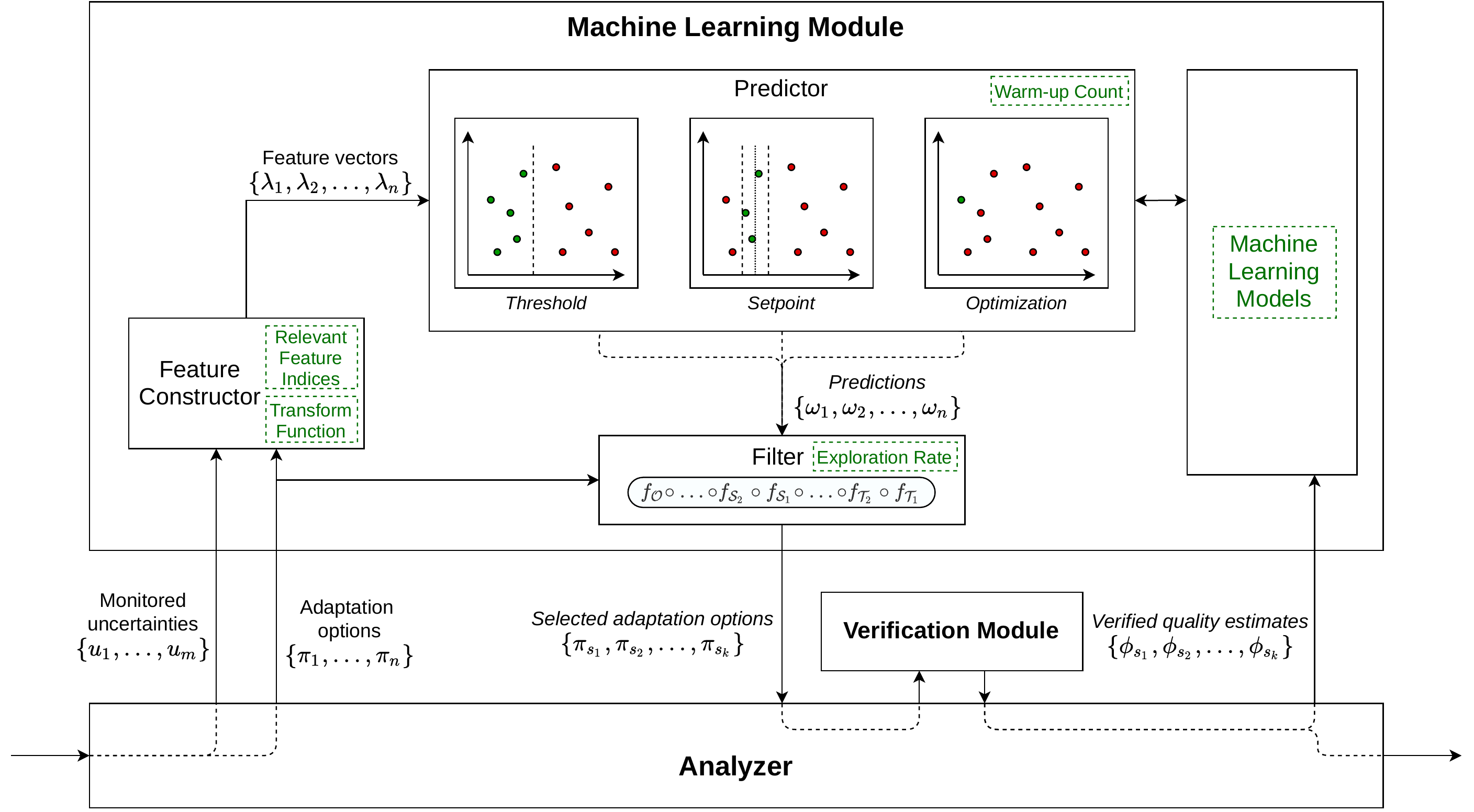}
    \caption{Architecture of the \MLM{}; configuration elements are marked in green dotted boxes.} 
    \label{fig:architecture-MLM}
\end{figure}

After completing the design stage activities, the \MLM{} can be set up and configured. Figure~\ref{fig:architecture-MLM} shows the architecture of the \MLM{} that comprises four main components. 

The \textit{Feature constructor} is responsible for assembling and extracting feature vectors. It takes as input a set of adaptation options and the values of uncertainties. The feature constructor combines this input using \textit{Feature composition} (we explain this runtime activity below) and \textit{Feature engineering}. The output is a set of feature vectors obtained from the runtime data. The feature constructor is configured using two specific parameters: the indices of relevant features $\{ind\}$ (determined in \textit{Feature selection}) and a \textit{Transform} function (determined in \textit{Feature engineering}).

The \textit{Machine Learning Models} that are determined during the design stage are maintained in a data repository. Conceptually, the learning models are part of the \MLM{}. However, in practice, the models may be stored in the \textit{Knowledge} repository of the MAPE-K feedback loop.

The \textit{Predictor} is responsible for making predictions of the adaptation options (i.e., the feature vectors produced by the feature constructor). In particular, the predictor makes predictions about the satisfaction of adaptation goals of the adaptation options (as specified by the $Predict$ function), leveraging on the machine learning models. The output of the predictor is a set of predictions for the different adaptation options that need further filtering. The predictor is configured using the internal parameter \textit{warm-up count} that determines the period that is used for training of the machine learning models. We explain the predictor below in the section about testing. 

Finally, the \textit{Filter} is responsible for filtering the adaptation options based on the predictions for the adaptation goals made by the predictor. Besides determining relevant adaptation options, the filter selects a subset of additional features to be explored based on the \textit{exploration rate} parameter. The output of the filter is a  reduced set of adaptation options that are used for verification. The verification results are then used for online learning of the machine learning models. We explain filtering further in the section about testing below. 
 


\subsection{Runtime Stage of the \techniqueNameShort{} Workflow in Detail: Training}\label{subsec:runtime-stage-training}

The runtime stage consists of two sub-stages:  \textit{Training} followed by \textit{Testing}. In contrast to the design stage activities, the runtime stage activities work fully automatic and require no human input.

\begin{figure}
    \centering
    \includegraphics[width=.95\textwidth]{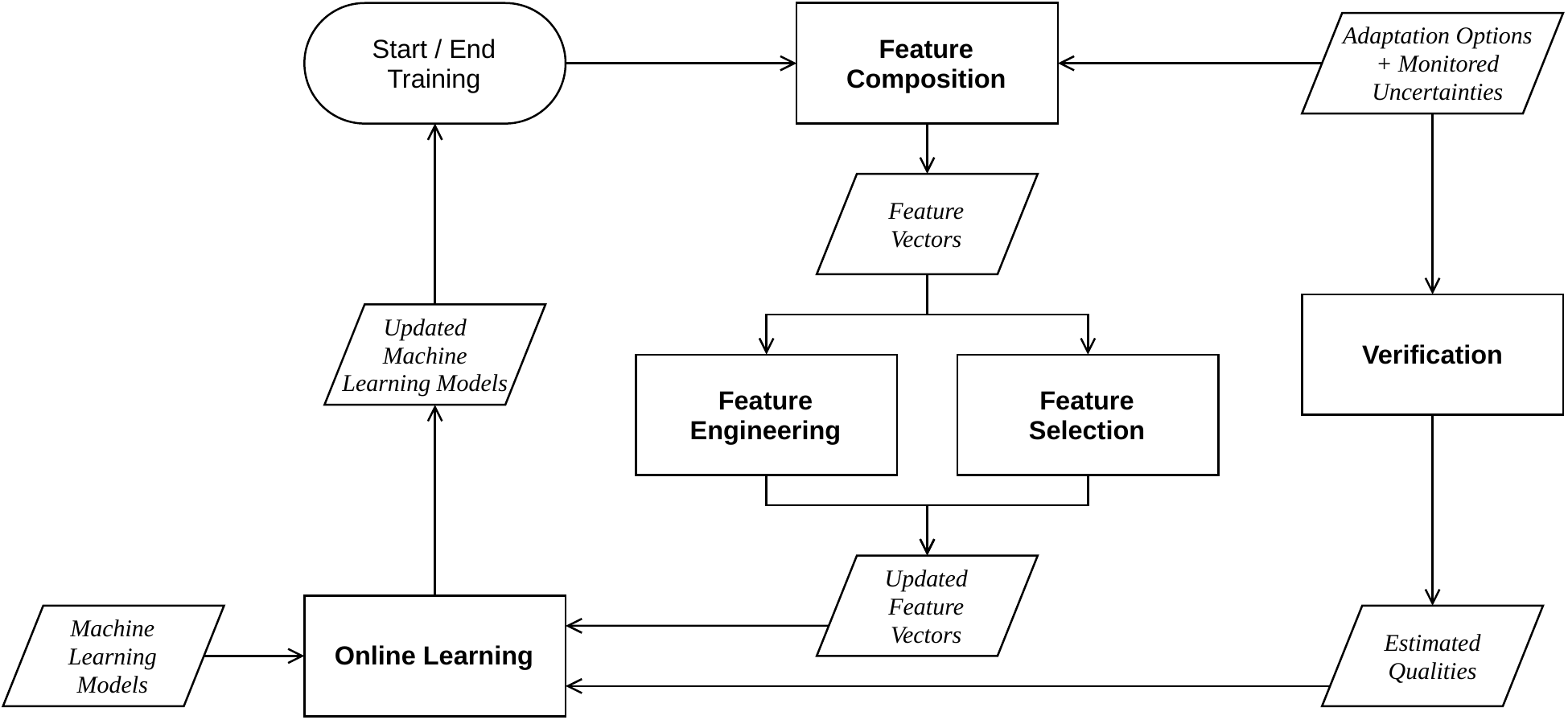}
    \caption{Workflow of the runtime stage activities in training cycles for \techniqueNameShort{}.}
    \label{fig:workflow-runtime-MLM-training}
\end{figure}

We start with \textit{Training}. Training takes place immediately after deployment, when the \MLM{} has not yet learned nor gathered enough data of the system and its environment to make accurate predictions about the system qualities. 
Figure~\ref{fig:workflow-runtime-MLM-training} gives a detailed overview of the workflow of the runtime stage activities during training. Training is applied for a number of cycles, based on the \textit{warm-up count} that was determined during the design stage. 

During training, the \MLM{} is not reducing the adaptation space yet. Instead, the available adaptation time of the system is used to formally verify as many adaptation options as possible, and the verification results are used to train the learning models of the \MLM{}. 
If not all adaptation options can be verified within a single time window that is available to make an adaptation decision, different strategies can be applied to select and verify adaptation options. A simple strategy selects adaptation options randomly. Another more balanced approach applies a round-robin strategy to select adaptation options one by one in consecutive time windows. Yet another strategy applies active learning to choose adaptation options for verification such that the \MLM{} can learn more efficiently by e.g., selecting options with maximum entropy~\cite{settles2009active}. \techniqueNameShort{} is flexible and does not prescribe any particular strategy.  
%

We use the following basic definitions to formally describe the activities of the runtime stage:

\begin{itemize}
    \item[] $\Omega_i = \{\omega_1, ..., \omega_n\}$: A set of predictions made by a learning model.
    \item[] $\Omega = \{\Omega_1, ..., \Omega_n\}$: The set of all sets of predictions made by a learning model.
    \item[] $\mathbb{Z}$: The set of all sets of predictions.  
\end{itemize}

It is important to note that \techniqueNameShort{} currently focuses on handling discrete adaptation options. System designers can however discretize a continuous adaptation space to apply \techniqueNameShort{}.

In the first activity of  training, feature vectors are composed, meaning, the set of possible adaptation options are combined with the set of uncertainties monitored by the system. Formally $ComposeFeatures$ is defined as follows: 
\begin{itemize}
    \item[] $ComposeFeatures: \Pi \times U \rightarrow \Lambda$
    \item[] $ComposeFeatures(\Pi_i, U_i) = \Lambda_i = \{\lambda_{\pi_1, U_i}, \lambda_{\pi_2, U_i}, ...\}$
\end{itemize}

Features composition generates a feature vector that combines the features representing a system configuration (adaptation option) with the features representing the monitored uncertainties. 
The composed features then undergo selection and engineering before they are used for online learning (see below), resulting in updated feature vectors. 
%
\vspace{5pt}\\
\ex{Table~\ref{tab:approach-example-compose-features-sbs} illustrates the composition of features as well as feature extraction, i.e., feature selection and feature engineering. The monitored uncertainties include the workload and available bandwidth of the machines. The different settings of the distribution of service requests represents here the adaptation options, i.e.,  system configurations. Based on feature extraction, the feature $ABW$ $M2$ is not included in the updated feature vectors.} \bigskip

\begin{table}
	\centering
	\caption{Example of composed features and application of feature extraction for the service-based application.}
	\begin{tabular}{|c|c|c|c|c|c|} \hline
		\textbf{Distribution} & \textbf{Workload M1} & \textbf{Workload M2} & \textbf{ABW M1} & {\color{cRed} \sout{\textbf{ABW M2}}} \\ \hline
		0.1 & 0.4 & 0.2 & 0.5 & {\color{cRed} \sout{0.3}} \\ \hline
		0.2 & 0.4 & 0.2 & 0.5 & {\color{cRed} \sout{0.3}} \\ \hline
		0.3 & 0.4 & 0.2 & 0.5 & {\color{cRed} \sout{0.3}} \\ \hline
		0.4 & 0.4 & 0.2 & 0.5 & {\color{cRed} \sout{0.3}} \\ \hline
		0.5 & 0.4 & 0.2 & 0.5 & {\color{cRed} \sout{0.3}} \\ \hline
		... & ... & ... & ... & ... \\ \hline
	\end{tabular}
	\vspace{5pt}
	\label{tab:approach-example-compose-features-sbs}
\end{table}

To enable online learning, i.e., which is actually a continued training activity, adaptation options need to be verified, preferably as many as possible (as explained above). We define  \textit{Verify} as follows:\\

\begin{itemize}
    \item[] $QM = \{qm_1, ..., qm_n\}$: A set of formal quality models used to estimate system qualities.
    \item[] $\mathbb{Q}$: The set of all sets of formal quality models.

    \item[] $Verify: \Pi \times U \times \mathbb{Q} \rightarrow \Pi \times \Phi$
    \item[] $Verify(\{\pi_1, ..., \pi_n\}, \{u_1, ..., u_m\}, \{qm_1, ..., qm_k\}) =\ <\{\pi_1, ..., \pi_n\}, \{\phi_1, ..., \phi_n\}>$
\end{itemize}

Verification generates a set of quality values (one per goal) for each adaptation option. It is important to note that the quality values for the different adaptation options are estimates. The accuracy of these estimates is determined by the precision of the quality models used, the measurements of the uncertainties, and the verification method applied. 
\vspace{5pt}\\
\ex{Table~\ref{tab:approach-example-formal-verification-sbs} illustrates the verification results of a quality model for response time of a sample of adaptation options from our example service-based application.}\bigskip

\begin{table}
	\centering
		\caption{Example of formal verification of a set of adaptation options, marked in blue.}
	\begin{tabular}{|c|c|c|c|c:c|} \hline
		\textbf{Distribution} & \textbf{Workload M1} & \textbf{Workload M2} & \textbf{ABW M1} & \textbf{ABW M2} & \textbf{Estimated RT} \\ \hline
		10 & 40 & 20 & 50 & 30 & \exampleHighlight{3ms} \\ \hline
		20 & 40 & 20 & 50 & 30 & \exampleHighlight{8ms} \\ \hline
		30 & 40 & 20 & 50 & 30 & \exampleHighlight{6ms} \\ \hline
		40 & 40 & 20 & 50 & 30 & \exampleHighlight{11ms} \\ \hline
		50 & 40 & 20 & 50 & 30 & \exampleHighlight{13ms} \\ \hline
		... & ... & ... & ... & ... & ... \\ \hline
	\end{tabular}
	\vspace{5pt}
	\label{tab:approach-example-formal-verification-sbs}
	\vspace{-15pt}
\end{table}

Lastly, we use the updated feature vectors and the estimated quality vectors to train the machine learning models. More specifically, we employ online learning to continuously update and refine the machine learning models during the testing cycles. Online learning, also referred to as incremental learning, allows a learner to incrementally learn from newly provided data samples. 
We refer the interested reader to the following articles~\cite{Cauwenberghs2000,Kuzborskij2013}. $LearnOnline$ is defined as follows:
\begin{itemize}
    \item[] $LearnOnline: \Lambda \times \Phi \times M \rightarrow M$
    \item[] $LearnOnline(ExtractFeatures(\Lambda_i, ind), \Phi_i, m) = m^{updated}$
\end{itemize}

Online learning is defined for a single learning model and hence needs to be repeated for all models. Initially, online learning starts from the model selected during the design stage ($m^{selected}$). Online learning then uses the features extracted from the composite feature vectors that are derived from the runtime data, the quality vectors associated with the adaptation options obtained from verification, and the model that is subject to training. The result is an updated learning model.

\subsection{Runtime Stage of the \techniqueNameShort{} Workflow in Detail: Testing}\label{subsec:runtime-stage-testing}

\begin{figure}
    \centering
    \includegraphics[width=.9\textwidth]{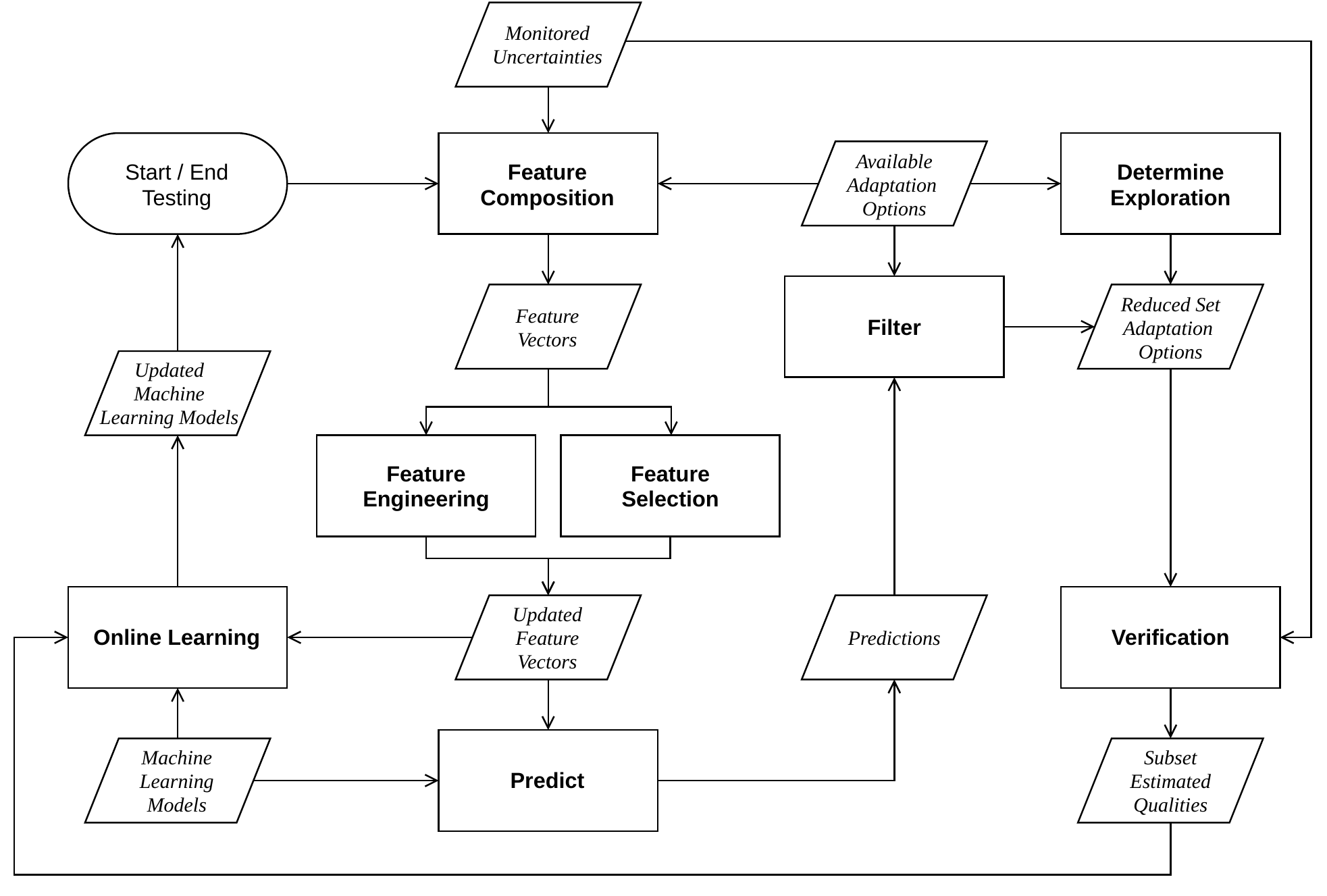}
    \caption{Workflow of the runtime stage activities in testing cycles for \techniqueNameShort{}.}
    \label{fig:workflow-runtime-MLM-testing}
\end{figure}

Once the machine learning models are trained (based on the \textit{warm-up count} determined during the design stage), the \MLM{} switches to \textit{Testing}. As opposed to training, during testing cycles, the \MLM{} uses the machine learning models to make effective predictions about the qualities of adaptation options. These predictions can then be used to reduce the adaptation space, improving the efficiency of the analysis.  Figure~\ref{fig:workflow-runtime-MLM-testing} shows the workflow with the activities of the testing cycles. 

Similar to training, feature vectors are composed from the adaptation options and the monitored uncertainties. These feature vectors undergo selection and engineering resulting in updated feature vectors. These feature vectors are then used together with  machine learning models to make predictions about the qualities of the adaptation options.
We define $Predict$ as follows:

\begin{itemize}
    \item[] $Predict: M \times \Lambda \rightarrow \Lambda \times \Omega$
    \item[] $Predict(m, \{\lambda_1, ..., \lambda_n\}) =\ < \{\lambda_1, ..., \lambda_n\}, \{\omega_1, ..., \omega_n\}>$
\end{itemize}

Predictions are made per machine learning model. Prediction takes as input a learning model and a set of feature vectors. The result is a set of predictions associated with the adaptation options (represented as feature vectors). The types of predictions depend on the machine learning model at hand. For instance, a classifier uses classes as representations, while a regressor uses values. 
\vspace{5pt}\\
\ex{Table~\ref{tab:approach-example-prediction-sbs} illustrates predictions for a sample of feature vectors done by a classifier that predicts the classes for each feature vector. Class 1 and 0 refer to predictions for the satisfaction and violation of a threshold goal respectively, as defined in the examples of the design stage.} \bigskip

\begin{table}[!htb]
	\centering
		\caption{Example of predictions made for feature vectors taken from Table~\ref{tab:approach-example-compose-features-sbs}.}
	\begin{tabular}{|c|c|c|c:c|} \hline
		\textbf{Distribution} & \textbf{Workload M1} & \textbf{Workload M2} & \textbf{ABW M1} & \textbf{Predicted class} \\ \hline
		0.1 & 0.4 & 0.2 & 0.5 & \exampleHighlight{1} \\ \hline
		0.2 & 0.4 & 0.2 & 0.5 & \exampleHighlight{1} \\ \hline
		0.3 & 0.4 & 0.2 & 0.5 & \exampleHighlight{0} \\ \hline
		0.4 & 0.4 & 0.2 & 0.5 & \exampleHighlight{0} \\ \hline
		0.5 & 0.4 & 0.2 & 0.5 & \exampleHighlight{0} \\ \hline
		... & ... & ... & ... & ... \\ \hline
	\end{tabular}
	\vspace{5pt}
	\label{tab:approach-example-prediction-sbs}
\end{table}

The predictions made by the machine learning models are then used to filter the adaptation options. Only the adaptation options that are predicted to meet the adaptation goals are included for verification. For a full explanation and formal foundation of the filter operation we refer the interested reader to~\ref{app:filter}. To summarize the filter operation: first adaptation options that are predicted to not meet any of the threshold or setpoint goals are filtered out. Second, out of the remaining adaptation options, the adaptation space is further reduced according to the specified granularity value (typically denoted by the letter $g$ from this point on) and the predictions affiliated with the quality of the optimization goal of the system (if present).

A consequence of the filtering approach is that adaptation options that do not meet all the adaptation goals are not selected for verification (since these options are filtered beforehand). 
However, these adaptation options may be of interest to the system. To deal with this issue we introduced the internal parameter \textit{exploration rate}; the internal parameter that we discussed in \textit{Model selection} during the design stage. The \textit{exploration rate} specifies the percentage of adaptation options that are included for verification, despite these options being filtered out by the filter. The \textit{exploration rate} offers the \MLM{} the ability to relearn and correct potentially outdated predictions. For a formal definition of the selection of explored adaptation options, we refer to the~\ref{app:exploration}. 
\vspace{5pt}\\
\ex{Table~\ref{tab:approach-example-filter-exploration-sbs} illustrates the application of filtering based on the predictions made by the \MLM{}, as well as extending this set with a selection of additional adaptation options based on the \textit{exploration rate}.}

\begin{table}[t!]
	\centering
		\caption{Example of the selection of adaptation options for verification based on predictions made in Table~\ref{tab:approach-example-prediction-sbs} and an additional set of options to explore (in the excerpt, only one such option is shown).}
	\begin{tabular}{|c|c|c|c|c:c|c|} \hline
		\textbf{Distribution} & \textbf{Workload M1} & \textbf{Workload M2} & \textbf{ABW M1} & \textbf{Predicted class} & \textbf{Verify?} \\ \hline
		0.1 & 0.4 & 0.2 & 0.5 & 1 & {\color{cGreen} Verify} \\ \hline
		0.2 & 0.4 & 0.2 & 0.5 & 1 & {\color{cGreen} Verify} \\ \hline
		0.3 & 0.4 & 0.2 & 0.5 & 0 & {\color{cGreen} Exploration} \\ \hline
		0.4 & 0.4 & 0.2 & 0.5 & 0 & {\color{cRed} Discard} \\ \hline
		0.5 & 0.4 & 0.2 & 0.5 & 0 & {\color{cRed} Discard} \\ \hline
		... & ... & ... & ... & ... & ... \\ \hline
	\end{tabular}
	\vspace{5pt}
	\label{tab:approach-example-filter-exploration-sbs}
\end{table}

\section{Algorithms, Models, and Metrics for Evaluating \techniqueNameShort{}}\label{sec:metrics}

Before we present the evaluation of \techniqueNameShort{}, we give an overview of the algorithms and models we used for the design of the learning modules, and we define the metrics that we use for the evaluation in Section~\ref{sec:evaluation}.

\subsection{Algorithms and Models for the Design of the \MLM{}s}

For the design of the \MLM{}s of the evaluation cases, we evaluated different machine learning algorithms and models. \review{The selection of the algorithms and models is based on their common use in the community; for some recent examples see~\cite{Deshpande2020,Lekshmy22,ahmad2022supervised}. In addition, the algorithms and models are supported by the widely used scikit-learn implementation kit~\cite{scikit-learn} (see for instance~\cite{jamshidi2018learning,VanDerDonckt19,Diallo2021}) that we also used for implementing the \MLM{}.} We summarize now the algorithms and models that we used in the different design steps.

\paragraph{Feature Extraction Algorithms}\label{metrics:feature-extraction}

Feature extraction involves two steps: \textit{Feature selection} and \textit{Feature engineering}. For \textit{Feature selection} we have used \textit{Extremely Randomized Tree} algorithms\,\cite{Geurts2006} to determine the importance of individual features based on their influence on the target values, i.e., the qualities of the system. The algorithms are based on random forest algorithms (composed of an ensemble of classical decision trees). On top of the random forest algorithms, the extremely randomized tree algorithms introduce extra randomness with the objective of reducing variance of the machine learning algorithm further (reducing overfitting). We utilized two implementations of the extremely randomized tree algorithms: an implementation of the algorithm for classification and an implementation of the algorithm for regression. After applying the algorithms and detecting relevant features, we adjusted the collected data accordingly for the subsequent activities.
For \textit{Feature engineering} we considered 4 scaling algorithms: no scaling algorithm, min-max scaling, max-abs scaling and standard scaling. We described the min-max and standard scaler briefly in Section~\ref{subsubsec:approach-featureengineering}. The max-abs scaler rescales the feature values in the range between 0 and 1 relative to their absolute value (e.g. the maximally encountered absolute feature value rescales to a value of 1).

\paragraph{Machine Learning Models}

For our evaluation, we considered classification and regression machine learning models from the scikit-learn library~\cite{scikit-learn} that are commonly used and support online learning. More specifically, we evaluated Stochastic Gradient Descent classifiers, Passive-Aggressive classifiers, Perceptron classifiers, Stochastic Gradient Descent regressors and Passive-Aggressive regressors. 
The Perceptron classifier is a single layer classifier that utilizes the broadly known Perceptron algorithm (written and published by Frank Rosenblatt~\cite{Rosenblatt1958}). 
The Stochastic Gradient Descent classifier and regressor both use an SGD learning routine to train their internal models. This learning routine tries to approximate the true gradient of the regularized training error of the model by analyzing a single training data sample at a time (based on \cite{RobbinsMonro1951}). 
The Passive-Aggressive classifier and regressor~\cite{Crammer2006} both utilize a more aggressive strategy compared to the previously mentioned Perceptron or SGD models by correcting its model in case the internal loss exceeds a threshold, regardless of the step-size required to amend the model.  


It is important to note that \techniqueNameShort{} supports any other type of machine learning model (classification or regression) that supports online learning and fits within the architecture as described in Figure~\ref{fig:architecture-MLM}.

\subsection{Metrics for Evaluating the Learning Models of \techniqueNameShort{}}\label{metrics:machine-learning-metrics}

Learning models need to be evaluated both during the design stage and the runtime stage. During design, evaluation is used to select the best model. For hyper-parameter tuning, we varied a number of parameters, in particular the loss function, penalty function (if applicable), scaler, exploration rate, and warmup-count for the learners that we evaluated\footnote{For scenarios with two threshold goals experimental evaluation revealed that using a single classifier to predict the satisfaction of the two threshold goals was slightly more accurate than using separate classifiers. Therefore, for such settings, we use a single classifier that classifies the adaptation options in four classes as follows: 
C\textsubscript{0}: No threshold goals are satisfied; C\textsubscript{1}: Only the first threshold goal is satisfied; C\textsubscript{2}: Only the second threshold goal is satisfied; and C\textsubscript{3}: Both threshold goals are satisfied. We use these classes in the evaluation in Section~\ref{sec:evaluation}.} (for the list see Section~\ref{subsubsec:approach-modelevaluation}). During runtime, we monitored the selected learning models to validate that they perform well after deployment. 
Table~\ref{tab:classfication-metric-values} and Table~\ref{tab:regression-metric-values} show the metrics we used for the evaluation of the learning models for classification and regression respectively.


\begin{table}
    \caption{List of used machine learning metrics when evaluating classification machine learning models.}
    \centering
    \begin{tabular}{|M{3cm}|m{9cm}|c|}
        \hline
        \textbf{Name} & \textbf{Description} & \textbf{Objective} \\\hline
        F1-score & A combined metric of recall (percentage of samples that were retrieved using the classifier) and precision (percentage of samples that were correctly predicted), defined in the interval [0, 1]. & Maximize \\\hline
        MCC & The Matthew's correlation coefficient: a metric representing how well the classifier performs compared to making random predictions, defined in the interval [-1, 1] (-1 representing completely incorrect predictions, 0 representing on par predictions with random predictions and 1 representing perfect predictions). & Maximize \\\hline
    \end{tabular}
    \label{tab:classfication-metric-values}
\end{table}

F1-score combines recall (the percentage of samples that were retrieved by prediction from a specific class) and precision (the fraction of samples that have been predicted to be of a specific class that are actually part of that class). The F1-score is defined as a value in the interval [0, 1]. A higher F1-score means in general a better performing classifier. \review{The F1-score is commonly used to judge the performance of classifiers,  e.g.,~\cite{Ayoub2018, Maimo2018}.}  The Matthews correlation coefficient has a value in the range [-1, 1], where 1 represent perfect predictions, 0 represent predictions that are equal to random predictions, and -1 represents incorrect predictions. 
\review{Hence, this metric enabled us to compare the predictions made by the machine learning model with an approach that predicts based on random selections.}

\begin{table}
    \caption{List of used machine learning metrics when evaluating regression machine learning models.}
    \centering
    \begin{tabular}{|M{3cm}|m{9cm}|c|}
        \hline
        \textbf{Name} & \textbf{Description} & \textbf{Objective} \\\hline
        R2-score & A metric representing how well the model predicts the target value by looking at the variance of the predictions, defined in the interval [0, 1]. & Maximize \\\hline
        MSE & The mean of the squares of errors on predictions made by the regressor. & Minimize \\\hline
        MAE & The median absolute error on predictions made by the regressor, less susceptible to outliers. & Minimize \\\hline
        ME & The maximum error on predictions made by the regressor. & Minimize \\\hline
    \end{tabular}
    \label{tab:regression-metric-values}
\end{table}

The R2-score represents how well the predictions of the model fit the actual quality values by looking at the variance of the predictions compared to the actual quality values. The R2-score is defined as a value in the interval [0, 1]. A higher R2-score indicates that the model is a better fit for the quality under consideration. The mean squared error refers, as the name suggests, to the mean of the squares of the errors in predictions made by the regressor. The median absolute error offers an alternative to the mean squared error, which is not as susceptible to outliers in the predictions. Lastly, the maximum error gives a good indication of the worst-case prediction made by the regressor. \review{The R2-score serves as a good general metric for evaluating learning models (e.g., used in~\cite{Samir2019}). The other metrics can provide useful insights depending on the domain and context of the application at hand, see e.g.,~\cite{Flores1986}.}

\subsection{Metrics for Evaluating 
Utility Penalty and Efficiency at Runtime} \label{subsec:evaluation-metrics}

Table~\ref{tab:evaluation-metric-values} summarizes the metrics for utility penalty and efficiency. We refer to these metrics as \textit{Quantitative Metrics} from this point on, since they address the evaluation of the \textit{Negligible Utility Penalty} and \textit{Efficiency} requirements, which are both quantitative by nature (see Section~\ref{sec:problem-description}).
The utility penalty metric is used to address the \textit{Negligible Utility Penalty} requirement of \techniqueNameShort{}.
In the utility penalty formula, $n$ equals the total number of adaptation cycles, $q_{o}^{i}$ represents the quality value in cycle $i$ which would have been chosen in an optimal situation and $q_{c}^{i}$ represents the quality value in cycle $i$ chosen by our proposed solution.

\begin{wrapfigure}{r}{.4\textwidth}
    \centering
    \includegraphics[width=\linewidth]{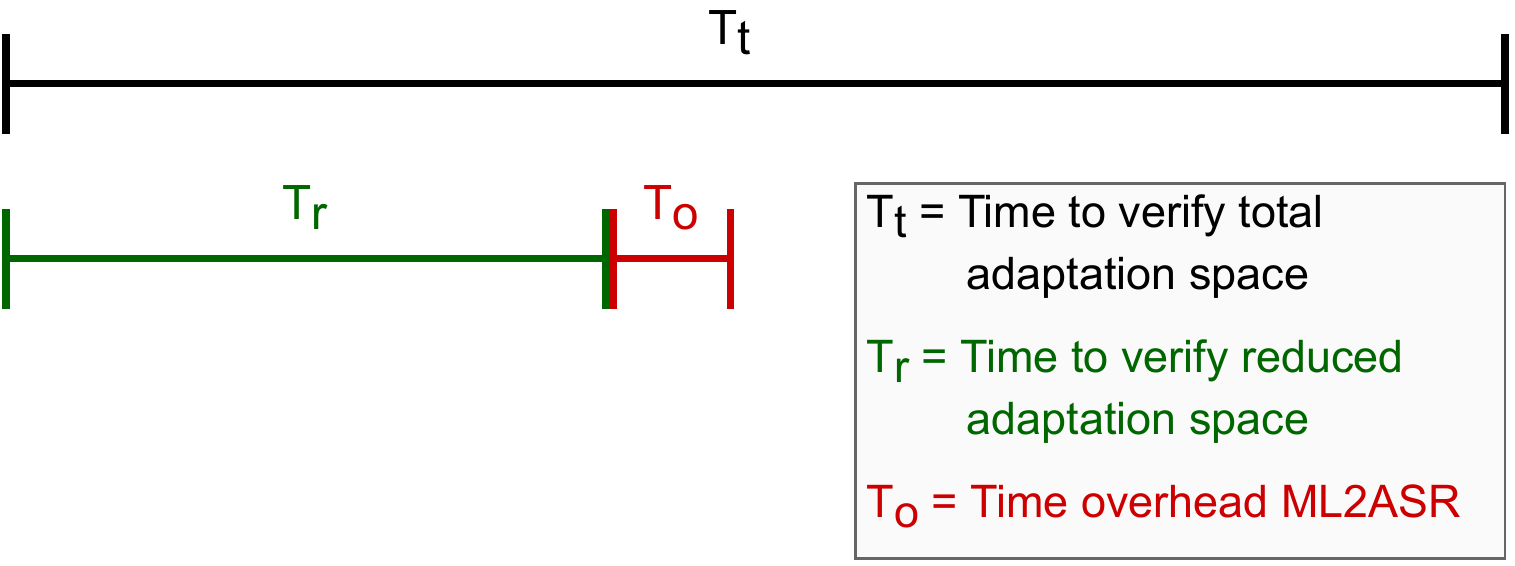}
    \caption{Time variables used in the efficiency metrics.}
    \label{fig:time-definitions}
\end{wrapfigure}

To address the \textit{Efficiency} requirement of \techniqueNameShort{}, we define three metrics: average adaptation space reduction, learning time overhead, and overall time saved. 
In the formula of average adaptation space reduction (AASR), $selected$ represents the average number of adaptation options selected by learning over multiple adaptation cycles, and $total$ represents the average of the total number of adaptation options over these adaptation cycles.
For the remaining formulas, the parameters $T_x$ refer to one of the time units as defined in Figure~\ref{fig:time-definitions}.

\begin{table}
    \caption{The evaluation metrics used throughout the evaluation section.}
    \centering
    \begin{tabular}{|M{3cm}|m{7cm}|c|c|}
        \hline
        \textbf{Name} & \textbf{Description} & \textbf{Formula} & \textbf{Objective} \\\hline
        \textit{Utility Penalty} & The average difference in value of quality properties of the system obtained by applying the reference approach, \dlaser{}, and \techniqueNameShort{}. & $\frac{\sum_{i=0}^{n} |q_{o}^{i} - q_{c}^{i}|}{n}$ & Minimize \\\hline
        \textit{Average Adaptation Space Reduction} & The average proportion of adaptation options that were filtered by \dlaser{} and \techniqueNameShort{}. & $(1 - \frac{selected}{total}) \times 100$ & Maximize \\\hline
        \textit{Learning Time Overhead} & The average proportion of additional time introduced by \dlaser{} and \techniqueNameShort{} at runtime. & $(1 - \frac{T_r + T_o}{T_t}) \times 100$ & Minimize \\\hline
        \textit{Overall Time Saved} & The average proportion of total time saved of \dlaser{} and \techniqueNameShort{} (taking into account overhead) compared to the reference approach. & $(\frac{T_o}{T_o + T_r}) \times 100$ & Maximize \\\hline
    \end{tabular}
    \label{tab:evaluation-metric-values}
\end{table}

\section{Evaluation \techniqueNameShort{}}\label{sec:evaluation}

We evaluate and benchmark \techniqueNameShort{} on two cases from different domains: DeltaIoT~\cite{Iftikhar2017} and a \sbs{} that is based on TAS~\cite{weyns2015tele}. DeltaIoT is a small IoT system with only threshold and optimization goals and a rather small adaptation space of 216 adaptation options. The \sbs{} is a more challenging case with threshold, setpoint, and optimization goals and an adaptation space of 13500 adaptation options. 

We start with the evaluation with DeltaIoT and then look at the \sbs{}. We present the results of the evaluation for different scenarios. For the runtime stage, we focus on the evaluation of the requirements with quantitative metrics: utility penalty, average adaptation space reduction, overall time saved, and learning time overhead. We elaborate on the other requirements in the discussion in Section~\ref{sec:discussion}.

Both applications are evaluated using a simulator. Simulations are run on a computer system with an AMD Ryzen 7 Pro 3700u CPU with 13.7GB of RAM. For the learning approaches, we have used the implementations of the \textit{Scikit-Learn} algorithms (classifiers, regressors, scalers)~\cite{scikit-learn}. The full replication package is available online.\footnote{\url{https://people.cs.kuleuven.be/danny.weyns/material/ML2ASR/}} 

\subsection{Evaluation with DeltaIoT}

We start with introducing DeltaIoT. Then we present two evaluation scenarios and we explain the benchmarks we use. Next, we present the results of the design stage activities and finally the results of the runtime stage activities.

\subsubsection{DeltaIoT Application}

\begin{figure}
	\centering
	\includegraphics[width=\linewidth]{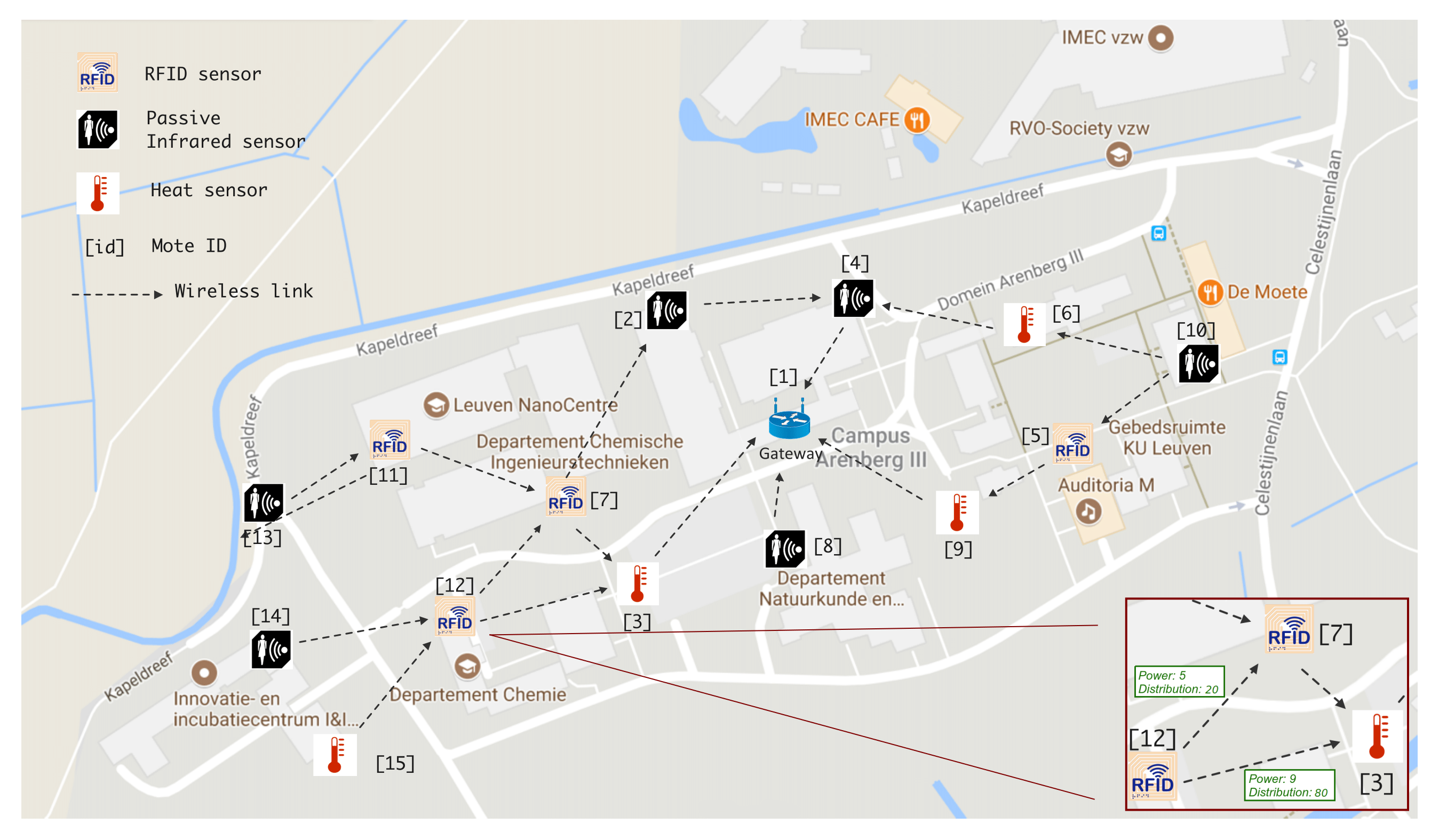}
	\caption{Deployment of DeltaIoT at the KU Leuven campus.}
	\label{fig:DeltaIoT-deployment}
\end{figure}

DeltaIoT is a small Internet-of-Things (IoT) application that offers a smart environment monitoring service. The application is developed by VersaSense.\footnote{\url{www.versasense.com}} 
The IoT network comprises 15 Long-Range (LoRa) motes that are deployed at the KU Leuven Computer Science Campus as shown in Figure~\ref{fig:DeltaIoT-deployment}. Each mote is equipped with a sensor (temperature, RFID and infrared) that periodically collects data and sends this data to a gateway. An end-user application processes the data allowing users to monitor the Campus area and take action when needed.

The network uses time-synchronized communication organized in cycles. Each cycle consists of a number of communication slots between a sender and a receiver mote. The slots are allocated from the leaf nodes of the network towards the gateway. Each mote has an internal buffer to store its own generated data and data received from other motes. When a mote is allocated a communication slot, it sends the data of the buffer to the receiving mote of the slot.

\begin{figure}[!htb]
    \centering
    \includegraphics[width=.7\linewidth]{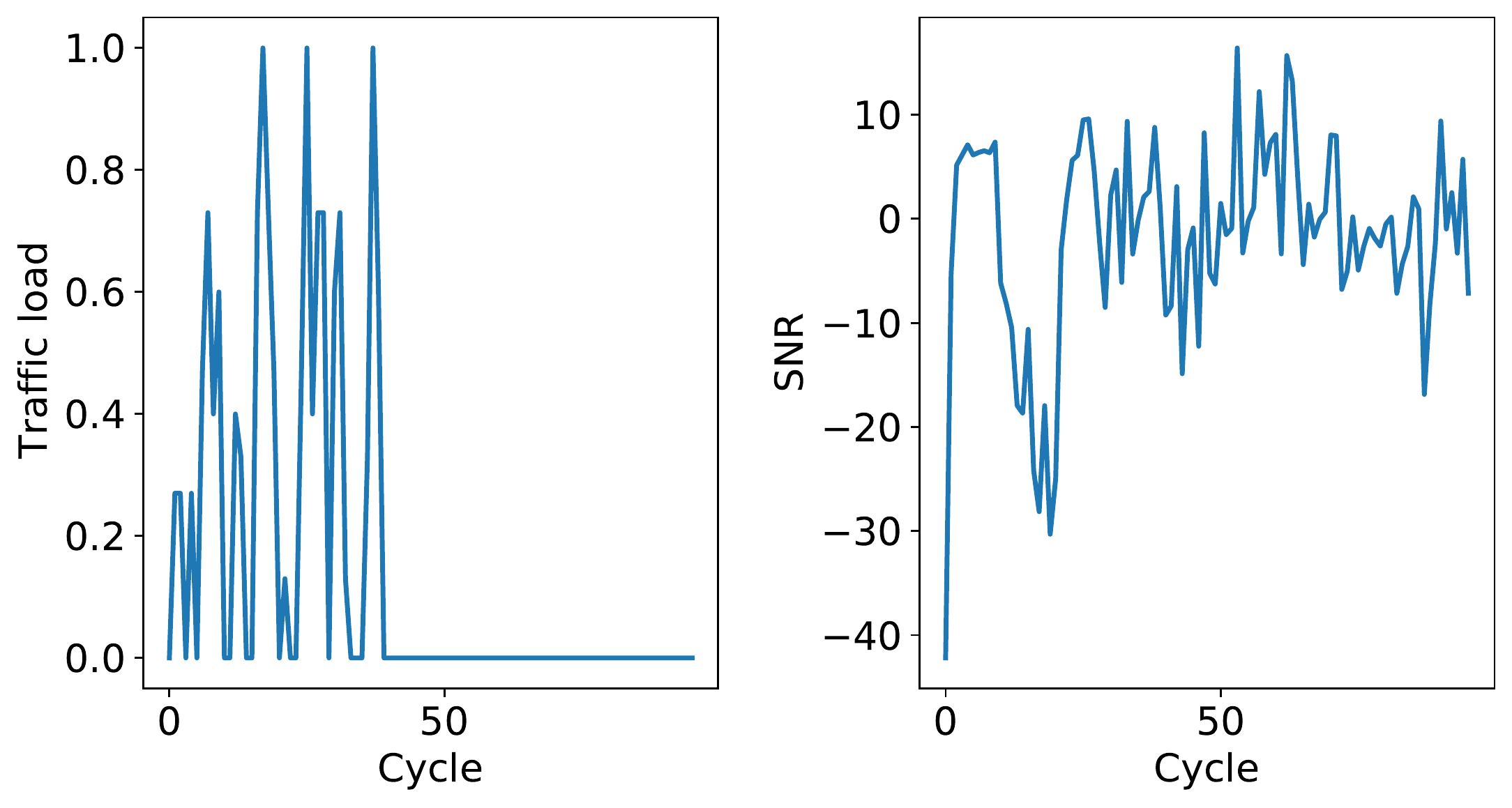}
	\caption{Example of traffic load generated by mote 13 and network interference on the link between mote 12 and mote 3 over 100 cycles.}
	\label{fig:SNRLoadProfiles}
\end{figure}

\paragraph{Uncertainties} We consider two important types of uncertainties: dynamics in the traffic load and interference of the wireless network. Dynamics in the traffic load result from variations in the frequency that sensors take samples and transmit data. For example: a temperature sensor collects and sends measurements periodically, while an RFID sensor only sends data when it is available, e.g., when a person scans an RFID badge. As a result, the load of packets that need to be sent to the gateway fluctuates. 
Interference of the wireless network arises from dynamic conditions in the environment, such as weather conditions or the presence of other wireless networks. Interference may result in the loss of packets communicated over the link. 
Figure~\ref{fig:SNRLoadProfiles} shows excerpts with data of both types of uncertainties over a period of time. This data is based on measurements of DeltaIoT in the field.\footnote{Network interference is represented as the Signal-to-Noise ratio (SNR). An SNR below 0 may lead to the loss of packets.} 

\paragraph{Quality Goals}

Besides \textit{what} the network should do, i.e., collecting data at the gateway, stakeholders of DeltaIoT also have demands on \textit{how} this is done, i.e., the quality of the transmission. We consider three quality goals of DeltaIoT: packet loss, latency, and energy consumption.~As explained above, packet loss depends on network interference. Latency depends on the traffic load in the network since only a limited number of packets can be transmitted during a time slot. The remaining packets remain in the buffers for communication in the next slot, causing delays in the transmission of data. Lastly, energy consumption depends on the number of packets that motes need to communicate and the power that is used to communicate packets. 
Evidently, stakeholders prefer to keep the packet loss, latency, and energy consumption low. However, these qualities are conflicting, for instance, using less energy (lower power) over a network link may result in higher packet loss as the signal may get lost in the noise along the link. 

\paragraph{Adaptation of the IoT Network}

To ensure the quality goals during operation, DeltaIoT offers a management interface that is connected with the gateway. This interface can be used to observe the behavior of the network (e.g., the interference along links, the packets lost over a time period, etc.) and change the settings of the motes in the network. Here, we consider two types of settings. First, the power used to transmit packets over an outgoing link of a mote can be set in a range $[0\ldots15]$ (0 is minimum power and 15 maximum power). Sending packets with a higher power setting reduces the chance of packets being lost over a noisy link, but it  consumes more power. Second, for motes with more than one outgoing link the distribution of the packets over these links can be set. This way, the transmission of packets along paths with high interference or high traffic can be reduced or avoided, yet the packets may follow a longer path requiring more energy. Since motes in DeltaIoT have at most two parent motes, we consider the following distribution settings for these motes: $0-100$, $20-80$, $40-60$, $60-40$, $80-20$, $100-0$. An example configuration is shown in Figure~\ref{fig:DeltaIoT-deployment} (bottom right corner). Here a power setting of 5 is used for the upper link that transmits 20\% of the packets, and a power setting of 9 is used for the bottom link that transmits 80\% of the packets.

Without self-adaptation, an operator is responsible for ensuring the quality goals by monitoring the network and adjusting the settings using the management interface. This is a tedious and costly task that is often not very efficient. To that end, we add a managing system (MAPE feedback loop) to the system that connects with the management interface to automate the adaptation of the settings. We use such setting for the evaluation of \techniqueNameShort{}.

\subsubsection{Evaluation Setup}\label{subsubsubsec:deltaiot-scenarios}

For the evaluation with DeltaIoT, we used a simulation of the network with 15 motes as shown in Figure~\ref{fig:DeltaIoT-deployment}. We applied 300 communication cycles of the network 
that correspond with a wall clock time of around three days. We used uncertainty profiles for traffic load of motes and network interference that are based on measurements of the physical network. For the traffic load, motes generate between 0 to 10 packets per cycle. The level of interference (SNR) fluctuates between -40dB and +15dB. Figure~\ref{fig:SNRLoadProfiles} shows two example profiles we used. 

\paragraph{Adaptation Goals}

We devised two evaluation scenarios with learning tasks for different adaptation goals summarized in Table~\ref{tab:deltaiot-scenarios}.
In scenario 1, learning needs to predict and filter adaptation options based on these two threshold goals. In scenario 2, learning needs to additionally predict and filter adaptation options for an optimization goal. Note that a threshold goal that should keep the average packet loss under 10\% over a period of 12 hours, implies that on average 90\% of the transmitted packets should be received by the gateway. On the other hand, a threshold goal that should keep the average latency under 5\% over a period of 12 hours, implies that on average at least 95\% of packets generated in a cycle should be received by the gateway within that cycle. 

\begin{table}
    \centering
        \caption{Evaluation scenarios for DeltaIoT with the adaptation goals considered for learning.}
    \label{tab:deltaiot-scenarios}
    \begin{tabular}{|M{1.65cm}|m{3.6cm}|m{3.6cm}|m{3.6cm}|}
        \cline{2-4}
        \multicolumn{1}{l|}{} & \textbf{Adaptation goal 1} & \textbf{Adaptation goal 2} & \textbf{Adaptation goal 3} \\ \hline
        \textbf{Scenario 1} & \textit{Threshold}: the average packet loss over 12 hours should not exceed 10\% of the messages sent. & \textit{Threshold}: the average latency over 12 hours should not exceed 5\% of the cycle time. & N/A \\ \hline
        \textbf{Scenario 2} & \textit{Threshold}: the average packet loss over 12 hours should not exceed 10\% of the messages sent. & \textit{Threshold}: the average latency over 12 hours should not exceed 5\% of the cycle time. & \textit{Optimization}: the average energy consumption over 12 hours should be minimized. \\ \hline
    \end{tabular}
\end{table}

\paragraph{Adaptation Settings}\label{subsubsubsec:deltaiot-settings}

Adaptation options are composed in each cycle following two steps. Firstly, the power setting is determined for each link of each mote. These settings are determined such that the current Signal to Noise ratio (SNR) over each link is at least 0dB. The adaptation options are then determined based on the possible distribution settings for outgoing links of motes with two parents ($0-100$, $20-80$, etc.).
As such, the complete adaptation space for the DeltaIoT case consists of $6^3 = 216$ adaptation options.\footnote{The power setting for each individual mote remains fixed within a single adaptation cycle. As such, adaptation options vary based on the different distribution settings. For the configuration of DeltaIoT, there are six variations of distribution settings for a mote with two parents, and there are three motes which have two parents, hence the total of $216$ adaptation options.}
The MAPE feedback loop and the quality models have been designed as networks of timed automata models. These models are directly executed at runtime using the ActivFORMS execution engine~\cite{IftikharLW16}. The analysis of the adaptation options is performed using the runtime models by applying statistical model checking at runtime using runtime statistical model checking with Uppaal-SMC~\cite{David2015}.

\paragraph{Benchmarks}\label{subsubsubsec:deltaiot-benchmarks} We benchmark \techniqueNameShort{} using \review{three} approaches. First, we use a baseline approach that analyzes the whole adaptation space without using machine learning. Second, we use a competing approach, called \dlaser{}, that applies a deep neural network to reduce adaptation spaces~\cite{VanDerDonckt2020}.\footnote{We selected~\cite{VanDerDonckt2020} since this approach is conceptually similar to \techniqueNameShort{}, relying on learning to provide first-class support for adaptation space reduction. Most other related approaches mix adaptation space reduction with decision-making, while our main focus is on adaptation space reduction. Since the preliminary version of ML2ASR~\cite{Quin2019} only supports one type of goal, we have not used it as a benchmark in this study.} We have rerun the results presented in~\cite{VanDerDonckt2020} to ensure that the same settings were used to compare \techniqueNameShort{} and \dlaser{}. \review{Third, as a sanity check, we used an approach that selects a subset of adaptation options randomly. We average the obtained results over 10 runs to reduce variability. We highlight the results of this random approach separately and focus on statistically relevant differences.}\\\vspace{-5pt}

\noindent In the next sections, we start with the evaluation results of the design stage. Then we present the results of the runtime stage. To conclude, we summarize the machine learning activities in both stages. 

\subsubsection{Design Stage Evaluation with DeltaIoT} \label{subsec:deltaiot-design-stage-activities}

\paragraph{Data Collection}

We collected data of 300 cycles of DeltaIoT to derive the machine learning modules for both scenarios, \review{each cycle containing 216 data points}. Experiments showed that 300 cycles for the design stage activities ensured that the learners performed well during runtime. As explained in Section~\ref{sec:approach}, the collected data consists of a set of feature vectors that represent adaptation options with uncertainties, and quality vectors that represent the qualities of the corresponding adaptation options. 

\paragraph{Feature Extraction}

After collecting the data, we applied \textit{Feature extraction}. The first activity, \textit{Feature selection}, removes features from the collected feature vectors that do not have an influence on the resulting qualities in the system. Based on feature extraction 
34 of the original 65 individual features were selected as relevant. For instance all features related to SNR were selected. An example of a feature that was not selected is the load of motes that generate a constant number of packets, for instance motes that periodically track the temperature in the environment. 
Next, we use the pruned data to perform the second activity of feature extraction: \textit{Feature engineering}. For both scenarios, we selected the Min-Max scaler for threshold goals. For the optimization goal, no scaler was selected as this provided the best results. As \textit{Feature engineering} closely ties with \textit{Model selection} we explain the results below. Section~\ref{metrics:feature-extraction} describes how \textit{Feature extraction} was done. For detailed results, we refer to the website with the replication package. 

\paragraph{Machine Learning Model Identification}

In the first activity, \textit{Model evaluation}, we evaluated three different types of classifiers and two types of regressors. For the second activity, \textit{Model selection}, we closely examined the evaluation metrics for each model to make a decision on the learning models to be used at runtime; Section~\ref{metrics:machine-learning-metrics} describes how this was done.   Table~\ref{tab:deltaiot-model-selection-summary} summarizes the chosen machine learning models and their corresponding metric values obtained during the evaluation process. \ref{app:model-selection-deltaiot} provides a detailed description of the chosen machine learning models.

\begin{table}[h!]
    \centering
    \caption{Summary of the chosen machine learning models during \textit{Model evaluation} and \textit{Model selection} for DeltaIoT (abbreviations: \abbrevML{}, \abbrevDeltaIoT{}).}
    \begin{tabular}{|c:c|m{5cm}|m{5cm}|N}
        \cline{2-5}
        \multicolumn{1}{c|}{} & \textbf{Goal(s)} & \textbf{Model} & \textbf{Metrics} &\\\hline
        
        \scenarioTable{1}{S1} & \thresholdtext{< 10\%}{Pl}, \thresholdtext{< 5\%}{La} & SGD Classifier (log loss, l1 penalty)\newline MinMax Scaler & F1: $0.818$, MCC: $0.715$ &\\[8pt]\hline
        
        \scenarioTable[-8pt]{2}{S2} & \thresholdtext{< 10\%}{Pl}, \thresholdtext{< 5\%}{La} & SGD Classifier (log loss, l1 penalty)\newline MinMax Scaler & F1: $0.818$, MCC: $0.715$ &\\[8pt]\cline{2-4}
        
        & \optimizationtext{min}{energy consumption} & Passive Aggressive Regressor\newline \hspace*{5pt}(squared epsilon insensitive loss)\newline No Scaler & R2: $0.833$, \:MSE: $0.004$, \newline MAE: $0.043$, \:ME: $0.269$ &\\[8pt]\hline
    \end{tabular}
    \label{tab:deltaiot-model-selection-summary}
\end{table}

\paragraph{Exploration Rate and Warm-up Count} Finally, we tested different exploration rates (extra random adaptation options selected for verification) and warm-up counts (the number of training cycles to initialize the learning model) that are required for the runtime stage, see Section~\ref{metrics:machine-learning-metrics}. We selected 5\% as exploration rate and 45 cycles (of 300) as the warm-up count. For detailed results, see  Appendix~\ref{tab:deltaiot-modelselection}.

\subsubsection{Runtime Stage Evaluation with DeltaIoT} \label{subsec:deltaiot-runtime-stage-activities}

\paragraph{Hypothesis} For the evaluation of the runtime stage of \techniqueNameShort{} with DeltaIoT we use the following hypotheses:

\begin{itemize}
    \item[\textbf{H1}:] The utility penalties when applying \techniqueNameShort{} are negligible compared to the reference approach.
    \item[\textbf{H2}:] The utility penalties when applying \techniqueNameShort{} is not significantly higher compared to \dlaser{}.
    \item[\textbf{H3}:] \techniqueNameShort{} significantly reduces the adaptation space and hence the time required for verification compared to the reference approach.
    \item[\textbf{H4}:] The reduction of adaptation spaces with \techniqueNameShort{} is not significantly lower compared to \dlaser{}, nor does \techniqueNameShort{} require significantly more time for adaptation space reduction. 

\end{itemize}

\paragraph{Granularities for Adaptation Space Reduction with an Optimization Goal}

In scenario 2, \techniqueNameShort{}   predicts the energy consumption in the network (optimization goal), on top of predicting packet loss and latency (threshold goals). After filtering out options that are predicted to satisfy the threshold goals (be of class C\textsubscript{3}, see Section~\ref{metrics:machine-learning-metrics}), \techniqueNameShort{} reduces the adaptation space further based on the energy consumption predictions. We evaluate two cases: a reduction to at most 25 options and at most 10 options, corresponding to granularity values of 25 and 10, respectively.

\paragraph{Quality of the Learning Models}

Table~\ref{tab:deltaiot-machinelearning-metrics} summarizes the results for the quality of the machine learning models during runtime. The F1-score acquired is $0.757$ and the Matthews Correlation Coefficient is $0.646$ in scenario 1. This is in line with what we expect when comparing to the metrics retrieved during \textit{Model selection} in the design stage: an F1-score of $0.818$ and a Matthews Correlation Coefficient of $0.715$. For scenario 2 we obtained similar results for classification, albeit slightly worse results due to the increased reduction of the adaptation space. The regressor, which handles the energy consumption optimization goal, has an R2-score of $0.799$, a mean squared error of $0.0045$, a median absolute error of $0.0446$ and a maximum error $0.27$. Overall, the results in the runtime stage are good, showing similar results to other studies~\cite{Mori2020, Baresi2021}.

\begin{table}[h!]
    \caption{Values of the machine learning metrics for the runtime stage evaluation of the machine learning models of DeltaIoT (abbreviations: \abbrevDeltaIoT{}).}
    \label{tab:deltaiot-machinelearning-metrics}
    \centering
    \begin{tabular}{|c:c|c|c|N}
        \cline{3-5}
        \multicolumn{2}{c|}{} & \textbf{F1-score} & \specialcell{\textbf{Matthews correlation}\\\textbf{coefficient}} &\\[15pt]\hline
        
        \scenarioTable{1}{S1} & \thresholdtext{< 10\%}{Pl}, \thresholdtext{< 5\%}{La} & 0.757 & 0.646 &\\[8pt]\hline
        \scenarioTable{1}{S2} & \thresholdtext{< 10\%}{Pl}, \thresholdtext{< 5\%}{La} & 0.743 & 0.608 &\\[8pt]\hline
    \end{tabular}
    
    \bigskip
    
    \begin{tabular}{|c:c|c|c|c|c|N}
        \cline{3-6}
        \multicolumn{2}{c|}{} & \textbf{R2-score} & \specialcell{\textbf{Mean squared}\\\textbf{error}} & \specialcell{\textbf{Median absolute}\\\textbf{error}} & \textbf{Maximum error} &\\[15pt]\hline
        
        \scenarioTable{1}{S1} & \optimizationtext{min}{Ec} & 0.799 & 0.0045 & 0.0446 & 0.270 &\\[8pt]\hline
    \end{tabular}
\end{table}

\paragraph{Summary of Results for Quantitative Metrics}

Table~\ref{tab:deltaiot-evaluation-metrics} summarizes the results of the evaluation for the quantitative metrics. We discuss these results now in detail. 

\begin{table}[!ht]
    \caption{Values of the metrics for the runtime stage evaluation of requirements of DeltaIoT (abbreviations: \abbrevDeltaIoT{}).}
    \label{tab:deltaiot-evaluation-metrics}
    \centering
    \begin{tabular}{|c:c|c|c|c|c|c|c|N}
        \cline{3-8}
        \multicolumn{2}{c|}{} & \multicolumn{3}{c|}{\textbf{Utility penalties}} & \multirow{2}{*}{\textbf{AASR}} & \multirowcell{2}{\textbf{Overall time}\\\textbf{saved}} & \multirowcell{2}{\textbf{Time}\\\textbf{overhead}}  &\\[5pt]\cline{3-5}
        \multicolumn{2}{c|}{} & \textit{\small \textbf{Pl}} & \textit{\small \textbf{La}} & \textit{\small \textbf{Ec}} & & & &\\[1pt]\hline

        \rotatebox[origin=c]{45}{\textbf{S1}} & \textbf{N/A} & 0.045\% & 0.025\% & N/A & 56.5\% & 62.81\% & 0.05\% &\\\hline
        
        \scenarioTable{2}{S2} & \textbf{Granularity 25} & 0.091\% & 0.073\% & 0.008mC & 88.5\% & 90.82\% & 0.26\% &\\\cline{2-8}
        & \textbf{Granularity 10} & 0.515\% & 0.299\% & 0.019mC & 95.4\% & 96.37\% & 0.56\% &\\\hline

    \end{tabular}
\end{table}

\paragraph{Utility Penalties}

\begin{figure}[t!]
	\centering
	\includegraphics[width=.65\linewidth]{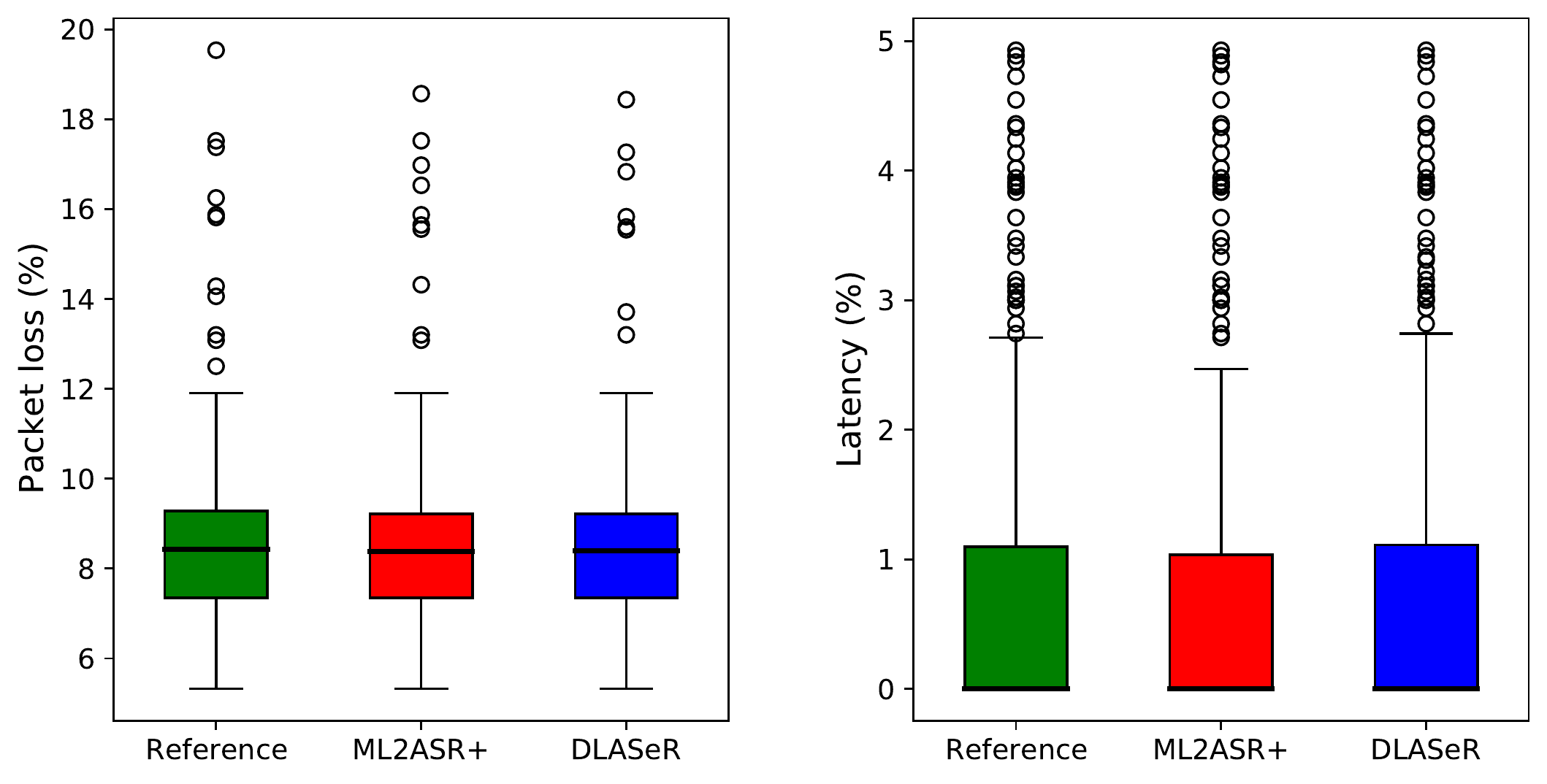}
	
	\bigskip
	
	\includegraphics[width=1\linewidth]{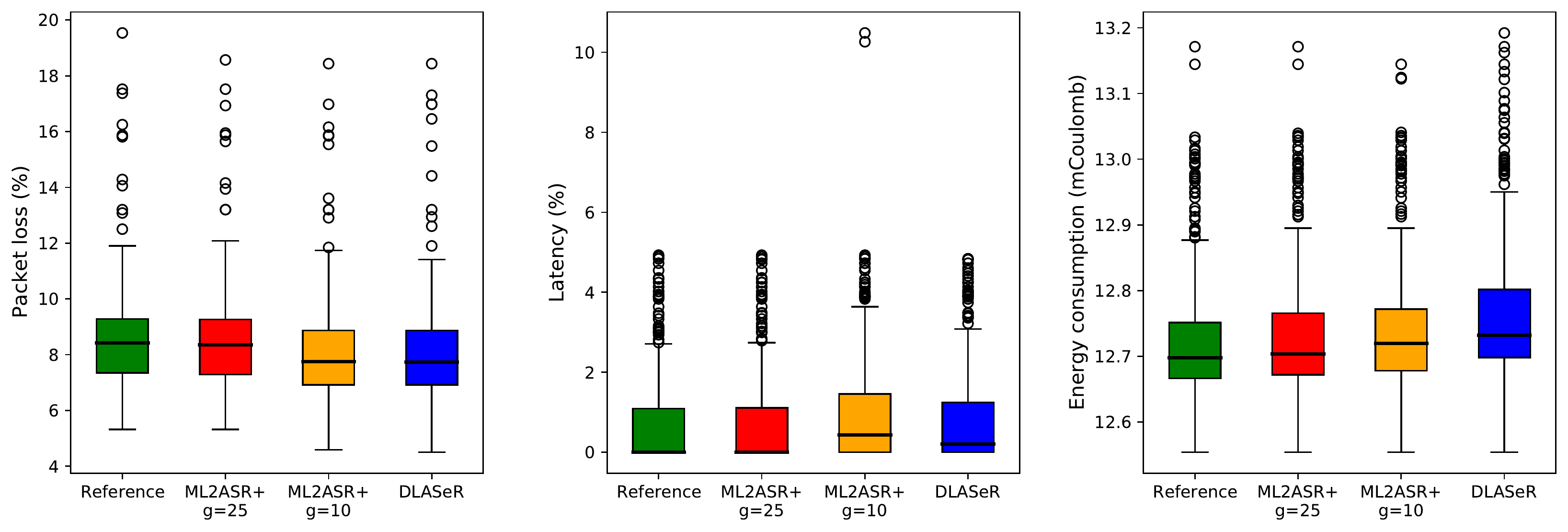}
	\caption{Utility penalties for scenario 1 (top) and scenario 2 (bottom) of DeltaIoT compared to the reference approach and \dlaser{}.}
	\label{fig:deltaiot-results-utility}
\end{figure}

Figure~\ref{fig:deltaiot-results-utility} shows the results for utility penalties. Note that the reference approach that exhaustively verifies all adaptation options provides optimal adaptation\footnote{Note that due to the nature of the application, in some adaptation cycles no adaptation option might exist that meets both threshold goals.}. 
First, we take a closer look at the threshold goals. Afterwards, we look at the optimization goal.

\paragraph{Threshold Goals}
When inspecting the results in detail, we notice that the values for the threshold goals with \techniqueNameShort{} are very close to those obtained with the reference approach. The average values of the penalties are respectively $0.045\%$ and $0.025\%$ for packet loss and latency in scenario 1, and $0.515\%$ and $0.299\%$ for the worst-case of scenario 2 with a granularity value of 10. The marginal increases that result from adaptation space reduction with \techniqueNameShort{} do not impede on the satisfaction of both goals compared to the reference approach in scenario 1 and scenario 2 with a granularity of 25. On the other hand, in scenario 2 with a granularity of 10 we notice that the threshold goals were not satisfied in 7 additional adaptation cycles after adaptation space reduction took place (cycles for which the reference approach does not violate the requirements). These results show that a lower granularity that substantially reduces the adaptation space for analysis may result in penalties for the quality properties of interest. This trade-off has to be carefully considered when making decisions about the granularity of adaptation space reduction.

Comparing the results of \techniqueNameShort{} with \dlaser{}, we observe that the satisfaction of the threshold goals are not impeded in both scenarios as opposed to the few violations in scenario 2 with a granularity value of 10 when using \techniqueNameShort{} (which corresponds to the strategy \dlaser{} employs: rank adaptation options based on predicted energy consumption, and subsequently look for adaptation options that meet both threshold goals).

\paragraph{Optimization Goal}
The results for the optimization goal show that the differences between the average values of energy consumption with the reference approach ($12.719mC$), \techniqueNameShort{} ($12.723mC$), and \dlaser{} ($12.724mC$) are marginal. This indicates that the reduced adaptation space most of the time also includes the adaptation option with the lowest energy consumption. 
For scenario 2 we observe values of $0.008mC$ and $0.019mC$ with \techniqueNameShort{} for granularity values of 25 and 10, respectively. Here we also notice a similar trade-off between granularity values and utility penalty: a more fine-grained reduction carries the risk of adapting the system less optimally compared to a less constrained strategy. Note that the penalty compared to the reference approach is still acceptable ($12.719mC$ mean energy consumption for the reference approach, $12.727mC$ for \techniqueNameShort{} with a granularity value of 25 and $12.739mC$ for \techniqueNameShort{} with a granularity value of 10) considering the significant time gain that both approaches offer (see below). For \dlaser{} we observe an average energy consumption of $12.769mC$ with a utility penalty for energy consumption of $0.038mC$, meaning that \techniqueNameShort{} performs quite well compared to the competing approach.

\begin{figure}
    \centering
    \includegraphics[width=.6\linewidth]{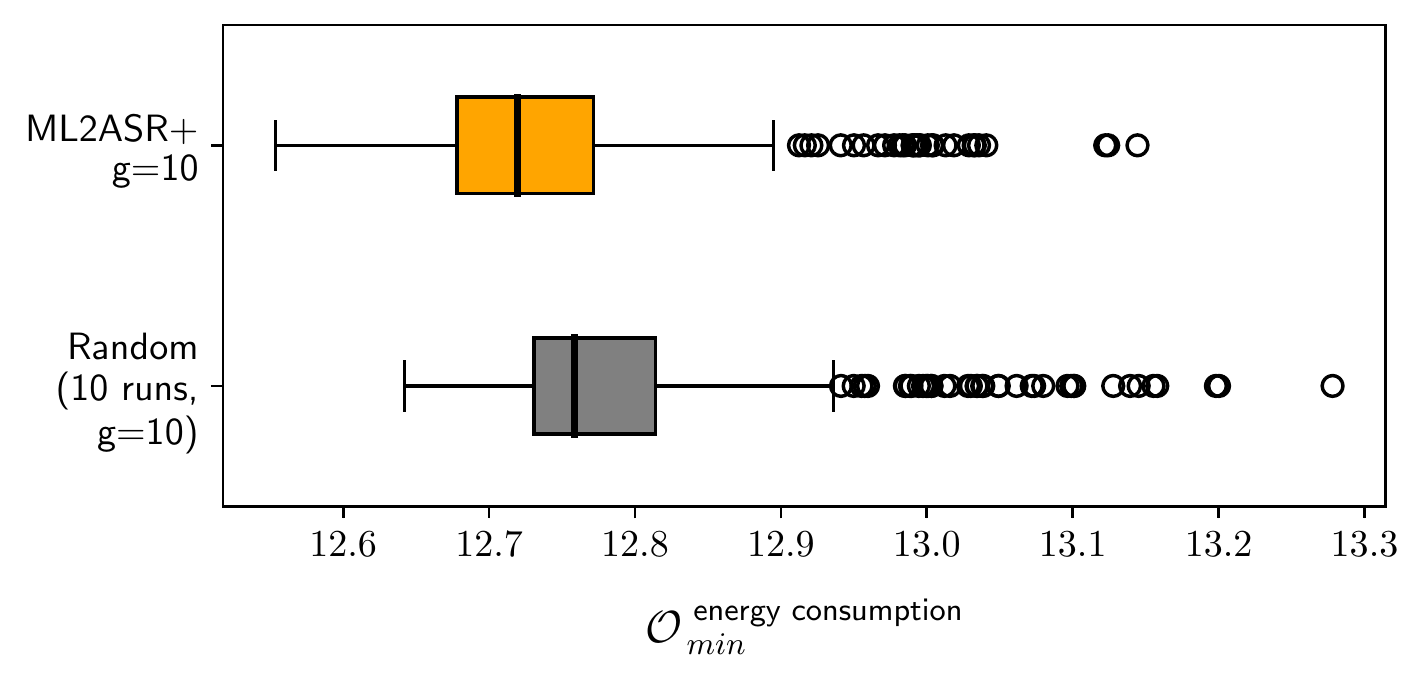}
    \caption{\review{Comparison of energy consumption for 10 random runs compared to \techniqueNameShort{} in scenario 2 of DeltaIoT.}}
    \label{fig:deltaiot-random-highlights}
\end{figure}

\paragraph{Sanity Check with Random Approach} \review{We compared \techniqueNameShort{} with a simple approach that randomly selects adaptation options (using an average of 10 random runs). 
For the threshold goals, packet loss and latency, both approaches satisfy the goals in both scenarios. However, the results show that the random approach violates the threshold goals for 28 adaptation cycles (of a total of 300 cycles), which is 16 cycles more compared to \techniqueNameShort{}. 
For the optimization goal of energy consumption in scenario 2, the Wilcoxon signed rank statistical test~\cite{Wohlin:2012} showed a significant difference between the random approach and \techniqueNameShort{} both for each random run and on the average of 10 random runs (for the latter we measured a p-value of $1.2e^{-15}$ with alpha level $0.05$)\footnote{\review{We add the caveat here that the results of these tests are not a general claim: the findings confirm a statistical difference in the 10 random runs we did, but this claim may not necessarily hold for other sets of 10 random runs.}}. Figure~\ref{fig:deltaiot-random-highlights} shows the distribution for the energy consumption of the IoT networks in scenario 2 with both approaches.
The average energy consumption with the random approach is $12.800mC$ compared to an energy consumption of $12.739mC$ for \techniqueNameShort{}. 
While the differences in absolute value of the  energy consumption is relatively small, the difference is statistically relevant. The second case will show that for more complex application scenarios, the impact is practically relevant.} \vspace{8pt}

\begin{tcolorbox}[colback=white]
\textbf{Hypotheses H1 (negligible utility penalties compared to reference approach) and H2 (utility penalties not significantly higher compared to \dlaser{})}. The results show that the utility penalties when applying \techniqueNameShort{} are negligible compared to the reference approach. \techniqueNameShort{} with a low granularity value in one scenario did not satisfy the threshold goals in all cycles, emphasizing the importance of a good selection of granularity. Comparing \techniqueNameShort{} with \dlaser{}, we notice that 
the utility penalties remain negligible 
for both threshold goals and the optimization goal. 
In conclusion, we can accept hypotheses H1 and H2. 
\end{tcolorbox}

\paragraph{Average Adaptation Space Reduction} 

\begin{figure}
    \centering
    \includegraphics[width=\linewidth]{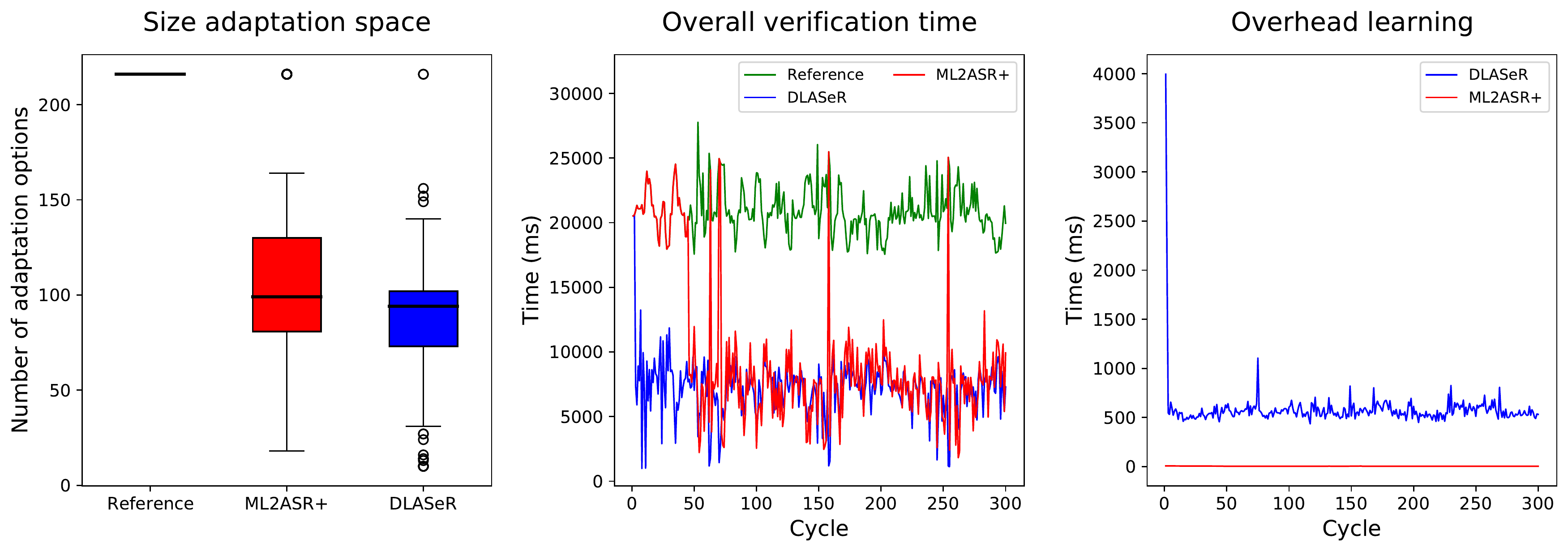}
    
    \vspace*{3pt}
    
    \includegraphics[width=\linewidth]{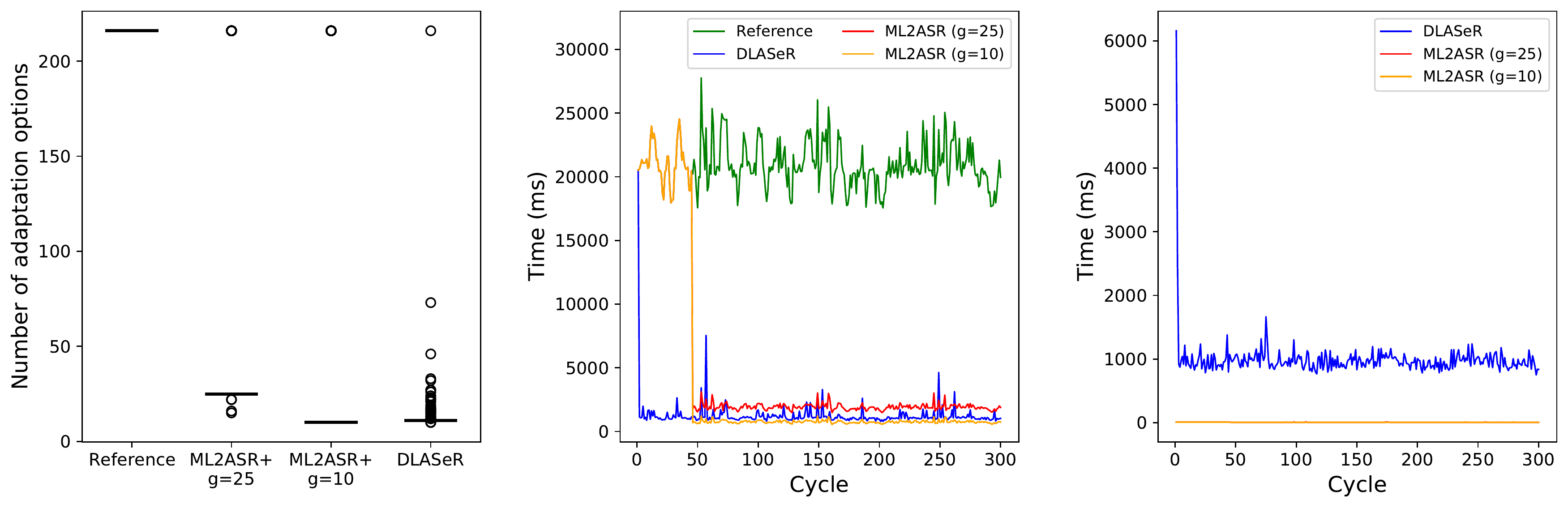}
    
    \caption{Number of verified adaptation options with \techniqueNameShort{} and \dlaser{} (left), overall time used (middle) and overhead (right) compared to the reference approach for scenario 1 (top row) and scenario 2 (bottom row) of DeltaIoT.}
    \label{fig:deltaiot-space-reduction-time}
\end{figure}

Figure~\ref{fig:deltaiot-space-reduction-time} (left) shows the size of the adaptation spaces for the three evaluated approaches. During the first 45 training cycles, when the \MLM{} of \techniqueNameShort{} is not exploited, all adaptation options are analyzed (multiple data points overlapping at 216 adaptation options for \techniqueNameShort{}). In the case of \dlaser{}, there is only a single entry at 216 adaptation options corresponding to the only training cycle. 

Applying \techniqueNameShort{} results in an Average Adaptation Space Reduction (AASR) of 56.5\% for the scenario 1, 88.5\% for scenario 2 with a granularity value of 25 and 95.4\% with a granularity value of 10. For scenario 1 this means that, on average, more than half of the adaptation options available in the adaptation space are filtered out before verification is applied. For scenario 2, we obtained results that match in most cases the granularity value.

For \dlaser{} we obtain an Average Adaptation Space Reduction of 58.8\% for scenario 1, a result similar to the one obtained with \techniqueNameShort{}. However, for scenario 2, \dlaser{} works differently: the approach relies on deep learning models starting with predicting the energy consumption of all adaptation options; then it iterates over the adaptation options (from low energy consumption predictions to high) until an adaptation option is found that meets both threshold goals. This way, \dlaser{} achieves an average adaptation space reduction of 94.19\% in scenario 2.

\paragraph{Learning Time Overhead} 
Figure~\ref{fig:deltaiot-space-reduction-time} (right) shows the overhead introduced by \techniqueNameShort{} (red and yellow lines) and \dlaser{} (blue line). The learning overhead is on average less than 1\% of the total time necessary to both reduce and verify the reduced adaptation space for \techniqueNameShort{} in both scenarios. Concretely, the overhead of \techniqueNameShort{} is at most 4.28ms, which is less than 10\% of the time required to verify a single adaptation option. We conclude that this overhead is negligible compared to the time necessary to verify the selected subset of adaptation options. 

\dlaser{} on the other hand introduces a slightly higher overhead of 8.34\% in scenario 1. Even though this number is higher, it is important to bear in mind that the overhead is still a minor part of the overall time required for verification of the reduced adaptation space. For scenario 2 however, we notice a significantly higher overhead of 45.96\% for  \dlaser{} due to the strategy \dlaser{} employs for adaptation space reduction. Even though \dlaser{} reduces the adaptation space to a small subset, the overhead is significantly higher
than \techniqueNameShort{}.

\paragraph{Overall Time Saved} 

Figure~\ref{fig:deltaiot-space-reduction-time} (middle) shows the overall time used to analyze all the adaptation options in each cycle with the reference approach (green), and the time used to reduce and verify the adaptation space with \techniqueNameShort{} (red and yellow) and \dlaser{} (blue). 
We observe that in scenario 1 \techniqueNameShort{} saves more than half of the time (62.81\%) for verifying the reduced adaptation space compared to the reference approach. This observation is in line with the average adaptation space reduction, resulting in a significant time gain compared to the reference approach. \dlaser{} shows results that are also in line with the average adaptation space reduction, albeit slightly worse due to the higher overhead introduced by the approach (62.59\% of the time saved). Similarly for scenario 2 we observe results closely aligned with the adaptation space reduction metric since the learning time is negligible: 90.82\% and 96.37\% for granularity values 25 and 10 respectively. For \dlaser{} in scenario 2 we notice an average time saved of 89.69\%.\vspace{8pt}

\begin{tcolorbox}[colback=white]
\textbf{Hypotheses H3 (significant reduction of adaptation spaces and time gain) and H4 (adaptation space reduction comparable to \dlaser{}).} \techniqueNameShort{} realizes a significant reduction of the adaptation space of 56.5\% for scenario 1 and over 90\% for scenario 2, resulting in an overall time saving for analysis of 62.81\% compared to the reference approach in scenario 1 and again over 90\% for scenario 2. \techniqueNameShort{} and \dlaser{} realize a similar adaptation space reduction in scenario 1 and scenario 2 with a granularity value of 10. Yet, the time required for adaptation space reduction with \techniqueNameShort{} is negligible ($<1\%$), and small to significantly larger for \dlaser{} ($8.34\%$ to $45.96\%$). In conclusion, we can accept hypotheses H3 and H4.
\end{tcolorbox}

\review{\subsubsection{Summary of design stage and runtime stage machine learning activities} Table~\ref{tab:deltaiot-chart-ML} summarizes the number of inputs, features, objective variables and metrics of the learning activities for DeltaIoT in each of the design stage and runtime stage activities.}

\begin{table}[h!]
    \centering
    \caption{\review{The number of inputs, features, objective variables and metrics for the activities of the machine learning pipeline of scenario 2 of DeltaIoT in the design and runtime stage (separated by the double line). The prediction column is marked in red to indicate that it is not be used yet in the training cycles. The number of inputs for  online learning is determined by the number of options that could be verified by the Verifier. Abbreviations: \abbrev{``f vectors''}{feature vectors}, \abbrev{``q vectors''}{quality vectors}, \abbrev{``Pl''}{packet loss}, \abbrev{``Ec''}{energy consumption}, \abbrev{``La''}{latency}.}}
    \label{tab:deltaiot-chart-ML}
    \footnotesize
    \begin{tabular}{|c|c|c|@{}|c|c|c|c|}
        \hhline{~|*{6}{-}}
        \multicolumn{1}{c|}{} 
            & \specialcell{\textbf{Feature}\\\textbf{Extraction}} 
            & \specialcell{\textbf{Machine Learning}\\\textbf{Model Identification}} 
            & \specialcell{\textbf{Feature}\\\textbf{Extraction}} 
            & \cellcolor{red!10}\textbf{Prediction} 
            & \textbf{Verification} 
            & \textbf{Online Learning} \\\hline
        
        \specialcell{\textbf{Number of}\\\textbf{inputs}} 
            & \specialcell{$300*216$ f vectors\\$300*216$ q vectors} 
            & \specialcell{$300*216$ f vectors\\$300*216$ q vectors} 
            & $216$ f vectors 
            & \cellcolor{red!10} $216$ f vectors 
            & $216$ f vectors 
            & \specialcell{$X$ f vectors\\$X$ q vectors} \\\hline 
        
        \specialcell{\textbf{Number of}\\\textbf{features}} 
            & 65 features 
            & 34 features 
            & 65 features 
            & \cellcolor{red!10} 34 features 
            & \cellcolor{gray!25}x 
            & 65 features 
            \\\hline 
        \specialcell{\textbf{Objective}\\\textbf{variables}} 
            & \specialcell{1. Pl + La class\\2. Ec value}
            & \specialcell{1. Pl + La class\\2. Ec value}
            & \cellcolor{gray!25}x
            & \cellcolor{red!10} \specialcell{1. Pl + La class\\2. Ec value}
            & \cellcolor{gray!25}x
            & \specialcell{1. Pl + La class\\2. Ec value}
            \\\hline 
        \textbf{Metrics} 
            & \cellcolor{gray!25}x
            & \specialcell{1. F1, MCC\\2. R2, MSE, MAE, ME}
            & \cellcolor{gray!25}x
            & \cellcolor{gray!25}x
            & \cellcolor{gray!25}x
            & \cellcolor{gray!25}x
            \\\hline 
    \end{tabular}
\end{table}

\subsection{Evaluation with the \sbs{}}

We present now the evaluation resuls of the second case: a service-based system. We follow the same structure: we start with introducing the application, evaluation scenarios and the benchmarks we use. Then, we present the results of the design stage activities and finally the results of the runtime stage activities.

\subsubsection{\sbs{} Application}

\begin{figure}[b!]
	\centering
	\includegraphics[width=\linewidth]{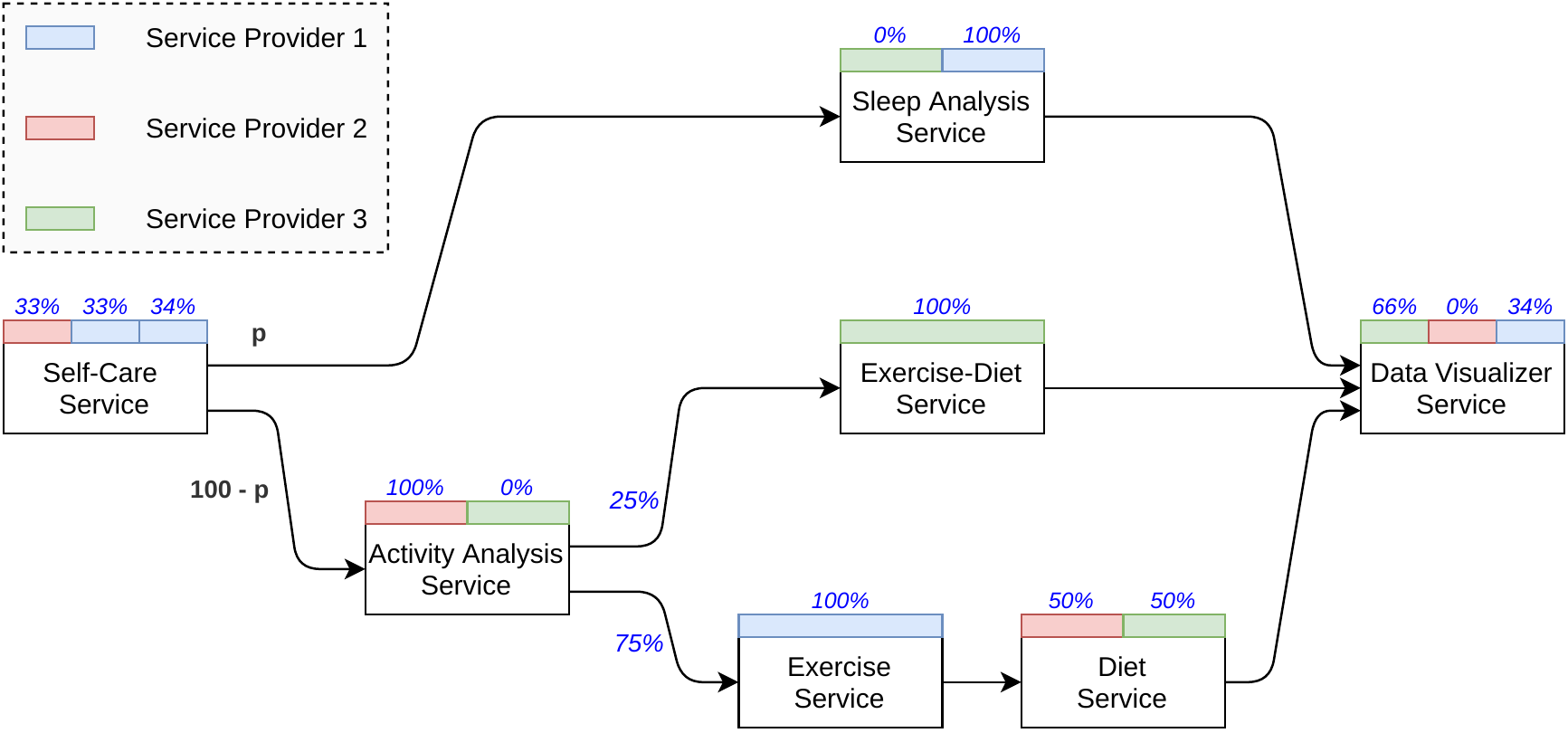}
	\caption{Workflow of the \sbs{} health monitoring and processing application. Blue percentages denote an example system configuration of the application (further elaborated on in Section~\ref{subsubsec:sbs-self-adaptation}).}
	\label{fig:sbs-workflow}
\end{figure}

Self-care enables patients to self-manage their illness~\cite{Bodenheimer2002,Riegel2012}. We consider a concrete example of a smartwatch application that analyzes data of patients and visualizes the result for the patient~\cite{weyns2015tele}. Our focus is on the underlying service-based system that processes the patient data. Figure~\ref{fig:sbs-workflow} describes the workflow of this system. 

The system consists of a set of services that perform tasks. The services are composed in a workflow with two main branches. The ``sleep branch'' analyzes the data of patients when they sleep that can be visualized for the patient afterwards. The ``awake branch'' analyzes data of different activities, processing the data using exercise and diet services and visualizing the results to the patient. Each branch fulfills its tasks using different services. For example, \textit{Exercise service} processes activity data regarding exercises and makes recommendations. Similarly, the \textit{Diet service} processes activity data regarding dietary information and makes recommendations. The \textit{Exercise-Diet service} combines both these responsibilities in a single service providing an alternative path in the workflow. 

The workflow defines service types that need to be instantiated. Three different service providers offer such service instances. These instances are marked in the workflow by small colored rectangles at the top of each service symbol. Service instances differ in the qualities they provide (e.g., response time) but also the cost for using them. During operation a concrete set of services instances is selected and used to handle incoming service requests.

\paragraph{Uncertainties} 

Each service provider is characterized by two parameters in our evaluation: its workload and the available bandwidth of its network. Both these parameters fluctuate at runtime representing uncertainties.
These fluctuations in turn affect the qualities of the service instances they provide, including the failure rate, response time and cost (see below). Figure~\ref{fig:sbs-uncertainties-influence} shows the models we used for the fluctuations of the qualities of service instances per service provider. Failure rate and cost increase with higher load, while response time decreases with lower bandwidth. 

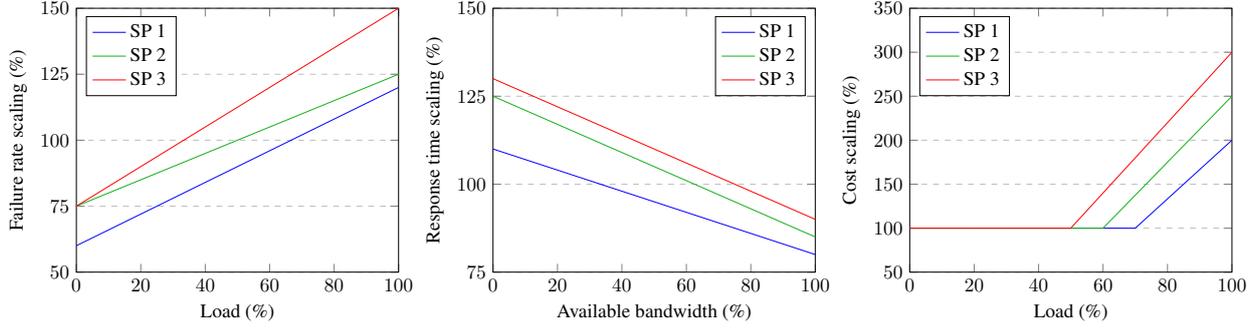
\begin{figure}
\centering
\resizebox{\linewidth}{!}{%
\begin{tikzpicture}
    \begin{axis}[
        width=0.45\linewidth,
        xlabel={Load (\%)},
        ylabel={Failure rate scaling (\%)},
        xmin=0, xmax=100,
        ymin=50, ymax=150,
        xtick={0,20,40,60,80,100},
        ytick={50,75,100,125,150},
        legend pos=north west,
        ymajorgrids=true,
        grid style=dashed,
        name=failurerate
    ]
    
        \addplot[color=blue]
            coordinates {
                (0, 60)(100, 120)
            };
        \addplot[color=darkgreen]
            coordinates {
                (0, 75)(100, 125)
            };
        \addplot[color=red]
            coordinates {
                (0,75)(100,150)
            };
            
        \legend{SP 1, SP 2, SP 3}
    \end{axis}
    
    \begin{axis}[
        width=0.45\linewidth,
        xlabel={Available bandwidth (\%)},
        ylabel={Response time scaling (\%)},
        xmin=0, xmax=100,
        ymin=75, ymax=150,
        xtick={0,20,40,60,80,100},
        ytick={75,100,125,150},
        legend pos=north east,
        ymajorgrids=true,
        grid style=dashed,
        name=responsetime,
        at=(failurerate.right of south east),
        anchor=left of south west
    ]
    
        \addplot[color=blue]
            coordinates {
                (0,110)(100,80)
            };
        \addplot[color=darkgreen]
            coordinates {
                (0,125)(100,85)
            };
        \addplot[color=red]
            coordinates {
                (0,130)(100,90)
            };
            
        \legend{SP 1, SP 2, SP 3}
    
    \end{axis}
    
    \begin{axis}[
        width=0.45\linewidth,
        xlabel={Load (\%)},
        ylabel={Cost scaling (\%)},
        xmin=0, xmax=100,
        ymin=50, ymax=350,
        xtick={0,20,40,60,80,100},
        ytick={50,100,150,200,250,300,350},
        legend pos=north west,
        ymajorgrids=true,
        grid style=dashed,
        name=cost,
        at=(responsetime.right of south east),
        anchor=left of south west
    ]
    
        \addplot[color=blue]
            coordinates {
                (0,100)(70,100)(100,200)
            };
        \addplot[color=darkgreen]
            coordinates {
                (0,100)(60,100)(100,250)
            };
        \addplot[color=red]
            coordinates {
                (0,100)(50,100)(100,300)
            };
            
        \legend{SP 1, SP 2, SP 3}
    \end{axis}
    
\end{tikzpicture}%
}
\caption{Effect of load and available bandwidth of individual service providers (SPs) on system qualities.}
\label{fig:sbs-uncertainties-influence}
\end{figure}

Besides fluctuations in work load and the available bandwidth, the system has to deal with an additional uncertainty, namely the distribution of service requests for sleep analysis (sleep branch) or activity analysis (awake branch). This distribution, denoted by the value \texttt{p} in the workflow, may change over time depending on the patient's behavior.

\paragraph{Quality Goals} 

In the evaluation, we consider three key qualities for stakeholders of the service-based system: the failure rate of service invocations, the response time, and the cost of invocations. Each service instance is characterized by a specific failure rate, response time and cost. 
Hence, the overall qualities for service requests are determined by the individual service instances that are selected to handle these requests. 
In particular, the overall failure rate is determined by the multiplication of each failure rate associated with the selected service instances. As an example, assume we invoke two service instances, each characterized by a failure rate of 5\%. The overall failure rate of a service request then corresponds to $1 - ( 0.95 * 0.95 ) = 0.0975$, i.e., 9.75\%. 
The overall response time of service requests is simply determined by the sum of the individual response times associated with these selected service instances. Similarly, the overall cost is determined as the sum of the costs associated with individual service instances.

Clearly, stakeholders want to keep the failure rate, response time, and cost as low as possible. Yet, these qualities conflict. Invoking a service with lower failure rate and/or lower response time will usually imply a higher cost. However, the selection of services is complicated by uncertainties. For instance, the cost to invoke a service of a service provider may increase when the service provider is under heavy load. Similarly, the failure rates and the response times of the provided service instances fluctuate in time. 

\paragraph{Adaptation of the \sbs{}}\label{subsubsec:sbs-self-adaptation}

Given the fluctuations in load and available bandwidth of service providers and changes of patient behavior, the selection of service instances may be changed dynamically based on the changing conditions. To that end, the system can be configured such that the requests are distributed in a particular way over different instances. In the evaluation setting, we use  service types with 2 and 3 instances. For services with 2 instances the system offers 3 possible configurations: 0/100\%, 50/50\% and 100/0\%. For services with 3 instances there are 10 possible configurations: 0/0/100\%, ..., 0/33/67\%, ...  100/0/0\%. This way, preference can be given to services with better actual quality values, or services can even be (temporally) avoided if necessary.  In addition, the parameter \texttt{\textalpha} that determines which path is taken in the awake branch (distinct services for the exercise and diet tasks or a combined service) can be set to one of four values: 0\%, 25\%, 50\%, 75\% and 100\%. Figure~\ref{fig:sbs-workflow} shows (on top of the general workflow) an example configuration of the workflow (with concrete selections for service instances and \texttt{\textalpha} set to 25\%). 

Without self-adaptation, it is practically infeasible for an operator to change the service selection dynamically. Hence, the only option for an operator would be to allocate a predefined set of possible service instances to the system and perform a coarse-grained adaptation. However, this would result in a sub-optimal solution or even worse in case particular services would fail unexpectedly. To that end, we add a managing system (MAPE-based feedback loop) to the system that monitors the changing conditions and adapts the service instances of the workflow dynamically when needed to maintain the stakeholder goals (failure rate, response time, and cost).

\subsubsection{Evaluation Setup}

We used a simulation of the setup as shown in Figure~\ref{fig:sbs-workflow}. We considered 30.000 service requests that are generated sequentially and processed individually by the system. Adaptation is triggered every 100 requests, resulting in 300 feedback loop iterations. The work load and available bandwidth are modelled as stochastic variables that gradually change during operation of the system (between 0 and 100\%). The change in values occurs by sampling a normal distribution with a standard deviation of 1.7, increasing or decreasing the bandwidth and work load of service providers. The factor \texttt{p} is initially set at 50\% and is modelled similarly to the work load and available bandwidth. 

\paragraph{Adaptation Goals}
We devised two scenarios of the service-based application as illustrated in Table~\ref{tab:sbs-scenarios}. In scenario 1 we consider three adaptation goals: two threshold goals and an optimization goal. Learning first filters out adaptation options based on the threshold goals, and subsequently orders and reduces the adaptation space according to the optimization goal. In scenario 2, learning has to deal with all three types of goals (threshold, setpoint, and optimization).

\begin{table}[!bth]
    \caption{System scenarios with their adaptation goals for the self-care application.}
    \label{tab:sbs-scenarios}
    \centering
    \begin{tabular}{|M{1.65cm}|m{3.4cm}|m{3.7cm}|m{3.7cm}|}
        \cline{2-4}
        \multicolumn{1}{l|}{} & \textbf{Goal 1} & \textbf{Goal 2} & \textbf{Goal 3} \\ \hline
        \textbf{Scenario 1} & \textit{Threshold}: the average \newline failure rate should not \newline exceed 10\%. & \textit{Threshold}: the average \newline response time should not exceed 10ms. & \textit{Optimization}: the average cost should be minimized. \\ \hline
        
        \textbf{Scenario 2} & \textit{Threshold}: the average \newline failure rate should not \newline exceed 10\%. & \textit{Setpoint}: the average \newline response time should be kept at 10ms. & \textit{Optimization}: the average cost should be minimized. \\ \hline
    \end{tabular}
\end{table}

\paragraph{Adaptation Settings}

The adaptation space for the \sbs{} is fixed. The adaptation options are determined based on the distribution of available service instances per service type and the setting of $\alpha$. Concretely, the total adaptation space comprises $5 * 10 * 10 * 3 * 3 * 3 = 13500$ adaptation options.\footnote{The numbers are composed following the description in Section~\ref{subsubsec:sbs-self-adaptation}: 5 represents the number of instantiations for $\alpha$, 10 represents the number of options for services with 3 instances (in total 2) and 3 represents the number of options for services with 2 instances (in total 3).} 
Similarly to the DeltaIoT case, we designed a MAPE feedback loop and quality models as networks of timed automata models that are directly executed using ActivFORMS~\cite{IftikharLW16}. For the analysis of the parameterized quality models, the feedback loop applies statistical model checking at runtime using Uppaal-SMC~\cite{David2015}.

\paragraph{Benchmark}
We benchmark \techniqueNameShort{} again with a reference approach that analyzes the whole adaptation space without using learning and with \dlaser{}~\cite{VanDerDonckt2020}, a state of the art approach. \review{As a sanity check, we use again an approach that selects adaptation options randomly over a set of runs to reduce variability}. It is important to note that the results for the different approaches are obtained from identical configurations and parameters settings of the application. For \techniqueNameShort{} and \dlaser{} the data is collected during simulation. Yet, for the reference approach the data is \mbox{collected during the design stage since analyzing the complete adaptation space for one cycle takes around 2 hours.} \\\vspace{-5pt}

\noindent In the next sections, we start with the results of the design stage. Then we present the results of the runtime stage. To conclude, we summarize the machine learning activities in both stages.

\subsubsection{Design Stage Evaluation with the \sbs{}}
\paragraph{Data Collection, Feature Extraction, Machine Learning Model Identification}
The design stage activities for the \sbs{} followed the same procedure as the activities for DeltaIoT (see Section~\ref{subsec:deltaiot-design-stage-activities}). First, we collected data from the system used to derive the machine learning modules for both scenarios. This data consisted of a set of feature vectors (composed of adaptation options and uncertainties in the application) and a set of quality vectors. We collected data for 100 adaptation cycles corresponding to 10.000 service requests, \review{each adaptation cycle containing 13500 data points}.
Then we performed \textit{Feature extraction}. During \textit{Feature selection}, all 22 features were selected as relevant, e.g., all features concerning the distribution of service request over service instances and all the features concerning the load of the service providers. 
During \textit{Feature engineering}, we determine which scaling algorithms to use to adjust feature values, following Section~\ref{metrics:feature-extraction}. 
Lastly, for \textit{Model evaluation} and \textit{Model selection}, we evaluated and selected machine learning models based on the criteria listed in Section~\ref{metrics:machine-learning-metrics}.

Table~\ref{tab:sbs-model-selection-summary} summarizes the results for both scenarios, including the selected scaling algorithms, the selected classifier and regressor models and their corresponding machine learning metric values obtained during model evaluation. We refer to \ref{app:model-selection-sbs} for a detailed description of the chosen machine learning models.

\begin{table}
    \centering
    \caption{Summary of the chosen machine learning models during \textit{Model evaluation} and \textit{Model selection} in the \sbs{}; Abbre- viations: \abbrevML{}, \abbrevSbs{}}
    \begin{tabular}{|c:c|m{5cm}|m{5cm}|N}
        \cline{2-5}
        \multicolumn{1}{c|}{} & \textbf{Goal(s)} & \textbf{Model} & \textbf{Metrics} &\\\hline
        
        \scenarioTable[-12pt]{2}{S1} & \thresholdtext{< 10\%}{Fr}, \thresholdtext{< 10ms}{Rt} & SGD Classifier\newline \hspace*{5pt}(hinge loss, l1 penalty)\newline No Scaler & F1: $0.895$, \newline MCC: $0.812$ &\\[8pt]\cline{2-4}
        
        & \optimizationtext{min}{C} & Passive Aggressive Regressor\newline \hspace*{5pt}(squared epsilon insensitive loss)\newline No Scaler & R2: $0.906$, \:MSE: $1.753$, \newline MAE: $0.901$, \:ME: $5.981$ &\\[8pt]\hline

        \scenarioTable[-25pt]{3}{S2} & \thresholdtext{< 10\%}{Fr} & SGD Classifier \newline\hspace*{5pt}(hinge loss, elasticnet penalty)\newline Standard Scaler & F1: $0.933$, \newline MCC: $0.866$ &\\[8pt]\cline{2-4}
        
        & \setpointtext{10ms, \epsilon=0.25ms}{Rt} & Passive Aggressive Regressor\newline \hspace*{5pt}(squared epsilon insensitive loss)\newline No Scaler & R2: $0.860$, \:MSE: $0.035$, \newline MAE: $0.123$, \:ME: $0.976$ &\\[8pt]\cline{2-4}
        
        & \optimizationtext{min}{C} & Passive Aggressive Regressor\newline \hspace*{5pt}(epsilon insensitive loss)\newline No Scaler & R2: $0.906$, \:MSE: $1.753$, \newline MAE: $0.901$, \:ME: $5.981$ &\\[8pt]\hline
    \end{tabular}
    \label{tab:sbs-model-selection-summary}
\end{table}

\paragraph{Exploration Rate and Warm-up Count} Finally we selected 5\% as exploration rate (extra random adaptation options selected for verification) and 60 cycles (of 300) as the warm-up count (the number of training cycles to initialize the learning model). For detailed results, see~\ref{app:model-selection-sbs}.

\subsubsection{Runtime Stage Evaluation with the \sbs{}}

\paragraph{Hypothesis} For the evaluation of the runtime stage activities of \techniqueNameShort{} we use the same hypotheses H1 to H4 as for DeltaIoT, see 
Section~\ref{subsec:deltaiot-runtime-stage-activities}. However, we test hypothesis H2 (the utility penalties when applying \techniqueNameShort{} is not significantly higher compared to \dlaser{}) and H4 (the reduction of adaptation spaces with \techniqueNameShort{} is not significantly lower compared to \dlaser{}, nor does \techniqueNameShort{} requires significantly more time for adaptation space reduction) only for scenario 1 as DLASeR does not support setpoint goals yet.

\paragraph{Granularities for Adaptation Space Reduction with an Optimization Goal}
 
In both scenarios, \techniqueNameShort{} has to deal with an optimization goal to keep the cost in the application minimal. After filtering out adaptation options based on the predicted satisfaction of threshold and setpoint goals in the system, \techniqueNameShort{} further reduces the adaptation space based on the cost predictions. We evaluate two cases for each scenario: a reduction to at most 1000 adaptation options and a reduction to at most 100 adaptation options. This corresponds to granularity values 1000 and 100.

\paragraph{Quality of the Learning Models}

Table~\ref{tab:sbs-machinelearning-metrics} shows the results for the quality of the learning models at runtime. 
We highlight the most important metrics. The classifier used to make predictions for both threshold goals in scenario 1 (failure rate and response time) has an F1-score of $0.841$, and the classifier used to predict the failure rate threshold goal in scenario 2 has an F1-score of $0.935$. The regressor used to predict the optimization goal in scenario 1 (cost) has an R2-score of $0.862$. For scenario 2, the regressor used to predict the setpoint goal (response time) has an R2-score of 0.902 and the regressor used to predict the optimization goal (cost) has an R2-score of $0.913$. These results confirm that the machine learning models can make accurate predictions for the quality properties of the system.

\begin{table}[h!]
    \caption{Values of the machine learning metrics for the runtime stage evaluation of the machine learning models of the \sbs{} (abbreviations: \abbrevSbs{}).}
    \label{tab:sbs-machinelearning-metrics}
    \centering

    \begin{tabular}{|c:c|c|c|N}
        \cline{3-5}
        \multicolumn{2}{c|}{} & \textbf{F1-score} & \specialcell{\textbf{Matthews correlation}\\\textbf{coefficient}} &\\[15pt]\hline
        
        \scenarioTable{1}{S1} & \thresholdtext{< 10\%}{Fr}, \thresholdtext{< 10\%}{Rt} & 0.841 & 0.737 &\\[8pt]\hline
        \scenarioTable{1}{S2} & \thresholdtext{< 10\%}{Fr} & 0.935 & 0.863 &\\[8pt]\hline
    \end{tabular}
    
    \bigskip
    
    \begin{tabular}{|c:c|c|c|c|c|N}
        \cline{3-6}
        \multicolumn{2}{c|}{} & \textbf{R2-score} & \specialcell{\textbf{Mean squared}\\\textbf{error}} & \specialcell{\textbf{Median absolute}\\\textbf{error}} & \textbf{Maximum error} &\\[15pt]\hline
        
        \scenarioTable{1}{S1} & \optimizationtext{min}{C} & 0.862 & 4.680 & 1.266 & 15.059 &\\[8pt]\hline
        \scenarioTable{2}{S2} & \setpointtext{10ms, 0.25ms}{Rt} & 0.913 & 2.970 & 1.123 & 11.251 &\\[8pt]\cline{2-7}
          & \optimizationtext{min}{C} & 0.902 & 0.039 & 0.130 & 1.212 &\\[8pt]\hline
    \end{tabular}
\end{table}

\paragraph{Summary of Results for Quantitative Metrics} Table~\ref{tab:sbs-evaluation-metrics} summarizes the evaluation results for the quantitative metrics for the \sbs{}. We discuss these results now in detail. 

\begin{table}
    \caption{Values of the metrics for the runtime stage evaluation of requirements of the \sbs{} (abbreviations: \abbrevSbs{}).}
    \label{tab:sbs-evaluation-metrics}
    \centering
    \begin{tabular}{|c:c|c|c|c|c|c|c|N}
        \cline{3-8}
        \multicolumn{2}{c|}{} & \multicolumn{3}{c|}{\textbf{Utility penalties}} & \multirow{2}{*}{\textbf{AASR}} & \multirowcell{2}{\textbf{Overall time}\\\textbf{saved}} & \multirowcell{2}{\textbf{Time}\\\textbf{overhead}}  &\\[5pt]\cline{3-5}
        \multicolumn{2}{c|}{} & \textit{\small \textbf{Fr}} & \textit{\small \textbf{Rt}} & \textit{\small \textbf{C}} & & & &\\[1pt]\hline

        \scenarioTable{2}{S1} & \textbf{Granularity 1000} & 0.134\% & 0.107ms & 1.381c & 92.59\% & 92.60\% & 0.04\% &\\\cline{2-8}
          & \textbf{Granularity 100} & 0.190\% & 0.229ms & 2.653c & 99.26\% & 99.26\% & 0.38\% &\\\hline

        \scenarioTable{2}{S2} & \textbf{Granularity 1000} & 0.138\% & 0.001ms & 1.589c & 93.13\% & 93.13\% & 0.05\% &\\\cline{2-8}
        & \textbf{Granularity 100} & 0.157\% & 0.003ms & 1.764c & 99.26\% & 99.26\% & 0.48\% &\\\hline
        
    \end{tabular}
\end{table}

\paragraph{Utility Penalties}

\begin{figure}[t!]
    \centering
    \includegraphics[width=1\linewidth]{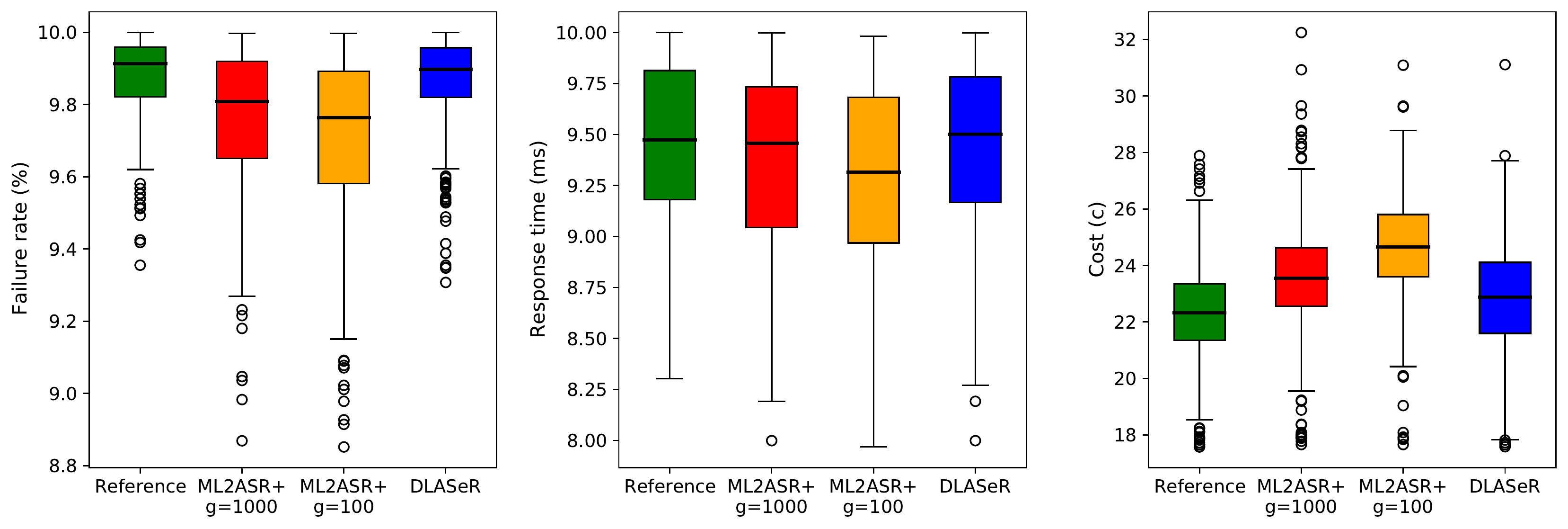}
    
    \bigskip
    
    \includegraphics[width=1\linewidth]{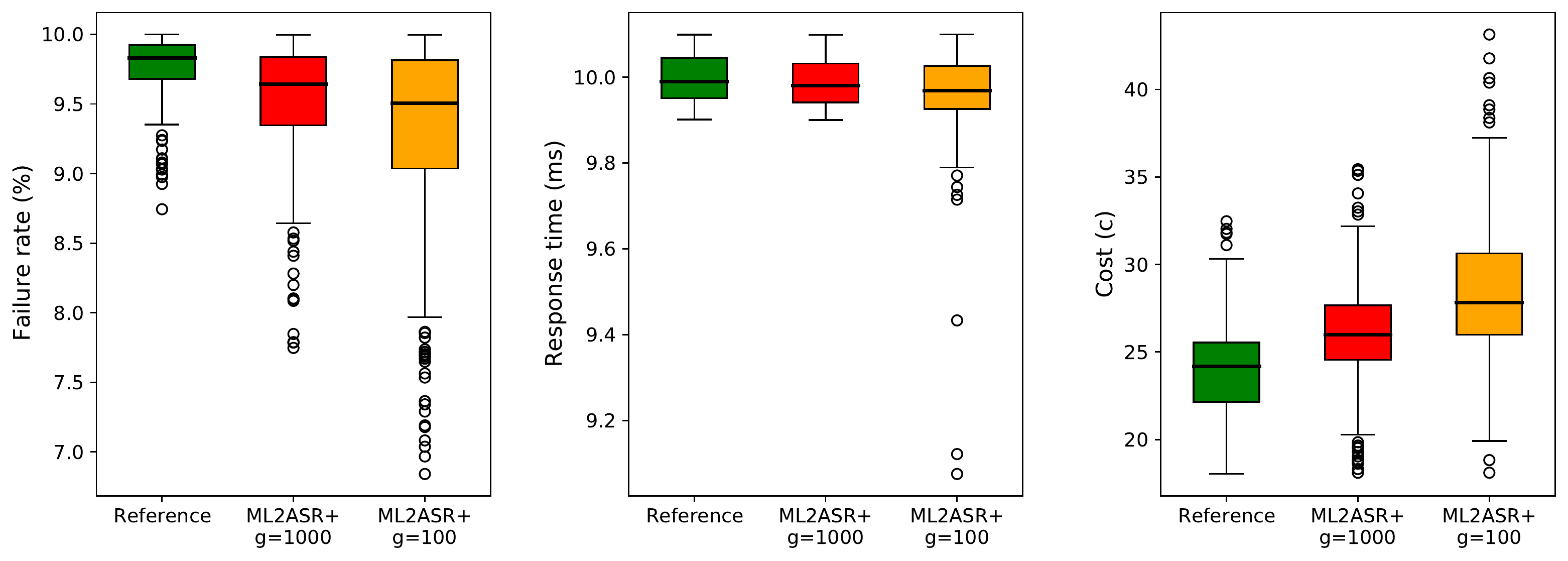}
    \caption{Utility penalties for scenario 1 (top) and 2 (bottom) of the \sbs{} compared to the reference approach and \dlaser{}.}
    \label{fig:sbs-results-utility}
\end{figure}

Figure~\ref{fig:sbs-results-utility} shows the results for the utility penalties for both scenarios when applying \techniqueNameShort{} with a granularity value of 1000 in red and a granularity value of 100 in orange, and \dlaser{} in blue (only for scenario 1). Subsequently, we zoom in on the threshold goals, setpoint goal, and optimization goals. 

\paragraph{Threshold Goals}

The graphs show us that each approach satisfies the threshold goals for all chosen adaptation options in both scenarios. For \techniqueNameShort{} with granularity values 1000 and 100, we notice utility penalties for the failure rate in the interval $[0.13\%, 0.19\%]$ and utility penalties for the response time in the interval $[0.1ms, 0.23ms]$. With \dlaser{} we notice (for scenario 1) generally lower utility penalties lower than \techniqueNameShort{}: $0.002\%$ for failure rate and $0.04ms$ for response time. 
Note that a higher utility penalty value for failure rate or response time value here is not necessarily relevant or negative since the threshold goals remain satisfied after adaptation space reduction. 

\paragraph{Setpoint Goal}

For the setpoint goal defined in scenario 2, we similarly notice that none of the chosen adaptation options violate the goal. The utility penalty for response time of this goal with \techniqueNameShort{} lies in the interval $[0.001ms, 0.003ms]$, showing that the effect of adaptation space reduction is negligible.

\paragraph{Optimization Goals}

Looking at the cost optimization goal, we see penalties that lay in the interval $[1.38c, 2.66c]$. In scenario 1, we notice a utility penalty of 0.436c for \dlaser{}, which is slightly better than \techniqueNameShort{}. 
We also observe that the results of  \techniqueNameShort{} with a granularity value of 100 are slightly worse compared to a granularity with value 1000. This can be explained by the additional restriction put on the reduced adaptation space size: the resulting adaptation space is on average 10 times smaller compared to a granularity value of 1000, leaving fewer adaptation options to be selected from to apply self-adaptation.

\begin{figure}[h!]
    \centering
    \includegraphics[width=.6\linewidth]{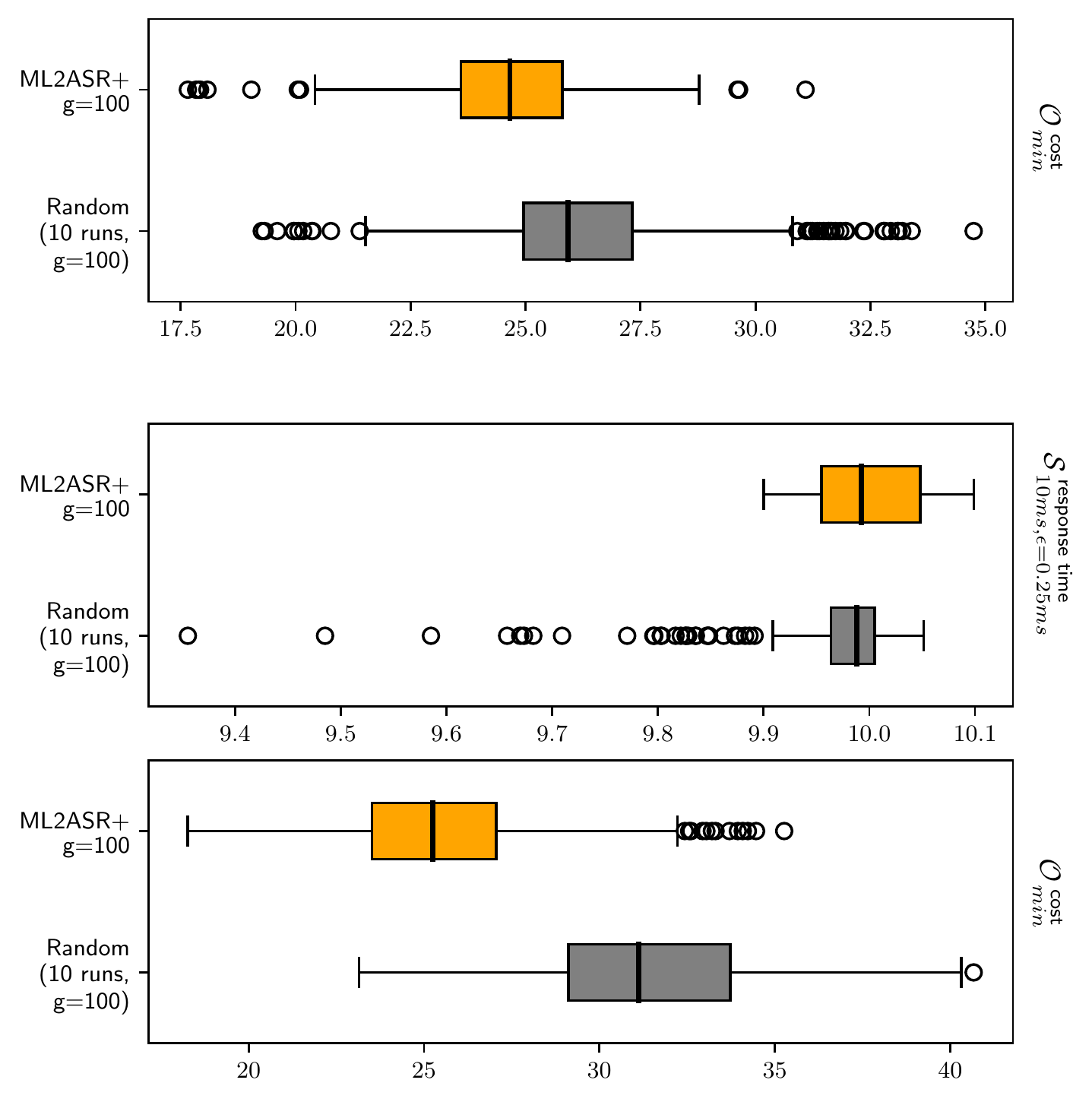}
    \caption{\review{Comparison of results for \techniqueNameShort{} and random selection of adaptation options (10 runs) in scenario 1 (optimization goal top graph) and scenario 2 (setpoint goal middle graph, and optimization goal bottom graph) of the \sbs{}.}}\vspace{-5pt}
    \label{fig:sbs-random-highlights}
\end{figure}

\paragraph{Sanity Check with Random Approach} \review{We compared \techniqueNameShort{} with batches of 10 runs of an approach that randomly selects adaptation options.  
For the threshold goals, failure rate and response time in scenario 1 and failure rate only in scenario 2, the random approach manages to always select at least one adaptation option that satisfies the goals. 
For the optimization goal in both scenarios and the setpoint goal in scenario 2, the Wilcoxon signed rank tests showed statistically significant results between \techniqueNameShort{} and the batches of 10 runs with the approach that randomly selects adaptation options (p-values of $4.88e^{-21}$ for cost in scenario 1, $1.44e^{-4}$ for response time in scenario 2, and $2.55e^{-66}$ for cost in scenario 2). Note that we could not identify statistical relevant differences for the response time in scenario 2 for all individual runs with the approach that randomly selects adaptation options.  
Figure~\ref{fig:sbs-random-highlights} shows the distributions for the optimization and setpoint goals of the IoT network in the two scenarios.\footnote{\review{Note that similar to the results for DeltaIoT, we cannot generalize the statistical differences we observe between \techniqueNameShort{} and the random approach to other sets of random runs.}} 
The average cost (optimization goal) for scenario 1 is $24.58c$ with \techniqueNameShort{} compared to $26.38c$ with random selection (Random), a difference of $1.8c$ (6.82\%). For scenario 2, the values are $25.57c$ with \techniqueNameShort{} compared to $31.51c$ with Random, a difference of $5.94c$ (18.85\%). For the average response time (setpoint goal at 10ms$\pm$0.25ms), we noticed that the approach that randomly selects adaptation options violated on average the goal in 48 cycles (of a total of 300 cycles) compared to no cycles for  \techniqueNameShort{}. These results show that \techniqueNameShort{} performs substantially better for more complex adaptation scenarios compared to an approach that randomly selects adaptation options.}\vspace{8pt}

\begin{tcolorbox}[colback=white]
\textbf{Hypotheses H1 (negligible utility penalties compared to reference approach) and H2 (utility penalties not significantly higher compared to \dlaser{}).} %
The results show that the utility penalties incurred by \techniqueNameShort{} are negligible compared to the reference approach. Specifically, the penalties for cost (optimization goal) are very low in both scenarios (at most $1.589mC$ and $2.653mC$ with granularity values 1000 and 100 respectively). A smaller granularity value reduces the adaptation space significantly but implies higher utility penalties. The satisfaction of the threshold and setpoint goals remains unaffected with \techniqueNameShort{}. The slight increase in cost is acceptable, especially considering that it is not feasible to use the reference approach in practice due to time constraints.
In scenario 1, \dlaser{} shows slightly better results compared to \techniqueNameShort{} with a granularity value of 1000 (with a penalty of cost of 1.381c vs 0.436c). However this cost is acceptable considering that \dlaser{} does not support all types of adaptation goals yet. In conclusion, we can accept hypotheses H1 (for scenario 1 and 2) and H2 (for scenario 1) in the \sbs{} application.
\end{tcolorbox}

\paragraph{Average Adaptation Space Reduction}

\begin{figure}
    \centering
    \includegraphics[width=\linewidth]{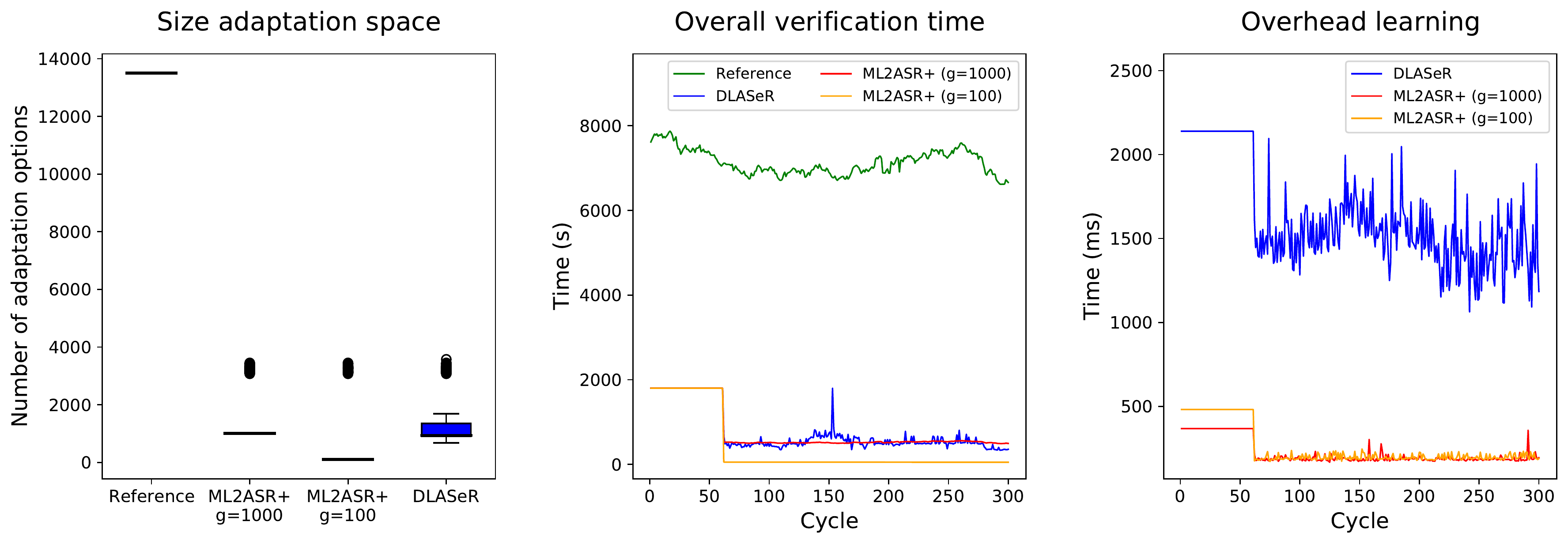}
    
    \vspace*{3pt}
    
    \includegraphics[width=\linewidth]{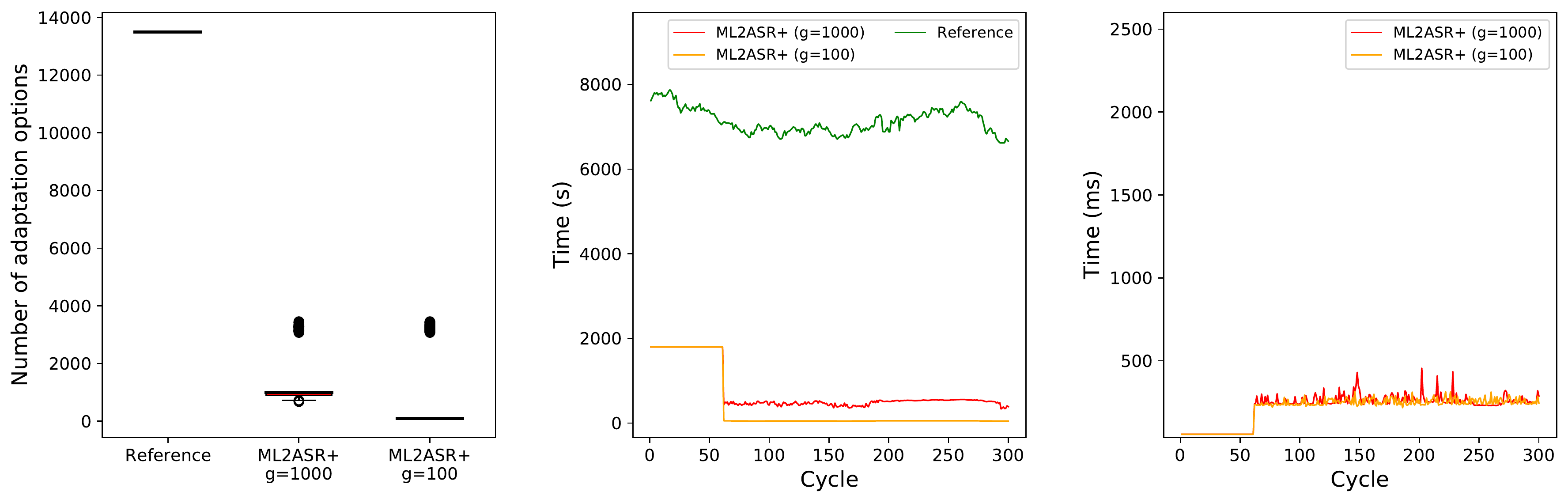}
    
    \caption{Number of verified adaptation options when using \techniqueNameShort{} (left), overall time used (middle), and overhead (right) compared to the reference approach and \dlaser{} for scenario 1 (top row) and scenario 2 (bottom row) of the \sbs{}.}
    \label{fig:sbs-space-reduction-time}
\end{figure}

Figure~\ref{fig:sbs-space-reduction-time} (left) shows the number of adaptation options remaining after reduction. We used a \textit{warm-up count} of 60 cycles for both \techniqueNameShort{} and \dlaser{}. The total number of options in these training cycles is limited by the available time for adaptation (30m for the \sbs{}). Note that the reference approach is not subject to this time restriction for the purpose of evaluation; i.e., the reference approach fully analyzes the whole adaptation space with 13500 adaptation options, which is infeasible in practice due to time constraints. 
During testing, the number of adaptation options with \techniqueNameShort{} is restricted by the granularity value, here 1000 and 100. This results in an Average Adaptation Space Reduction of 92-93\% and 99\% respectively in both scenarios. For \dlaser{} in scenario 1, we observe a similar Average Adaptation Space Reduction of 92.66\%.

\paragraph{Learning Time Overhead}

Figure~\ref{fig:sbs-space-reduction-time} (right) shows the learning time overhead introduced by \techniqueNameShort{} with the two evaluated granularity values, and \dlaser{} (scenario 1).
The overhead of \techniqueNameShort{} is very small compared to the overall verification time (the overhead for learning is denoted in ms, while overall verification time is denoted in s). Specifically, the overhead for both granularity values accounts for less than 0.5\% of the time required to reduce and verify the adaptation space. In absolute terms, \techniqueNameShort{}'s overhead is capped at approximately 500ms during online training cycles.
The learning time overhead of \dlaser{} in scenario 1 is substantially higher compared to \techniqueNameShort{} with a granularity value of 1000 (with close to equal Average Adaptation Space Reduction). Yet, the overhead remains minor compared to the overall verification time; the overhead of \dlaser{} is 0.30\% of the overall verification time. In absolute terms, the overhead of \dlaser{} is capped at approximately 2200ms during the training cycles as well.

\paragraph{Overall Time Saved}

Figure~\ref{fig:sbs-space-reduction-time} (middle) shows the overall time used to analyze the (selected) adaptation options. We can clearly see that the overall verification time is significantly reduced, closely aligned to the corresponding Average Adaptation Space Reduction. Concretely, we observe an overall time saved of approximately 92-92\% and 99\% using \techniqueNameShort{} with granularity values 1000 and 100 respectively. 
For \dlaser{}, we notice an overall time saved of 92.62\%, which is in line with \techniqueNameShort{} for a granularity value of 1000 in scenario 1.

\review{\subsubsection{Summary of design stage and runtime stage machine learning activities}}

\review{Table~\ref{tab:sbs-chart-ML} summarizes, similarly to the DeltaIoT case before, the number of inputs, features, objective variables and metrics for the \sbs{}. Note that no features are removed by \textit{Feature Extraction} since all features were deemed to be relevant (hence the table cells being marked in gray).}\vspace{8pt}

\begin{table}
    \centering
    \caption{\review{The number of inputs, features, objective variables and metrics for the machine learning pipeline of scenario 2 of the \sbs{} in the design and runtime stage (separated by the double line). The prediction column is marked in red to indicate that it is not be used yet in the training cycles. The number of inputs for online learning is determined by the number of options that could be verified by the Verifier. Abbreviations: \abbrev{``f vectors''}{feature vectors}, \abbrev{``q vectors''}{quality vectors}, \abbrev{``Pl''}{packet loss}, \abbrev{``Ec''}{energy consumption}, \abbrev{``La''}{latency}.}}
    \label{tab:sbs-chart-ML}
    \footnotesize
    \begin{tabular}{|c|c|c|@{}|c|c|c|c|}
        \hhline{~|*{6}{-}}
        \multicolumn{1}{c|}{} 
            & \specialcell{\textbf{Feature}\\\textbf{Extraction}} 
            & \specialcell{\textbf{Machine Learning}\\\textbf{Model Identification}} 
            & \cellcolor{gray!25}\specialcell{\textbf{Feature}\\\textbf{Extraction}} 
            & \cellcolor{red!10}\textbf{Prediction} 
            & \textbf{Verification} 
            & \specialcell{\textbf{Online}\\\textbf{Learning}} \\\hline
        
        \specialcell{\textbf{Number of}\\\textbf{inputs}} 
            & \specialcell{$100*13500$ f vectors\\$100*13500$ q vectors} 
            & \specialcell{$100*13500$ f vectors\\$100*13500$ q vectors} 
            & \cellcolor{gray!25}$13500$ f vectors 
            & \cellcolor{red!10} $13500$ f vectors 
            & $13500$ f vectors 
            & \specialcell{$X$ f vectors\\$X$ q vectors} \\\hline 
        
        \specialcell{\textbf{Number of}\\\textbf{features}} 
            & 22 features 
            & 22 features 
            & \cellcolor{gray!25}22 features 
            & \cellcolor{red!10} 22 features 
            & \cellcolor{gray!25}x 
            & 22 features 
            \\\hline 
        \specialcell{\textbf{Objective}\\\textbf{variables}} 
            & \specialcell{1. Fr class\\2. Rt value\\3. C value}
            & \specialcell{1. Fr class\\2. Rt value\\3. C value}
            & \cellcolor{gray!25}x
            & \cellcolor{red!10} \specialcell{1. Fr class\\2. Rt value\\3. C value}
            & \cellcolor{gray!25}x
            & \specialcell{1. Fr class\\2. Rt value\\3. C value}
            \\\hline 
        \textbf{Metrics} 
            & \cellcolor{gray!25}x
            & \specialcell{1. F1, MCC\\2. R2, MSE, MAE, ME\\3. R2, MSE, MAE, ME}
            & \cellcolor{gray!25}x
            & \cellcolor{gray!25}x
            & \cellcolor{gray!25}x
            & \cellcolor{gray!25}x
            \\\hline 
    \end{tabular}
\end{table}

\begin{tcolorbox}[colback=white]
\textbf{Hypotheses H3 (significant reduction of adaptation spaces and time gain) and H4 (adaptation space reduction comparable to \dlaser{}).}
The evaluation shows that \techniqueNameShort{} significantly reduces the adaptation space: up to a reduction of 99\% depending on the specified granularity value. Paired with this, up to 99\% of the time used by the reference approach is saved, closely aligned with the average adaptation space reduction, since the overhead introduced by \techniqueNameShort{} is minimal (constituting less than 0.5\%).
For scenario 1, \dlaser{} obtains a similar Average Adaptation Space Reduction (92.66\% vs 92.59\%) and overall time saved (92.62\% vs 92.60\%) compared to \techniqueNameShort{} with granularity value 1000. \techniqueNameShort{} with granularity value 1000 outperforms \dlaser{} on learning overhead with a value of 0.04\% vs 0.30\%. 
As such, we can accept hypothesis H3 for scenarios 1 and 2, and H4 for scenario 1 in the \sbs{}.
\end{tcolorbox}\vspace{10pt}

\section{Discussion}\label{sec:discussion}

In Section~\ref{sec:problem-description}, we described the research question targeted in this work and we listed the desirable requirements for an approach to tackle the research question. To that end, we proposed \techniqueNameShort{}. In the evaluation, we assessed the quantitative requirements. We now discuss the remaining qualitative requirements, we answer the research question, we highlight insights obtained from this research endeavor, and conclude with a discussion of threats to validity.

\subsection{Qualitative Requirements}

\noindent\paragraph{\textbf{Reusability}}
With reusability we refer to the ability of \techniqueNameShort{} to be instantiated and  applied over multiple application domains. To demonstrate that we have covered this requirement, we demonstrated the applicability of \techniqueNameShort{} to the Internet of Things domain and the Service-Based Systems domain. In both applications we analyzed the performance of \techniqueNameShort{} in two evaluation scenarios, while also assessing different granularity values. From the results, we can conclude that \techniqueNameShort{} has the ability to handle both applications and the different evaluation scenarios.

\noindent\paragraph{\textbf{Automatic Operation at Runtime}}

To evaluate the second requirement, we make a distinction between the design stage and the runtime stage of the approach. In the design stage, the system requires manual input from the system developer(s) to properly configure the \MLMi{}, as described in section~\ref{subsec:design-stage}. Hence this step is not completely automated. 
Once the \MLMi{} is deployed, no further input or intervention is necessary from the system operators. This is also demonstrated in the evaluation: during operation \techniqueNameShort{} reduces the adaptation space without any input from an operator or system developer. \techniqueNameShort{} thus satisfies this requirement.

\noindent\paragraph{\textbf{Modularity Adaptation Goals}}

To evaluate the ability of \techniqueNameShort{} to deal with different combinations of adaptation goals, we specifically investigate whether \techniqueNameShort{} is able to deal with threshold, setpoint, and optimization goals. In our evaluation, we defined four scenarios that combine different types of goal types for two different applications. This way, we ensured  that different combinations of the three types of adaptation goal types are assessed. We conclude that \techniqueNameShort{} supports  dealing with all the goal types in the evaluated application scenarios.

\noindent\paragraph{\textbf{Granularity of Adaptation Space Reduction}}
With granularity we refer to the degree with which \techniqueNameShort{} is able to reduce adaptation spaces, i.e., selecting a specified number or percentage of adaptation options from the original adaptation space.
\techniqueNameShort{} allows the specification of a granularity value that constraints the size of the reduced adaptation space. We have demonstrated this for both applications in different evaluation scenarios with granularity values of 10, 25, 100 and 1000. We can thus conclude that \techniqueNameShort{} satisfies this last requirement as well. \vspace{8pt}


\begin{tcolorbox}[colback=white]
\textbf{Answer to research question "\textit{How can machine learning be used to reduce large adaptation spaces of self-adaptive systems with different types of adaptation goals to perform more efficient analysis without compromising the goals}?"} 
This work demonstrates how classic supervised machine learning techniques can be used to reduce the adaptation space to a more manageable subset. After designing the \MLM{}, \techniqueNameShort{} initializes the learning models, trains the models during warm-up, and then uses the models to make predictions about the satisfaction of adaptation goals for individual adaptation options. \techniqueNameShort{} uses classification to predict the satisfaction of a threshold or setpoint goals, and regression to predict the quality value associated with an adaptation option. \techniqueNameShort{} provides the means to reduce the adaptation space with the specified granularity value; this flexibility enables the approach to adjust with the available time window to perform adaptation. Empirical evaluation shows that \techniqueNameShort{} drastically reduces the time required for analysis with a negligible effect on the satisfaction of the adaptation goals in our evaluated systems. 
\end{tcolorbox}

\subsection{Insights}

We share a number of insights we obtained during the design and evaluation of \techniqueNameShort{}: 

\begin{itemize}
    \item When handling multiple adaptation goals, there is a risk that errors in learning models propagate further with each prediction. Accumulating prediction errors may ultimately reduce the efficacy of the approach.
    \item The overhead introduced by the learning approach directly links with the selected granularity value. Even for small granularity values, the gain in time required for analysis is significant. Yet, such a setting may also have a significant impact on the utility penalties. Hence, the right choice for setting the granularity value is important and requires experimentation.
    \item It is important to highlight that the use of linear machine learning models is not a ``one size fits all'' solution. The effectiveness of the approach depends on the underlying relation between input data (features) and the output (qualities). If this relation cannot be properly modeled in a linear way, other approaches such as \dlaser{} that rely on deep neural networks may be preferable as these approaches capture these intrinsic relations better. It is however important to keep in mind that other approaches may follow different workflows and carry their own drawbacks. For instance, \dlaser{} follows different steps in its design stage and runtime stage workflows and the approach introduces a larger learning overhead compared to \techniqueNameShort{}.
    \item \techniqueNameShort{} relies on the assumption that the (formal) models that are used to estimate quality properties of the underlying system provide reliable and correct results. A quality model that cannot handle concept drift nor evolution of the system may yield data that does not capture the real system accurately. This can affect the performance of the machine learning models. Yet, extracting data directly from the real system rather than the (formal) model is not a solution as this data is inherently very limited since only a single adaptation option can be applied each cycle. However, exploiting the data retrieved from the real system to detect issues with the model and adapt or evolve the models dynamically is an option; we leave this as future work.
    \item For the validation of \techniqueNameShort{}, we trained the learning models both during the design stage and the runtime stage. In principle, it would be possible to generate all the data and train machine learning models completely during design stage. However, since the space of adaptation options combined with the uncertainties leads to a very huge space of possible configurations, generating all the data and training machine learning models would be very time consuming. Therefore, we used a representative sample of the data to apply design stage training and then collect additional data after startup to continue the training according to the actual system configuration and uncertainties at runtime.
    \item In this work, we considered setpoint goals based on a small window $\epsilon$, resembling similarities with steady state error in control-based approaches~\cite{7929422}. Any configuration within this window complies with the goal. An interesting option for future work is to refine this view and consider the option closest to the setpoint as the optimal one. Combined with an optimization goal this will lead to a multi-objective optimization problem. 
\end{itemize}

\subsection{Threats to Validity}

The empirical evaluation of \techniqueNameShort{} is subject to threats to validity. For each threat, we discuss potential critiques of this study and we explain how we dealt with those.


\noindent\paragraph{\textbf{Internal Validity}} To make sure that we can draw a causal conclusion based on the study, we took several measures. Concerning the contribution, we specified the approach formally, providing a basis to define that the approach works as described. Concerning the evaluation, we have applied the same settings of the simulator with the same settings for the application parameters when comparing \techniqueNameShort{} with the other approaches. This is particularly relevant in settings with stochastic behavior. As such we provide a basis for deriving the conclusions of comparing the approaches. We also provide a replication package~\cite{reproduction-package} for other researchers to validate the results. 

% ...

\noindent\paragraph{\textbf{External Validity}} External validity concerns the generalization of the results beyond the scope of the study. This study contributes an architectural approach for adaptation space reduction in self-adaptive systems that is centred on the \MLMi{} with an according workflow. This approach uses classical supervised machine learning techniques to support the adaptation process. Since we have applied and evaluated \techniqueNameShort{} to a limited set of scenarios with particular characteristics and types of uncertainties, we cannot make general claims about the efficacy of the approach in other settings. To mitigate this threat to some extent, we have evaluated the approach in two distinct domains with different challenges regarding adaptation space reduction for different combinations of adaptation goals.
%


\noindent\paragraph{\textbf{Construct Validity}} With construct validity we analyze whether we have obtained the right measures to answer the proposed research question. To minimize threats to construct validity we provided an explicit definition of six requirements to be evaluated. For several of these requirements we defined concrete metrics that enabled us to evaluate the performance of the approach empirically (in terms of efficiency and overhead). Several of these metrics are based on established practice for the evaluation of learning approaches. In addition, the formal specification of \techniqueNameShort{} provides a rigorous description of how the approach works. Nevertheless, we acknowledge that other metrics may have been considered for evaluating the appropriateness of adaptation space reduction.

\noindent\paragraph{\textbf{Conclusion Validity}} Threats to conclusion validity concerns reaching an incorrect conclusion about a relationship in the observations. To mitigate conclusion validity threats, we applied \techniqueNameShort{} in different scenarios of different domains with different characteristics. Based on a set of well-defined metrics, the results confirm the observation that \techniqueNameShort{} is effective for adaptation space reduction in self-adaptive systems. In addition, we have made all code and experimental data publicly available~\cite{reproduction-package} to reproduce the experiments in order to confirm the findings.

\section{Conclusion}\label{sec:conclusions}

In this paper we presented \techniqueNameShort{}, a novel approach to analyze large adaptation spaces more effectively by exploiting classic supervised machine learning techniques to reduce adaptation spaces on the fly. \techniqueNameShort{} extends the basic MAPE-K architecture with a \MLMi{} that supports the \textit{Analyzer} component by reducing the adaptation space to a manageable subset. In particular, the \MLMi{} filters  adaptation options that are predicted to not meet the adaptation goals in the system. We have demonstrated the effectiveness and viability of \techniqueNameShort{} in our evaluation in two different application domains. We evaluated the effectiveness of \techniqueNameShort{} in reducing the adaptation space as well as the overhead introduced by the approach. The results showed that the overhead introduced by \techniqueNameShort{} is minimal compared to the time required to verify the remaining subset of filtered adaptation options. On top of this, the penalty in system qualities is negligible when choosing a new system configuration from the reduced adaptation space. In future work, we plan to investigate adaptation space reduction in decentralized self-adaptive systems where multiple feedback loops need to coordinate the analysis. In the long term, we plan to expand our study on the use of machine learning and self-adaptive systems, and investigate how evolutionary learning can be used to support self-adaptation in systems that are exposed to unanticipated changes, requiring system evolution. First ideas in this direction are reported in~\cite{abs-2108-08802}.

\bibliographystyle{ACM-Reference-Format}
\bibliography{main}

\newpage
\appendix

\section{Auxiliary Formal Definitions}




\subsection{Model training (split train-test)}\label{app:split}

To enable the evaluation of learning models, one options is to split the data set in 2 parts: a training data set and a testing data set. We define a \textit{split} function as follows ($tr$ short for training and $te$ short for testing):

\begin{itemize}
	\item[] $Map: \Lambda_i \rightarrow \Phi_i$ is a function that maps a system state, represented by a feature vector, to the qualities of the system, represented by a quality vector. 
    \item[] $Split: \Lambda \times \Phi \times W \rightarrow \Lambda \times \Phi \times \Lambda \times \Phi$
    \item[] $Split(\Lambda_{o}, \Phi_{o}, w) = \  <\Lambda_{tr}, \Phi_{tr}, \Lambda_{te}, \Phi_{te}>$ with \\
    \hspace*{5pt} $\Lambda_{tr} \cup \Lambda_{te} = \Lambda_{o}$ \quad and\quad $\Phi_{tr} \cup \Phi_{te} = \Phi_{o}$ \\
    \hspace*{5pt} and\; $|\Lambda_{tr}| = w * |\Lambda_{o}|$ \quad and\quad $|\Phi_{tr}| = w * |\Phi_{o}|$ \\
    \hspace*{5pt} and\; $\forall \lambda_i \in \Lambda_{tr}$: $Map(\lambda_i) = \phi_{i}$ with $\phi_{i} \in \Phi_{tr} $\ \ and\ \ $\forall \lambda_j \in \Lambda_{te}$: $Map(\lambda_j) = \phi_{j}$ with $\phi_{j} \in \Phi_{te}$
    
    
\end{itemize}

The training data set is used to train the machine learning models, while the testing data set is used to test and validate the trained machine learning models. This testing is conducted by comparing the predictions made to the actual quality values in the form of machine learning evaluation metrics.

\subsection{Exploration}\label{app:exploration}

We formally define the selection of explored adaptation options as follows:

\begin{itemize}
    \item[] $DetermineExploration: \Pi \times E \rightarrow \Pi$
    \item[] $DetermineExploration(\{\pi_1, ..., \pi_n\}, e) = \Pi_e$ \\
    \hspace*{5pt} with $\Pi_e \{\,\pi_i \in \{\pi_1, ..., \pi_n\} \mid \pi_i \notin \Pi_{filtered} \,\}$ and $|\Pi_e| = e * |\{\pi_1, ..., \pi_n\}|$
\end{itemize}

The set $\Pi_{filtered}$ refers to the set of adaptation options that were predicted by the machine learning models to satisfy the adaptation goals. Hence, we explore adaptation options outside the set of  adaptation options that were already selected for verification. It is important to note that adaptation options that are predicted to meet all system goals should be given priority in case of insufficient time to verify all the included adaptation options. The logic is that the explored adaptation options should not hinder the verification of adaptation options with greater promise.

\subsection{Filter}\label{app:filter}

We formally define filtering as follows:

\begin{itemize}
    \item[] $G = \{g\}$: The granularity that defines an upper bound on the size of the filtered adaptation space.
    \item[] $\mathbb{G}_s \in \mathbb{G}$: The specific set of adaptation goals of the system.
    \item[] $Filter: \Pi \times \mathbb{Z} \times G \times \mathbb{G} \rightarrow \Pi$
    \item[] $Filter(\Pi_i, \{\Omega_1, ..., \Omega_n\}, g, \mathbb{G}_s) = \Pi_j \,\text{ where }\, \Pi_j \subseteq \Pi_i \:\,\text{and}\:\, |\Pi_j| \leqslant g$
\end{itemize}

Filtering takes a set of adaptation options, a set of quality predictions, a granularity value that puts a bound on adaptation space reduction, and an adaptation goal. The result is a reduced set of adaptation options. 

The criteria for filtering adaptation options varies depending on the type of quality goals that are evaluated. In particular, 
filtering handles three types of operations, one for each type of adaptation goals. The first type of filter operation filters adaptation options that do not comply with a threshold goal in the system. Formally, the filter operation for a threshold goal $\threshold{} \in \mathbb{T}$ with a threshold value $\bar{x}$ for any quality value $q$ is defined as follows:
%
%
\begin{align*}
  f_{\threshold{<\bar{x}}} = \{\pi_1, \pi_2, ..., \pi_n\} \mapsto \{\pi_{f_1}, \pi_{f_2}, ..., \pi_{f_m}\}  &\text{ where } \threshold{<\bar{x}}(q_k) = True, \:k \in \{f_1, f_2, ..., f_m\} \\
  f_{\threshold{>\bar{x}}} = \{\pi_1, \pi_2, ..., \pi_n\} \mapsto \{\pi_{f_1}, \pi_{f_2}, ..., \pi_{f_m}\}  &\text{ where } \threshold{>\bar{x}}(q_k) = True, \:k \in \{f_1, f_2, ..., f_m\}
\end{align*}

The second type of filter operation relates to setpoint goals in the system. We define the filter operation that filters adaptation options according to setpoint goal $\setpoint{}$  with target $\mu$ and error margin $\epsilon$ for any quality value $q$ as follows: 
\begin{align*}
    f_{\setpoint{\mu, \epsilon}} = \{\pi_1, \pi_2, ..., \pi_n\} \mapsto \{\pi_{f_1}, \pi_{f_2}, ..., \pi_{f_m}\} \text{ where } \\
m \leq g \text{ and } \sum_{i = 0}^{m} \abs{q_{f_i} - \mu} = min(\{\sum_{k \in K} \abs{q_k - \mu} \mid K \subseteq \Pi\;\, \text{and}\;\, \abs{K} = m\})
\end{align*}

Lastly, the filter deals with up to one optimization goal. We define the filter operation for an optimization goal $\optimization{}$ for quality values $q$ as follows:
\begin{align*}
    f_{\optimization{min}} = \{\pi_1, \pi_2, ..., \pi_n\} \mapsto \{\pi_{f_1}, \pi_{f_2}, ..., \pi_{f_m}\} \text{ where } \\ 
m \leq g \text{ and } \sum_{i = 0}^{m} q_{f_i} = min(\{\sum_{k \in K} q_k\} \mid K \subseteq \Pi\;\, \text{and}\;\, \abs{K} = m)
\end{align*}

\begin{align*}
    f_{\optimization{max}} = \{\pi_1, \pi_2, ..., \pi_n\} \mapsto \{\pi_{f_1}, \pi_{f_2}, ..., \pi_{f_m}\} \text{ where } \\ 
m \leq g \text{ and } \sum_{i = 0}^{m} q_{f_i} = max(\{\sum_{k \in K} q_k\} \mid K \subseteq \Pi\;\, \text{and}\;\, \abs{K} = m)
\end{align*}

In our research, we use filters that combine the different filter operations in a predefined order. In particular, the filter first filters adaptation options that violate threshold goals. Next it filters adaptation options that violate the setpoint goals of the system. Finally, it filters the options based on a single optimization goal. We restrict filtering to a single optimization goal to avoid conflicting scenarios when multiple optimization goals are specified in the system. Equation~\ref{eq:filter-main} specifies how we define the main filter operation:

\begin{equation}
    \mathcal{F} = f_{\optimization{}} \circ ... \circ f_{\setpoint{2}} \circ f_{\setpoint{1}} \circ ... \circ f_{\threshold{2}} \circ f_{\threshold{1}}
    \label{eq:filter-main}
\end{equation}

In case any of the types of adaptation goals are not applicable, that type is ignored by the filter.

\section{Additional Machine Learning Material}

Table~\ref{tab:deltaiot-modelselection} and Table~\ref{tab:sbs-modelselection} summarize the scalers and models selected for evaluation scenarios in both applications. The numbers between square brackets indicate the boundaries of the evaluation metric values for the alternative options that were not selected. For both applications and scenarios, the \textit{warm-up count} is selected from 30, 45 and 60, and the \textit{exploration rate} is selected from 5\% and 10\%.

\label{app:model-selection-deltaiot}

\begin{table}
    \centering
    \caption{Design stage model selection for the DeltaIoT application in the two system scenarios. MCC: Matthews Correlation Coefficient; MSE:  Mean Squared Error; MAE:  Median Absolute Error; ME:  Maximum Error}
    \label{tab:deltaiot-modelselection}
    \begin{tabular}{|c|m{2.2cm}|m{2.2cm}|N}
        \cline{2-3}
        \multicolumn{1}{l|}{\textit{Scenario 1}} & \thresholdtext{< 10\%}{packet loss} & \thresholdtext{< 5\%}{latency} &\\[8pt]\hline
        \textbf{ML algorithm} & \multicolumn{2}{l|}{SGD Classifier} &\\\hline
        \textbf{Loss function} & \multicolumn{2}{l|}{Log} &\\\hline
        \textbf{Penalty function} & \multicolumn{2}{l|}{l1} &\\\hline
        \textbf{Scaler type} & \multicolumn{2}{l|}{MinMax Scaler} &\\\hline
        \textbf{Exploration rate} & \multicolumn{2}{l|}{5\%} & \\\hline
        \textbf{Warm-up count} & \multicolumn{2}{l|}{45} & \\\hline
        
        \textbf{Metrics} & \multicolumn{2}{p{4.4cm}|}{
            \begin{tabular}{@{}l@{\hspace{8pt}}l@{\hspace{5pt}}l@{\hspace{-11.7pt}}}
                \textit{F1-score}: & 0.818 & \interval{0.022}{0.818} \\
                \textit{MCC}: & 0.715 & \interval{-0.004}{0.716} \\
            \end{tabular}
        } &\\\hline
    \end{tabular} 
    
    \bigskip
    
    \vspace*{0.2cm}
    \begin{tabular}{|c|m{2.2cm}|m{2.2cm}|m{4.6cm}|N}
        \cline{2-4}
        \multicolumn{1}{l|}{\textit{Scenario 2}} & \thresholdtext{< 10\%}{packet loss} & \thresholdtext{< 5\%}{latency} & \optimizationtext{min}{energy consumption}  &\\[8pt]\hline
        \textbf{ML algorithm} & \multicolumn{2}{l|}{SGD Classifier} & Passive Aggressive Regressor &\\\hline
        \textbf{Loss function} & \multicolumn{2}{l|}{Log} & Squared Epsilon Insensitive &\\\hline
        \textbf{Penalty function} & \multicolumn{2}{l|}{l1} & N/A &\\\hline
        \textbf{Scaler type} & \multicolumn{2}{l|}{MinMax Scaler} & None &\\\hline
        \textbf{Exploration rate} & \multicolumn{2}{l|}{5\%} & 5\% & \\\hline
        \textbf{Warm-up count} & \multicolumn{2}{l|}{45} & 45 & \\\hline
        
        \textbf{Metrics} & \multicolumn{2}{p{4.4cm}|}{
            \begin{tabular}{@{}l@{\hspace{8pt}}l@{\hspace{5pt}}l@{\hspace{-11.7pt}}}
                \textit{F1-score}: & 0.818 & \interval{0.022}{0.818} \\
                \textit{MCC}: & 0.715 & \interval{-0.004}{0.716} \\
            \end{tabular}
        } & {
            \begin{tabular}{@{}l@{\hspace{8pt}}l@{\hspace{5pt}}l@{\hspace{-6.5pt}}}
                \textit{R2-score}: & 0.833 & \interval{-1.091}{0.854} \\
                \textit{MSE}: & 0.004 & \interval{0.004}{$8.8e^{24}$} \\
                \textit{MAE}: & 0.043 & \interval{0.040}{$4e^{12}$} \\
                \textit{ME}: & 0.269 & \interval{0.241}{$9e^{12}$} \\
            \end{tabular}
        } &\\\hline
    \end{tabular}
\end{table}

\label{app:model-selection-sbs}

\begin{table}
    \caption{Design stage model selection for the \sbs{} application in the two system scenarios.}
    \label{tab:sbs-modelselection}
    \centering
    \begin{tabular}{|M{1.9cm}|m{2.1cm}|m{2.1cm}|m{4.3cm}|N}
        \cline{2-4}
        \multicolumn{1}{l|}{\textit{Scenario 1}} & \thresholdtext{< 10\%}{failure rate} & \thresholdtext{< 10ms}{response time} & \optimizationtext{min}{cost} &\\[8pt]\hline
        \textbf{ML algorithm} & \multicolumn{2}{l|}{SGD Classifier} & Passive Aggressive Regressor &\\\hline
        \textbf{Loss function} & \multicolumn{2}{l|}{Hinge} & Squared Epsilon Insensitive &\\\hline
        \textbf{Penalty function} & \multicolumn{2}{l|}{l1} & N/A &\\\hline
        \textbf{Scaler type} & \multicolumn{2}{l|}{None} & None &\\\hline
        \textbf{Exploration rate} & \multicolumn{2}{l|}{5\%} & 5\% &\\\hline
        \textbf{Warm-up count} & \multicolumn{2}{l|}{60} & 60 &\\\hline
        
        \textbf{Metrics} & \multicolumn{2}{p{4.4cm}|}{
            \begin{tabular}{@{}l@{\hspace{8pt}}l@{\hspace{5pt}}l@{\hspace{-11.7pt}}}
                \textit{F1-score}: & 0.895 & \interval{0.000}{0.895}\\
                \textit{MCC}: & 0.812 & \interval{-0.112}{0.812}\\
            \end{tabular}
        } & \begin{tabular}{@{}l@{\hspace{8pt}}l@{\hspace{5pt}}l@{\hspace{-6.5pt}}}
                \textit{R2-score}: & 0.906 & \interval{$\text{-}7.8e^{23}$}{0.908} \\
                \textit{MSE}: & 1.753 & \interval{1.706}{$1.4e^{25}$} \\
                \textit{MAE}: & 0.901 & \interval{0.859}{$2.7e^{12}$} \\
                \textit{ME}: & 5.981 & \interval{5.981}{$1.3e^{13}$} \\
            \end{tabular}
        &\\\hline
    \end{tabular}
    
    \bigskip
    
    \begin{tabular}{|M{1.8cm}|m{3.5cm}|m{3.7cm}|m{3.7cm}|N}
        \cline{2-4}
        \multicolumn{1}{l|}{\textit{Scenario 2}} & \thresholdtext{< 10\%}{failure rate} & \setpointtext{10ms, \epsilon=0.25ms}{response time} & \optimizationtext{min}{cost}  &\\[8pt]\hline
        \textbf{ML algorithm} & SGD Classifier & Passive Aggressive\newline Regressor & Passive Aggressive\newline Regressor &\\\hline
        \textbf{Loss function} & Hinge & Squared Epsilon\newline Insensitive &  Epsilon Insensitive &\\\hline
        \textbf{Penalty function} & Elasticnet & N/A & N/A &\\\hline
        \textbf{Scaler type} & Standard Scaler & None & None &\\\hline
        \textbf{Exploration rate} & 5\% & 5\% & 5\% &\\\hline
        \textbf{Warm-up count} & 60 & 60 & 60 &\\\hline
        
        \textbf{Metrics} & 
            \begin{tabular}{@{\hspace{-2pt}}l@{\hspace{7pt}}l@{\hspace{3pt}}l@{\hspace{-11.7pt}}}
                \textit{F1}: & 0.933 & \interval{0.293}{0.934} \\
                \textit{MCC}: & 0.866 & \interval{-0.028}{0.867} \\
            \end{tabular} & 
            \begin{tabular}{@{\hspace{-2pt}}l@{\hspace{7pt}}l@{\hspace{3pt}}l@{\hspace{-6.5pt}}}
                \textit{R2}: & 0.860 & \interval{$\text{-}5.0e^{25}$}{0.868} \\
                \textit{MSE}: & 0.035 & \interval{0.033}{$1.4e^{25}$} \\
                \textit{MAE}: & 0.123 & \interval{0.119}{$2.8e^{12}$} \\
                \textit{ME}: & 0.976 & \interval{0.975}{$1.5e^{13}$} \\
            \end{tabular} &
            \begin{tabular}{@{\hspace{-2pt}}l@{\hspace{8pt}}l@{\hspace{5pt}}l@{\hspace{-6.5pt}}}
                \textit{R2}: & 0.906 & \interval{$\text{-}7.8e^{23}$}{0.908} \\
                \textit{MSE}: & 1.753 & \interval{1.706}{$1.4e^{25}$} \\
                \textit{MAE}: & 0.901 & \interval{0.859}{$2.7e^{12}$} \\
                \textit{ME}: & 5.981 & \interval{5.981}{$1.3e^{13}$} \\
            \end{tabular}
        &\\\hline
    \end{tabular}
\end{table}

\end{document}